\documentclass[11pt]{article}
\usepackage{sint,epsfig,macros_static,cite}
\usepackage{color}
\usepackage{wrapfig}

\begin{document}
\begin{titlepage}

\begin{flushright}
DESY 06-206\\
SFB/CPP-06-51
\end{flushright}

\vskip 0.75cm
\begin{center}
{\Large\bf 
Non-perturbative QCD: \\renormalization, $\Oa$-improvement 
 and \\matching to Heavy Quark Effective Theory \footnote{
Lectures given at ILFTN Workshop on ``Perspectives in Lattice QCD'', Nara, Japan, 31 Oct - 11 Nov 2005.}
\\[0.5ex] 
}
\end{center}
\vskip 1.0cm
\begin{center}
{\large
Rainer Sommer
}
\vskip 1.0cm
Deutsches Elektronen-Synchrotron DESY, Zeuthen \\
Platanenallee~6, D-15738 Zeuthen, Germany\\[1ex]
{\em rainer.sommer@desy.de}
\vskip 0.875cm
{\bf Abstract}
\vskip 0.7ex
\end{center}

We give an introduction to three topics in lattice gauge theory: \\[2ex]
I. The \SF and $\Oa$ improvement.  \\[1ex]
$\Oa$ improvement has been reviewed several times. 
Here we focus on explaining the basic ideas in detail
and then proceed directly to an overview of the literature
and our personal assessment of what has been achieved and what 
is missing. 
\\[2ex]
II. The computation of the running coupling, running quark masses
and the extraction of the renormalization group invariants.   \\[1ex]
We focus on the basic strategy and on the large effort that has been invested in 
understanding the continuum limit. We point out what remains to be done.
\\[2ex]
III. Non-perturbative Heavy Quark Effective Theory. \\[1ex]
Since the literature on this subject is still rather sparse, 
we go beyond the basic ideas and discuss in some detail how 
the theory works in principle and in practice. 

\vskip 0.29cm
\vfill

\begin{center}
 November 2006
\end{center}

\eject
\vfill
\eject

\end{titlepage}

\section{Introduction}

Lattice QCD, the subject of this school, is our prime tool to make
quantitative predictions in the low energy sector of QCD. Also   
connecting this sector to the perturbative high energy regime 
requires non-perturbative control over the theory, which can be
achieved by lattice gauge theories. 
In these lectures the emphasis is on a non-perturbative treatment 
of the theory {\em including} its renormalization. Connecting
the perturbative and the non-perturbative regions is one 
of the main subjects (II.). 

Heavy quarks require special care on a lattice with spacing
$a$, simply because 
their mass is of the order of the cutoff, $a^{-1}$, or higher. 
Effective theories can be used in this situation, in particular
Heavy Quark Effective Theory (HQET) is appropriate for hadrons
with a single heavy quark. 
It allows to compute the expansion of their properties 
in terms of
the inverse quark mass (in practice the b-quark mass).
The renormalization of this effective theory introduces
couplings whose number grows with the order of the expansion.
In order to preserve the predictivity of the theory, these couplings
ought to be determined from the underlying theory, QCD.
Again this step can be seen as the renormalization
of the effective theory. As will be explained, 
non-perturbative precision is required 
if one wants to be able to take the continuum limit
of the lattice effective field theory.
Actually it is a general property of the expansion 
that a $\minv$ correction is only
defined once all parts including the matching are done 
non-perturbatively.

Renormalization is an ultraviolet phenomenon with 
relevant momentum scales of order $a^{-1}$. Since the QCD coupling
becomes weak in the ultraviolet, one may expect to be able to perform 
renormalizations perturbatively, i.e.
computed in a power series in the bare coupling
$g_0^2$ as one approaches the continuum 
limit $a \to 0$.\footnote{
%
% footnote
%
For simplicity we ignore here the cases of mixing of a given operator 
with operators of lower dimension where this statement does not hold.}  
However, one has to care about the following point.
In order to keep the numerical effort of a simulation tractable,
the number of degrees of freedom in the simulation may not be excessively 
large. This means that  the lattice spacing $a$ can not be taken very much smaller than the
relevant physical length scales of the observable that is considered.
Consequently the momentum scale $a^{-1}$ that is relevant for the renormalization is not 
always large enough to justify the truncation of the perturbative series.
In particular one has to remember that the bare coupling vanishes only
logarithmically as $a\to0$: $g_0^2 \sim 1/\log(a\Lambda_\mrm{QCD})$.
In order to obtain a truly non-perturbative answer, the renormalizations 
have to be performed non-perturbatively. 

Depending on the observable,
the necessary renormalizations are of different nature. I will use
this introduction to point out the different types,  
and in particular explain the problem that occurs in a non-perturbative
treatment of scale dependent renormalization. 

\subsection{Basic renormalization: hadron spectrum\label{s_hs}}

The calculation of the hadron spectrum starts by choosing certain values for the bare coupling,
$g_0$, and the bare masses of the quarks in units of the lattice spacing,
$a m_{0,i}$. The flavor index $i$ assumes values $i={\rm u,d,s,c,b}$ for the 
up, down, charm and bottom quarks that are sufficient to describe
hadrons of up to a few $\GeV$ masses.
We ignore the problem of simulating the b-quark for the moment, 
neglect isospin breaking and take 
the light quarks to be degenerate, 
$m_{0,\rm u}=m_{0,\rm d}=m_{0,\rm l}$.

Next, from MC simulations of suitable correlation functions, one computes 
masses of five different hadrons $H$, e.g. $H={\rm p},\pi,{\rm K,D,B}$
for the proton, the pion and the K-,D- and B-mesons,
\bes
  a m_H = a m_H(g_0,a m_{0,\rm l}, a m_{0,\rm s}, 
                 a m_{0,\rm c},a m_{0,\rm b}) \enspace .      \label{hadrons}
\ees
The theory is renormalized by first
setting $m_{\rm p}=m_{\rm p}^{\rm exp}$, where 
$m_{\rm p}^{\rm exp}$ is the experimental value of the proton mass. This determines the lattice spacing via
\bes
  a= (a m_{\rm p}) / {m_{\rm p}^{\rm exp}} \enspace .      \label{spacing}
\ees
Next one must 
choose the parameters $a m_{0,i}$ such that (\ref{hadrons}) is 
indeed satisfied
with the experimental values of the meson masses. Equivalently, one
may say that at a given value of $g_0$ one fixes the bare quark masses
from the condition
\bes 
(a m_H) / (a m_{\rm p}) = m_H^{\rm exp} / m_{\rm p}^{\rm exp} \, , \quad 
       H=\pi,{\rm K,D,B} 
                                           \enspace .      \label{mesons}
\ees
and the bare coupling $g_0$ then determines the value
of the lattice spacing through \eq{spacing}.

After this {\it renormalization}, namely {\it the elimination of the bare parameters
in favor of physical observables}, the theory is completely defined and
predictions can be made. E.g. the leptonic decay constant, $\fpi$, of the pion can be
determined,
\bes 
 \fpi=a^{-1} [a \fpi][1 + \Oa] \enspace .    \label{e:fpia}
\ees 
For the rest of this section, I assume that the bare parameters
have been eliminated  and consider the additional renormalizations
of more complicated observables.

\paragraph{Note.} Renormalization as described here is done without any reference to perturbation theory.
One could in principle use the perturbative formula for $(a \Lambda_\mrm{QCD})(g_0)$
for the renormalization of the bare coupling, where $\Lambda_\mrm{QCD}$ denotes the
$\Lambda$-parameter of the theory (in some scheme). Proceeding in this way, one obtains
a further prediction namely  $m_{\rm p}/\Lambda_\mrm{QCD}$ but at the price of 
introducing $\rmO(g_0^2)$ errors in the prediction of the observables.
As mentioned before, such errors decrease very slowly as one performs 
the continuum limit. A better method to compute the $\Lambda$-parameter
will be discussed later.

\subsection{Scale dependent renormalization and fundamental parameters of QCD\label{s_sdr}}
As we take the relevant length scales in correlation functions 
to be small  or take the energy scale in 
scattering processes to be high, QCD is better and better
approximated by weakly 
coupled quarks and gluons. The strength of the interaction may be 
measured for instance by the ratio of 
the production rate of three jets to the rate for two jets in high energy
$e^+ ~ e^-$ collisions\footnote{One should really use some rather 
inclusive process, e.g. one computable directly in the Euclidean theory. 
For explaining the principle we ignore this issue.} 
\bes
\alpha(\mu) &\propto& 
 {\sigma(e^+ ~ e^- \to q ~ \bar{q} ~g) \over \sigma(e^+ ~ e^- \to q ~ \bar{q})}
  \, , \quad \mu^2=q^2=(p_{e^-}+p_{e^+})^2 \gg  10 \GeV^2 \enspace . \label{e_jets}
\ees
We observe the following points. 
  \begin{itemize}
     \item{The perturbative renormalization group tells us that $\alpha(\mu)$       
          decreases logarithmically with growing energy $\mu$. In other words
           the renormalization from the bare coupling to a renormalized one
           is logarithmically scale dependent.} 
     \item{Different definitions of $\alpha$ are possible; but with 
           increasing energy, $\alpha$ 
           depends less and less on the definition (or the process).} 
     \item{In a similar way, one may define running quark masses $\mbar$ 
           from combinations of observables       
           at high energies.}    
     \item{Using a suitable definition (scheme), the $\mu$-dependence of 
           $\alpha$ and
           $\mbar$ can be determined non-perturbatively and at high energies
           the short distance parameters $\alpha$ and $\mbar$ can be converted
            to the renormalization group invariants using perturbation theory 
            in $\alpha$. Being defined non-perturbatively, the latter are the
	   natural fundamental parameters of QCD.}
  \end{itemize}
Explaining these points in detail is the main objective of the second lecture.

\subsection{Irrelevant operators \label{s_Io}}
Another category of renormalization is associated with the removal of 
lattice discretization errors such as the linear $a$-term in \eq{e:fpia}.
Following Symanzik's improvement program, this can be achieved order by order
in the lattice spacing by adding irrelevant operators, i.e. operators
of dimension larger than four, to the lattice Lagrangian~\myref{impr:Sym1}. 
The coefficients of these operators are easily determined at 
tree level of perturbation theory,
but in general they need to be renormalized. 
We will explain the general idea of the non-perturbative determination
of the coefficients arising at order $a$
and then briefly review the present status of $\Oa$ improvement. 

Note also the alternative approach of removing lattice artifacts 
order by order in the coupling constant but non-perturbatively in the 
lattice spacing $a$ described
in the lectures by Peter Hasenfratz. % \cite{nara:peter}. 
Linear effects in $a$ are automatically absent if the lattice regularization
has enough chiral symmetry. Indeed
chiral symmetry can be kept exactly in the discretized
theory~\cite{exactchi:neub,exactchi:perfect,exactchi:martin},
but these theories are rather expensive to simulate. On the other hand
also the ``twisted mass'' regularization~\cite{tmqcd:pap1,tmqcd:pap2} 
is automatically\footnote{``Automatically'' still means
that the standard mass term has to be tuned to zero, but that can
be done by the use of the PCAC relation.} $\Oa$-improved\cite{tmqcd:FR1} 
(see the appendix of \cite{tmqcd:FMPR} and Stefan Sint's lectures at this school
for a simple argument), but at the price of the violation
of isospin symmetry.  

\subsection{Heavy Quark Effective Theory \label{s_IHQET}}

This theory is very promising for B-physics. It 
approximates heavy-light bound state properties systematically
in an expansion of $\Lambda_\mrm{QCD}/\mbeauty$, a small expansion parameter.
A non-trivial issue is the renormalization of the theory. Already 
at the lowest order in $\minv$, the associated uncertainties are significant
if renormalization is treated perturbatively.  
At that order renormalization can be carried 
out by the methods 
discussed in the second lecture\cite{zastat:pap1,zastat:pap3,stat:zbb_pert}, 
but when one includes $\rmO(\minv)$ corrections one has to deal in addition
with the mixing of operators of different dimensions.\footnote{
Note that the computation of an order $\Lambda_\mrm{QCD}$ term in
the renormalized quark mass already constitutes a $\rmO(\minv)$ correction
to the leading term, although it is done in static approximation.} 
The continuum limit of the effective theory then exists only if the
{\em power divergent mixing coefficients} are
computed non-perturbatively. 

In the third lecture we will 
explain these issues in detail. 
We will formulate HQET non-perturbatively. The 
power divergent mixing coefficients can then be
determined by matching the theory to QCD. A 
possible strategy will be explained.
As an example we will show the computation of the b-quark mass
including $\minv$ corrections.

%%% Local Variables: 
%%% mode: latex
%%% TeX-master: "sect_Lambda"
%%% End: 

\lecture{I}{The \SF and $\Oa$-improvement of lattice QCD}
\renewcommand\thesection      {I.\arabic{section}}

\section{The Schr\"odinger functional (SF)\label{s:SF}}

For various applications, for instance scale dependent renormalization in QCD,
$\Oa$-improvement and Heavy Quark Effective Theory, 
we need QCD in a finite volume with boundary conditions
suitable for (easy) perturbative calculations and MC simulations. 
These are provided by the SF of QCD, which we
introduce below. For a while we restrict the discussion to the 
pure gauge theory. In this part 
the presentation follows closely \cite{SF:LNWW}; we refer to this work
for further details as well as proofs of the properties described below. 
 
%%%%%%%%%%%%%%%%%%%%%%%%%%%%%FIGURE%%%%%%%%%%%%%%%%%%%%%%%%%%%%%%%%%%%
\begin{figure}[ht]
\centerline{
\psfig{file=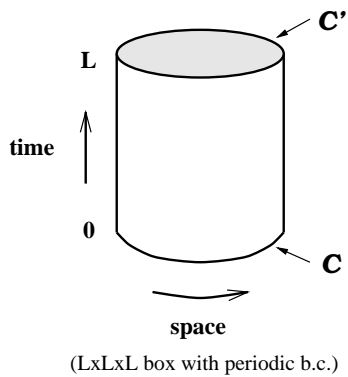,%
width=4.5cm}
}
\vspace{-0.0cm}
\caption{Illustration of the Schr\"odinger functional.\label{f_SF}}
\end{figure}
%%%%%%%%%%%%%%%%%%%%%%%%%%%%%%%%%%%%%%%%%%%%%%%%%%%%%%%%%%%%%%%%%%%%%%

\subsection{Definition}

Here, we give a formal definition of the SF in the
Yang-Mills theory in continuum space-time,
noting that a rigorous treatment is possible in the lattice regularized 
theory.

Space-time is taken to be a cylinder illustrated in \fig{f_SF}.
We impose  Dirichlet boundary conditions for the vector 
potentials\footnote{We use anti-hermitian vector potentials.
}
in time,
\bes
  A_k(x)&=& \left\{ \begin{array}{ll}
             \bvalue_k^{\Lambda}({\bf x}) & \mbox{at} \quad x_0=0 \\
             \bvalue'_k({\bf x})          & \mbox{at} \quad x_0=L
             \end{array}
          \right.  \enspace ,  \label{e_bcA}
\ees
where $C$, $C'$ are classical gauge potentials and $A^{\Lambda}$
denotes the gauge transform of $A$,
\bes          
  A_k^{\Lambda}({\bf x})&=&
  \Lambda({\bf x})A_k({\bf x})\Lambda({\bf x})^{-1}+
  \Lambda({\bf x})\partial_k\Lambda({\bf x})^{-1}, 
   \qquad \Lambda \in \SU \enspace . \label{e_gaugetrafo}
\ees
In space, we impose periodic boundary conditions,
\bes         
 A_k(x+L\hat k)&=& A_k(x), \qquad   \Lambda({\bf x}+L\hat k)= \Lambda({\bf x}) 
 \enspace .     
\ees
The (Euclidean) partition function with these boundary conditions 
defines the SF,
\bes
  { \schrodinger[\bvalue',\bvalue]} &\equiv&
  \int\rmD[\Lambda]\int\rmD[A]\, 
  \rme^{-\Sg[A]} \enspace,  \label{e_SF}   \label{e_SFdef}\\
  \Sg[A]&=&-{1\over2g_0^2}\int\rmd^4x\,
  \tr\left\{F_{\mu\nu}F_{\mu\nu}\right\}, \nonumber \\ 
  F_{\mu\nu}&=&\partial_{\mu}A_{\nu}-\partial_{\nu}A_{\mu}
  +[A_{\mu},A_{\nu}] \enspace ,\nonumber \\
  \qquad\rmD[A]&=&\prod_{{\bf x},\mu,a}\rmd A_\mu^a({x}),
  \qquad\rmD[\Lambda]=\prod_{\bf x}\rmd\Lambda({\bf x}) \enspace . \nonumber
\ees
Here $\rmd\Lambda({\bf x})$ denotes the Haar measure of $\SU$. 
It is easy to
show that the SF is a gauge invariant functional of the boundary fields,
\bes
  \schrodinger[\bvalue'^{\Omega'},\bvalue^{\Omega}] = 
  \schrodinger[\bvalue',\bvalue] \enspace ,
\ees
where also large gauge transformations are permitted. 
The invariance
under the latter is an automatic property of the SF
defined on a lattice, while in the continuum
formulation it is enforced by the integral over $\Lambda$
in \eq{e_SF}.

\subsection{Quantum mechanical interpretation}

The SF is the quantum mechanical transition amplitude 
from a state $|\bvalue\rangle$ to a state $|\bvalue'\rangle$
after a (Euclidean) 
time $L$.
To explain the meaning of this statement,
we introduce the 
Schr\"odinger representation. The Hilbert space consists of 
wave-functionals $\Psi[A]$ which are functionals of the spatial 
components of the vector potentials,
$A_k^a({\bf x})$. 
The canonically conjugate field variables are represented by functional 
derivatives,
$
  E_{k}^a({\bf x})
  ={1\over i}{\delta\over\delta A_k^a({\bf x})}
$, and a scalar product is given by
\bes
  \langle\Psi|\Psi'\rangle=\int\rmD[A]\,\Psi[A]^*\Psi'[A],
  \qquad\rmD[A]=\prod_{{\bf x},k,a}\rmd A_k^a({\bf x}) \enspace .
\ees
The Hamilton operator,   
\bes
  \ham
  =\int_0^L\rmd^3x\,\left\{
  {g_0^2\over2}E_{k}^a({\bf x})E_{k}^a({\bf x})+
  {1\over4g_0^2}F_{kl}^a({\bf x})F_{kl}^a({\bf x})\right\} \enspace ,
\ees
commutes with the
projector, $\projector$, onto the physical subspace of the 
Hilbert space (i.e. the space of gauge invariant states), where 
$\projector$ acts as 
 \bes
  \projector\psi[A]=\int\rmD[\Lambda]\,\psi[A^\Lambda]
 \enspace .
\ees
Finally, each 
classical gauge field defines a state 
$|\bvalue\rangle$ through
\bes
  \langle \bvalue|\Psi\rangle=\Psi[\bvalue] \enspace .
\ees
After these definitions, the 
quantum mechanical representation of the SF is given by
 \bes
  \schrodinger[\bvalue',\bvalue]&=&
  \langle \bvalue'|\rme^{-\ham T}\projector|\bvalue\rangle \nonumber \\
  &=&
  \sum_{n=0}^{\infty} \rme^{-E_nT}
  \Psi_n[\bvalue'] \Psi_n[\bvalue]^* \enspace . \label{e_QM1}
\ees
In Wilson's original lattice formulation, \eq{e_QM1} can be derived 
rigorously and is valid with real energy eigenvalues $E_n$.

\subsection{Background field \label{s_BF}}

A complementary aspect of the SF is that it allows a treatment
of QCD in a color background field in an unambiguous way.
Let us assume that we have a solution $B$ of the equations of motion,
which satisfies also the boundary conditions \eq{e_bcA}.
If, in addition,
\bes
S[A] > S[{B}] \label{e_minim}
\ees
for all gauge fields $A$ that are not equal to a gauge transform $B^\Omega$
of $B$, then we call $B$ the background field 
(induced by the boundary conditions).
Here, $\Omega(x)$ is a gauge transformation defined for all
$x$ in the cylinder and its boundary and $B^\Omega$ is the
corresponding generalization of \eq{e_gaugetrafo}.
Background fields $B$, satisfying these conditions are known; we 
will describe a particular family of fields, later.

Due to \eq{e_minim}, 
fields close to $B$ dominate the path integral 
for weak coupling $g_0$ and the 
effective  action,
\bes
  \effaction[\bfield] &\equiv&
  -\ln \schrodinger\left[\bvalue',\bvalue\right]  \enspace ,\label{e_eff_act}
\ees
has a regular 
perturbative expansion,  
\bes
  \effaction[\bfield] &=&
 {1 \over g_0^{2}}\effaction_0[\bfield]+
  \effaction_1[\bfield]+g_0^2\effaction_2[\bfield]
  +\ldots  \enspace ,\label{e_eff_act_pt} 
  \\
  \effaction_0[\bfield] &\equiv& g_0^2S[\bfield] \enspace . \nonumber
\ees
Above we have used that due to our assumptions, the background field, 
$\bfield$, 
and the boundary values $\bvalue,\bvalue'$ are in one-to-one correspondence
and have taken $\bfield$ as the argument of $\effaction$.

\subsection{Perturbative expansion}

For the construction of the SF as a renormalization scheme,
one needs to study the renormalization properties of the functional
$\schrodinger$. 
L\"uscher, Narayanan, Weisz and Wolff have performed a one-loop calculation for
arbitrary background field \cite{SF:LNWW}. The calculation is done
in dimensional regularization with 
 $3-2\varepsilon$ space dimensions and one time dimension. 
One expands the field $A$ in terms of the
background field and a fluctuation field, $q$, as
\bes
  A_{\mu}(x)=\bfield_{\mu}(x)+g_0\qfield_{\mu}(x) \enspace .
\ees
Then one 
adds a gauge fixing term (``background field gauge'') and 
the corresponding Fadeev-Popov term. Of course, care must be taken 
about the proper boundary conditions in all these expressions.
Integration over the quantum field and the ghost fields then
gives
\bes
  \effaction_1[\bfield]=\frac{1}{2}\ln\det\deltaonehat-
  \ln\det\deltazerohat \enspace ,
\ees
where $\deltaonehat$ is the fluctuation operator 
and $\deltazerohat$ the Fadeev-Popov operator defined in
\cite{SF:LNWW}.
The result can be cast in the form
\bes
  \effaction_1[\bfield]
  \mathrel{\mathop=_{\varepsilon\to0}}
  &&-{b_0 \over \varepsilon}
  \,\effaction_0[\bfield]
  +\rmO(1) \enspace , %%% \quad b_0={11 N \over 48\pi^2} 
\ees
with the important result that the only (for $\varepsilon\to0$) singular term 
is proportional to $\Gamma_0$.  

After renormalization of the coupling, i.e. the
replacement of the bare coupling by $\gbarMSbar$ via 
\bes
     g_0^2 =&& \bar{\mu}^{2\varepsilon} \gbarMSbar^2(\mu) 
         [ 1+ z_1(\varepsilon)\gbarMSbar^2(\mu)],
         \quad z_1(\varepsilon)=-{b_0 \over \varepsilon} \enspace ,
\ees
the effective action is finite,
\bes
   \effaction[\bfield]_{\varepsilon=0}
  &=&\left\{{ 1\over\gbarMSbar^2}-
  b_0
  \left[\ln\mu^2-\frac{1}{16\pi^2}\right]\right\}
  \effaction_0[\bfield]
  \nonumber \\
  &&-\frac{1}{2}\zetaprime{\deltaone}+\zetaprime{\deltazero}
  +\rmO(\gbarMSbar^2) \\
  \zetaprime{\Delta}&=&
  \left.{\rmd\over\rmd s}\zetafunc{s}{\Delta}\right|_{s=0}, \qquad
   \zetafunc{s}{\Delta}=\Tr\,\Delta^{-s} \enspace . \nonumber
\ees
Here, $\zetaprime{\Delta}$ is a complicated functional of $\bfield$,
which is not known analytically but can be
evaluated numerically for specific choices of $\bfield$.

The important result of this calculation is that (apart from
field independent terms that have been dropped everywhere)
the SF is finite after eliminating $g_0$ in favor of $\gbarMSbar$.
The presence of the boundaries does {\it not} introduce any
extra divergences. In the following section we argue 
that this property is correct in general, not just in
one-loop approximation.

\subsection{General renormalization properties}

The relevant question here is whether local quantum field
theories formulated on space-time manifolds {\it with boundaries}
develop divergences that are not present in the absence
of boundaries (periodic boundary conditions or infinite space-time).
In general the answer is ``yes, such additional divergences 
exist''.
In particular, Symanzik  studied the $ \phi^4$-theory with 
\SF boundary conditions \cite{SF:Sym,SF:martin}.
He presented arguments that to all orders of perturbation theory 
 the \SF is finite after
\begin{itemize}
 \item{renormalization of the self-coupling, $\lambda$,  and the
       mass, $m$,}
 \item{{\it and} the addition of the boundary counter-terms
\bes
  \int_{x^0=T}\rmd^3x\,
  \left\{Z_1\phi^2+Z_2\phi\partial_0\phi\right\}+
  \int_{x^0=0}\rmd^3x\,
  \left\{Z_1\phi^2-Z_2\phi\partial_0\phi\right\} \enspace .
\ees
  }
\end{itemize}

In addition to the standard renormalizations,
one has to add counter-terms formed by local composite fields 
integrated over the boundaries.
One expects that in general, all fields with dimension 
$d\leq3$ have to be taken into account.
Already Symanzik conjectured that counter-terms with this property are
sufficient to renormalize the SF of any quantum field theory
which is renormalizable when no boundaries are present. 

Since this conjecture forms
the basis for many applications of the  SF, 
we note a few points concerning its status.
\begin{itemize}
 \item{As mentioned, a proof to all orders of perturbation theory 
       does not exist. An application of power counting in momentum space 
       in order to prove the 
       conjecture is not
       possible due to the missing translation invariance.}
 \item{There is no gauge invariant local field with $d\leq3$ in the  Yang--Mills 
       theory. Consequently no additional
       counter-term is necessary in accordance with the 1-loop result described in 
       the previous
       subsection.}
 \item{In QCD it has been checked also by
       explicit 2--loop calculations~\myref{pert:2loop_SU2,pert:2loop_fin}. 
       MC simulations
       give further support for its validity beyond perturbation theory;
	we give examples in
       the second lecture.
       }
 \end{itemize}
Although a general proof is missing, there is little doubt that Symanzik's 
conjecture is valid in general. Concerning QCD, 
this puts us into the position 
to give an elegant definition of
a renormalized coupling in finite volume.

\subsection{Renormalized coupling \label{s:alphasf}}

For the definition of a running coupling we need a quantity which
depends only on one scale. We choose $L B$ such that it
depends only on one dimensionless variable $\bfieldparm$. 
In other words, the strength of the field is scaled as $1/L$.
The background field is assumed to fulfill the requirements of
\sect{s_BF}.
Then, following the above discussion, the derivative
\bes
  \effaction'[\bfield]=
  {\partial\over\partial\bfieldparm}
  \effaction[\bfield] \enspace ,
\ees
is finite when it is expressed in terms of a renormalized coupling
like $\gbarMSbar$ but $\effaction'$ is defined 
non-perturbatively.
From \eq{e_eff_act_pt} we read off immediately that a properly
normalized coupling is given by
\bes 
 \gbar^2(L)=\effaction'_0[\bfield]\bigm/\effaction'[\bfield] \enspace .
  \label{e_gbarsf}
\ees
Since there is only one length scale $L$, it is evident that $\gbar$
defined in this way runs with $L$.

A specific choice for the gauge group $\SUthree$ is
the abelian background field induced by the boundary values~\myref{alpha:SU3}
\bes
  C^{}_k = \frac{i}{L} \left(
          \begin{array}{ccc} 
               \phi^{}_1 & 0          & 0        \\
                0        & \phi^{}_2  & 0        \\      
                0        & 0          & \phi^{}_3 
         \end{array} \right) \, ,\quad
  C'_k = \frac{i}{L}
         \left( \begin{array}{ccc}
               \phi'_1   & 0          & 0        \\
                0        & \phi'_2    & 0        \\      
                0        & 0          & \phi'_3 
         \end{array} \right) \, ,
 \quad k=1,2,3, 
 \label{e_abelian}       
\ees                      
with
\bes
  \begin{array}{lll}
  \phi^{}_1  = \eta-\frac{\pi}{3},      
    &\quad& \phi'_1 = -\phi^{}_1-\frac{4\pi}{3},  \\[1ex]
  \phi^{}_2  = -\frac12 \eta,       
    &\quad& \phi'_2 = -\phi^{}_3+\frac{2\pi}{3},   \\[1ex]
  \phi^{}_3  = -\frac12\eta+\frac{\pi}{3}, 
    &\quad&\phi'_3  = -\phi^{}_2+\frac{2\pi}{3}.   
    \end{array}
     \label{e_bflds}
\ees
In this case, the derivatives with respect to 
$\eta$ are to be evaluated at $\eta=0$. The 
associated background field,
 \begin{equation}
  B_0=0,\qquad B_k=\left[x_0 C_k' + (L-x_0) C_k^{}\right]/L,
  \quad k=1,2,3 \enspace ,\label{e_BF}
\end{equation}
has a field tensor with  non-vanishing components
\bes
 G_{0k}=\partial_0 B_k=(C_k'-C_k)/L,\quad k=1,2,3 \enspace .\label{e_G0k}
\ees
It is a constant color-electric field. 

\subsection{Quarks \label{s_Q}}

In the end, the real interest is in the renormalization of
QCD and we need to consider the SF with  quarks.
We restrict our discussion to the original
formulation of S.~Sint~\cite{SF:stefan1}.

Special care has to be taken in formulating the 
Dirichlet boundary conditions for the quark fields; since the Dirac operator
is a first order differential operator, the Dirac equation has a unique 
solution when one half of the components of the fermion fields are
specified on the boundaries. Indeed, a detailed investigation shows
that the boundary condition
\bes
        P_+\psi|_{x_0=0} &=& \rho, \quad P_- \psi|_{x_0=L} = \rho'\, ,
          \qquad P_\pm = \frac{1}{2}(1\pm\gamma_0) \, ,\\
       \psibar P_-|_{x_0=0} &=& \rhobar, 
        \quad \psibar P_+|_{x_0=L} = \rhobar' \, ,
\ees
lead to a quantum mechanical 
interpretation analogous to \eq{e_QM1}. The SF 
\bes
  {\cal Z}[C',\rhobarprime,\rhoprime; C,\rhobar,\rho]=
  \int\rmD[A]\rmD[\,\psi\,]\rmD[\,\psibar\,]\,\rme^{-S[A,\psibar,\psi\,]}
  \label{e_sfqcd}
\ees 
involves an integration over all fields with the specified boundary
values.
The full action may be written as
\bes
S[A,\psibar,\psi\,]&=&\Sg[\psibar,\psi\,]+\Sf[A,\psibar,\psi\,] \nonumber \\ 
\Sf &=& \int \rmd^4 x \, \psibar(x)  [\gamma_\mu D_\mu +m]  \psi(x) 
    \label{e_fermact} \\
    && - \int \rmd^3 {\bf x} \,[ \psibar(x) P_- \psi(x)]_{x_0=0}
       - \int \rmd^3 {\bf x} \,[ \psibar(x) P_+ \psi(x)]_{x_0=L}
       \enspace , \nonumber
\ees
with  $\Sg$ as given in \eq{e_SFdef}. In \eq{e_fermact} we use standard 
Euclidean $\gamma$-matrices. The covariant derivative, $D_\mu$, acts as
$D_\mu \psi(x) = \partial_\mu \psi(x) + A_\mu(x) \psi(x)$.

Let us now discuss the renormalization of the SF with quarks.
In contrast to the pure Yang-Mills theory, gauge invariant 
composite fields of dimension three are present in QCD.
Taking into account the boundary conditions 
one finds \myref{SF:stefan1} that the counter-terms,
\bes
    \psibar P_- \psi|_{x_0=0}\, \, {\rm and} \, \, \psibar P_+ \psi|_{x_0=L} 
    \enspace ,
\ees
have to be added to the action with 
weight $1-Z_{\rm b}$ to obtain a finite
renormalized functional. These counter-terms 
are equivalent to a multiplicative renormalization of the
boundary values, 
\bes
  \rho_{\rm R} = Z_{\rm b}^{-1/2} \rho \, , \,\,\ldots \, \,\, , \,
  \rhobar'_{\rm R} = Z_{\rm b}^{-1/2} \rhobar' \enspace .
  \label{Zb}
\ees
It follows that -- apart from 
the renormalization of the coupling and the quark mass -- no
additional renormalization of the SF is necessary for {\it vanishing}
boundary values
$\rho, \ldots,\rhobar'$.
So, after imposing homogeneous boundary conditions for the fermion fields,
a renormalized coupling may be defined as in the previous subsection.

As an important aside,  we point out that the boundary conditions for the 
fermions
introduce a gap into the spectrum of the Dirac operator (at least for
weak couplings). One may hence simulate the lattice
SF for vanishing physical quark 
masses. It is then convenient to supplement the definition of the renormalized
coupling by the requirement $m=0$. In this way, one defines a mass-independent
renormalization scheme with simple renormalization group equations.
In particular, the $\beta$-function remains independent of the quark
mass.

\subsubsection{ Correlation functions}
are given in terms of the expectation values of
any product $\cal O$ of fields,
\bes
  \langle{\cal O}\rangle=\left\{{1\over{\cal Z}}
  \int\rmD[A]\rmD[\,\psi\,]\rmD[\,\psibar\,]\,{\cal O}\,
  \rme^{-S[A,\psibar,\psi\,]}\right\}_
  {\rhobarprime=\rhoprime=\rhobar=\rho=0} \enspace,
\ees
evaluated for vanishing boundary values
$\rho, \ldots,\rhobar'$. Apart from the gauge field and the quark and 
anti-quark fields integrated over, $\cal O$ may involve the 
``boundary fields"~\myref{impr:pap1}
\bes
  \zeta({\bf x})&=&{\delta\over\delta\rhobar({\bf x})},
  \qquad
  \zetabar({\bf x})=-{\delta\over\delta\rho({\bf x})},
  \nonumber \\
  \zeta'({\bf x})&=&{\delta\over\delta\rhobarprime({\bf x})},
  \qquad
  \zetabarprime({\bf x})=-{\delta\over\delta\rhoprime({\bf x})} \enspace .
\ees
An application of fermionic correlation functions 
including the boundary fields
is the definition of the renormalized quark mass in the
SF scheme to be discussed next.

\subsubsection{Renormalized mass \label{s_Rm}}

Just as in the case of the coupling constant, there is a great freedom in 
defining renormalized quark masses. A natural starting point 
is the PCAC relation which expresses the divergence of the axial current
\footnote{The reader is not to confuse $A_\mu^a(x)$, with the gauge vector potentials
$A_\mu(x)$.}
,
\bes            
  A_\mu^a(x) &=& 
            \psibar(x)\dirac{5}\dirac{\mu}\frac{1}{2}\tau^a\psi(x) \enspace ,
  \label{e_Amu}          
\ees
(for simplicity we have chosen just
$\nf=2$ degenerate flavors and
the Pauli matrix $\tau^a$ acts in this flavor space),
in terms of the associated pseudo-scalar density,
\bes            
  P^a(x) &=& 
            \psibar(x)\dirac{5}\frac{1}{2}\tau^a\psi(x) \enspace ,
  \label{e_density}          
\ees
via
\bes
 \partial_\mu A_{\mu}^a(x) = 2 m P^a(x)\enspace .
 \label{e_PCAC}
\ees
This operator identity is easily derived at the classical level 
(cf. \sect{s_curr}).
After renormalizing the operators,
\bes
 (\ar)_{\mu}^a &=& \za A_{\mu}^a \enspace ,\nonumber \\
  \pr^a &=& \zp P^a \enspace , \label{e_zazp}
\ees
a renormalized current quark mass may be defined by
\bes
 \mr=\mbar = {\za \over \zp } m \enspace . 
 \label{e_mbar}
\ees
Here, $m$, is to be taken from \eq{e_PCAC} inserted into an arbitrary 
correlation function and $\za$ can be determined from a proper
chiral Ward identity \cite{Boch,impr:pap4,impr:za_nf2}.
Note that $m$ does not
depend on which correlation function is used because the PCAC relation is an 
operator identity.
The
definition of $\mbar$ is completed by supplementing
\eq{e_zazp} with a specific normalization
condition for the pseudo-scalar density. The running
mass $\mbar$ then inherits its scheme-
and scale-dependence ($\mu$) from the corresponding dependence of
$\pr$. Such a normalization condition may be imposed through infinite
volume correlation functions. Since we want to be  able to compute
$\mbar(\mu)$ for large energy scales $\mu$,  we do, however, need a finite volume
definition (see \sect{s_Rls}). This is readily given in terms of
correlation functions in the SF.
%%%%%%%%%%%%%%%%%%%%%%%%%%%%%FIGURE%%%%%%%%%%%%%%%%%%%%%%%%%%%%%%%%%%%
\begin{figure}[ht]
\vspace{ -0.0cm}
\centerline{
\psfig{file=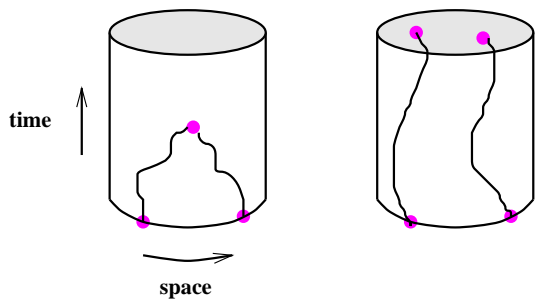,%
width=7cm}}
\vspace{-0.0cm}
\caption{$\fp$ (left) and $f_1$ (right) in terms of quark propagators.
\label{f_matrixelements}}
\end{figure}
%%%%%%%%%%%%%%%%%%%%%%%%%%%%%%%%%%%%%%%%%%%%%%%%%%%%%%%%%%%%%%%%%%%%

To start with, let us define (iso-vector) pseudo-scalar fields at the 
boundary of the SF,
\bes
  \op{}^a &=& \int \rmd^3 {\bf u} \int \rmd^3 {\bf v}
  \,\,\zetabar({\bf u})\dirac{5}\frac{1}{2}
  \tau^a\zeta({\bf v}), 
  \nonumber \\
  \opprime{}^a &=& \int \rmd^3 {\bf u} \int \rmd^3 {\bf v}
  \,\,\zetabarprime({\bf u})\dirac{5}\frac{1}{2}
  \tau^a\zeta'({\bf v}) 
  \enspace,  \label{e_boundops}
\ees
to build up the correlation functions 
\bes
 \fp(x_0) &=& - \frac13 \langle P^a(x)  \op{}^a\rangle \enspace ,\nonumber \\
 f_1 &=& \langle\opprime{}^a\op{}^a\rangle \enspace ,
\ees
which are illustrated in \fig{f_matrixelements}.

We then form the ratio 
\bes
\zp = {\rm const.} \sqrt{f_1} / \fp(x)|_{x_0=L/2}  
\enspace , \label{e_zp}
\ees
such that the renormalization of the boundary quark fields, \eq{Zb}, 
cancels out. 
The proportionality
constant is to be chosen such that $\zp=1$ at tree level. To define the scheme
completely one needs to further specify the boundary values $C,C'$ and the
boundary conditions for the quark fields in space. These details 
are of no importance, here. 

We rather mention some more basic points about this
renormalization scheme. 
Just like in the case of the
running coupling, the only physical scale that exists in our definitions 
\eq{e_mbar},~\eq{e_zp} is the linear dimension of the SF, the length scale, $L$.
So the mass $\mbar(\mu)$ runs with $\mu=1/L$.
We have already emphasized
that $\gbar$ is to be evaluated at zero quark mass. It is
advantageous to do the same for $\zp$. In this way
we define a mass-independent renormalization scheme, with simple
renormalization group equations.

By construction, the SF scheme is non-perturbative
and independent of a specific regularization. For a concrete non-perturbative
computation, we do, however, need to evaluate the expectation values
by a MC-simulation of the corresponding lattice theory. We proceed
to introduce the lattice formulation of the SF.

\subsection{Lattice formulation \label{s:latt}}
A detailed 
knowledge of the form of the lattice action is 
not required for an understanding of the following sections. 
Nevertheless, we 
give a definition of the SF in lattice regularization. This is done
both for completeness and
because it allows us to obtain a first impression about the
size of discretization errors.

We choose a hyper-cubic Euclidean lattice with spacing $a$.
A gauge field $U$ on the lattice is an assignment of a matrix 
$U(x,\mu)\in\SUn$ to every lattice point $x$ and direction 
$\mu=0,1,2,3$.
Quark and anti-quark fields, $\psi(x)$ and $\psibar(x)$,
reside on the lattice sites and  
carry Dirac, color and flavor indices as in the continuum.
To be able to write the quark action in
an elegant form it is useful to 
extend the fields, initially defined only inside the SF manifold (cf. \fig{f_SF})
to all times $x_0$ by ``padding" with zeros.
In the case of the quark field one sets
$$
  \psi(x)=0\quad\hbox{if $x_0<0$ or $x_0>L$},
$$
and
$$
  P_{-}\psi(x)|_{x_0=0}=
  P_{+}\psi(x)|_{x_0=L}=0,
$$
and similarly for the anti-quark field.
Gauge field variables that reside outside the manifold are set to 1.

We may then write the fermionic action as a sum
over all space-time points without restrictions for the time-coordinate,
\bes 
  \Sf[U,\bar\psi,\psi] &=& a^4\sum_{x}\bar\psi(D+m_0)\psi , \label{e_Sf}
\ees
and with the
standard Wilson-Dirac operator,
\begin{equation}
  D = \frac12 \sum_{\mu=0}^3
  \{\gamma_\mu(\nabla_\mu^\ast+\nabla_\mu^{})- a\nabla_\mu^\ast\nabla_\mu^{}\} \, .
\label{e_Dlat}
\end{equation}
Here, forward and backward covariant derivatives, 
\begin{eqnarray}
  \nabla_\mu^{}\psi(x)     &=& \frac1a [U(x,\mu)\psi(x+a\hat\mu)-\psi(x)],\\
  \nabla_\mu^{\ast}\psi(x) &=&
   \frac1a [\psi(x)-U(x-a\hat\mu,\mu)^{-1}\psi(x-a\hat\mu)] \enspace ,
\end{eqnarray}
are used and $m_0$ is to be understood as a diagonal matrix in flavor space
with elements $m_0^f$.

The gauge field action $\Sg$ is a sum over all oriented 
plaquettes $p$ on the lattice, with the weight factors $w(p)$, and 
the parallel transporters $U(p)$  around $p$,  
\bes
 \Sg[U]={1\over g_0^2}\sum_p w(p)\,\tr\{1-U(p)\} \enspace.
 \ees
The weights 
$w(p)$ are 1 for plaquettes in the interior and 
\begin{equation}
 w(p)=
 \left\{ 
 \begin{array}{ll}
         {\scriptstyle \frac12} c_s &
      \mbox{if $p$ is a spatial plaquette at $x_0=0$ or $x_0=L$},\\
      c_t &
     \mbox{if $p$ is time-like and attached to a boundary plane.} 
   \end{array}
   \right.
\end{equation} 
The choice $c_s=c_t=1$ corresponds to the standard Wilson action.
However, these parameters can be tuned in order to reduce lattice artifacts,
as will be briefly discussed below.

With these ingredients, the  
path integral representation of the Schr\"odinger functional 
reads~\myref{SF:stefan1},
\bes
  {\cal Z} &=& \int{\rm D}[\psi]{\rm D}[\bar\psi\,]{\rm D}[U]\,
                 {\rm e}^{-S}\,, \quad S=\Sf+\Sg \enspace , \\
           &&{\rm D}[U]  = \prod _{x,\mu} \rmd U(x,\mu) \enspace , \nonumber  
\ees
with the Haar measure $\rmd U$.

\subsubsection{Boundary conditions and the background field.}

The boundary conditions for the lattice gauge fields 
may be obtained from the continuum boundary values by forming the 
appropriate parallel transporters from $x+a\hat{k}$ to $x$ 
at $x_0=0$ and $x_0=L$. 
For the constant abelian boundary
fields $C$ and $C'$
that we considered before, they are simply
\begin{equation}
  U(x,k)|_{x_0=0}=\exp(aC_k^{}),\qquad U(x,k)|_{x_0=L}=\exp(aC'_k),
  \label{e_latbc}
\end{equation}
for $k=1,2,3$.
All other boundary conditions are as in the continuum.

For the case of \eq{e_abelian},\eq{e_bflds}, 
the boundary conditions~(\ref{e_latbc}) lead to 
a unique (up to gauge transformations) minimal action configuration $V$,
the lattice background field. It can be expressed 
in terms of $B$~(\ref{e_BF}), 
\begin{equation}
  V(x,\mu)=\exp\left\{a B_\mu(x)\right\} \enspace .
\end{equation}

\subsubsection{Lattice artifacts.}

Now we want to get a first impression about the dependence of the lattice
SF on the value of the lattice spacing. In other words we study
lattice artifacts.
At lowest order in the bare coupling we have, just like in the 
continuum,
\bes
  \effaction = 
 {1 \over g_0^{2}}\effaction_0[\bfieldlat]+ \rmO((g_0)^0)
  \, , \quad
  \effaction_0[\bfieldlat] \equiv g_0^2 \Sg[\bfieldlat] \enspace . 
  \label{e_eff_act_pt_lat}
\ees
Furthermore one easily finds the action for small lattice spacings, 
\bes
  \Sg[\bfieldlat]&=&\left[1+(1-\ct) \frac{2a}{L}\right] %%\times
  {3 L^{4}\over g_0^2}
  \sum_{\alpha=1}^N
  \left\{{2\over a^2}\sin\left[{a^2\over2L^2}
  \left(\phi'_{\alpha}-\phi_{\alpha}\right)
  \right]\right\}^2 \nonumber \\
  &=& 
 {3 \over g_0^2}
  \sum_{\alpha=1}^N
  \left(\phi'_{\alpha}-\phi_{\alpha}\right)^2 
  \left[1+( 1-\ct) { \frac{2a}{L} +\rmO(a^4)}\right] \enspace .
  \label{e_act_expanded}
\ees
We observe: at tree-level of 
perturbation theory, all linear lattice artifacts
are removed when one sets $\ct=1$. 
Beyond tree-level (and in the theory without quarks), one has to 
tune the coefficient $\ct$ as a function
of the bare coupling. We will show the effect, when this is done to first
order in $g_0^2$, below. Note that the existence of linear $\rmO(a)$ 
errors in the Yang-Mills theory is special to the SF; they originate
from dimension four operators $F_{0k}F_{0k}$ and $F_{kl}F_{kl}$
which are irrelevant terms (i.e. they carry an explicit factor of
the lattice spacing) when they are integrated over the surfaces.
$\cs$, which can be tuned to cancel the effects of $F_{kl}F_{kl}$,
does not appear for the electric field that we discussed above.

Once quark fields are present, there are more irrelevant operators that can 
generate $\rmO(a)$ effects as discussed in detail in \cite{impr:pap1}. 
Here we emphasize a different feature of \eq{e_act_expanded}:
once the $\rmO(a)$-terms are canceled, the remaining $a$-effects are tiny.
This special feature of the abelian background field is most welcome
for the numerical computation of the running coupling; it allows for
reliable extrapolations to the continuum limit.

\subsubsection{Explicit expression for $\effaction'$.}

Let us finally explain that $\effaction'$ is an observable that can easily
be calculated in a MC simulation. From its definition we find immediately
\bes
{\Gamma'} &=& -{\partial \over \partial\eta}
 \ln  \left\{ \int{ {\rm D}[\psi]{\rm D}[\bar\psi\,]{\rm D}[U]\,
  {\rm e}^{-S} } \right\} 
     =\left\langle {\partial \Sg \over \partial\eta} \right\rangle  
     +\left\langle {\partial \Sf \over \partial\eta} \right\rangle  \enspace .
\ees
The derivative $\frac{\partial \Sg}{\partial\eta}$ evaluates
to the (color 8 component of the) electric field
at the boundary,
\bes
{\partial \Sg \over \partial\eta}&=&
 - {2 \over g_0^2 \, L} a^3 \sum_{\bf x}\left\{E_k^8({\bf x}) - 
                (E_k^8)'({\bf x}) \right\} \enspace , \\
 E_k^8({\bf x})&=&  {1 \over a^2}\Re \tr \left\{ i \lambda_8 U(x,k) U(x+a\hat k,0) 
                  U(x+a\hat 0,k)^{-1} U(x,0)^{-1}
               \right\}_{x_0=0} \enspace ,\nonumber
\ees
where $\lambda_8={\rm diag}(1,-1/2,-1/2)$.
(A similar expression holds for $( E_k^8)'({\bf x}) $). 
The second term ${\partial \Sf \over \partial\eta}$, which is
only non-zero in the $\Oa$-improved formulation is numerically
less relevant. An explicit expression is given in \cite{mbar:nf2}.

The renormalized coupling is related to the expectation value
of a local operator; no correlation function is involved. 
This means that it is easy and fast in computer time
to evaluate it. It further turns out that a 
good statistical precision is reached with a moderate size statistical
ensemble.

\subsection{More literature \label{s:SF_others}}
We here give some guide for further reading on the SF.
Independently of the work of Symanzik,  G.~C.~Rossi
and M.~Testa discussed different boundary conditions imposed 
at fixed time\cite{SF:RT1,SF:RT2}. The renormalization
properties of that functional have not yet been discussed.
 
There are also rather recent developments. Different
formulations of the lattice \SF 
with overlap fermions satisfying the Ginsparg Wilson relation have 
been found by Y.~Taniguchi using an orbifold 
construction~\cite{SF:Taniguchi},
and by M.~L\"uscher using a general universality argument concerning
QFT's with boundaries~\cite{SF:martin}; see also
\cite{SF:Taniguchi2}. 
As the \SF breaks chiral symmetry by the boundary 
conditions, it is relevant into which direction in
flavor space the mass term is introduced. The 
\SF with a twisted mass term  \cite{tmqcd:pap1,tmqcd:pap2}
and the boundary conditions
specified above differs 
from the SF with a standard mass term (at finite quark mass). 
S.~Sint found a modification of the boundary conditions,
which yields the standard SF as the continuum limit of the lattice theory
with a twisted mass~\cite{lat05:stefan}. An even number of flavors 
is required in this formulation. 
It offers also advantages in
the massless limit, where ``automatic bulk $\Oa$-improvement'' is achieved
{\em after the tuning of one counter-term}.
It is discussed in detail
in S.~Sint's lectures at this school.  Another \SF
with automatic bulk $\Oa$-improvement is proposed in~\cite{tmqcd:FR2}. 

%%% Local Variables: 
%%% mode: latex
%%% TeX-master: "Nara"
%%% End: 

\section{Chiral symmetry and $\Oa$-improvement
         \label{s_curr}}

The main focus of this section is on the
$\Oa$ improvement of Wilson's lattice QCD.
However, we also mention the finite normalization of isovector currents. 
Both of these problems have the same
origin, namely that chiral symmetry is broken in Wilson's 
regularization and then also the same solution: 
chiral Ward identities. The possibility to use
these to normalize the currents has first been discussed
by Refs.~\cite{Boch,MaMa}. 
Here, we describe their application
in the computation of the $\Oa$-improved
action and currents. A difference to the aforementioned
work is that we insist that only on-shell improvable correlation functions
are used in the normalization conditions in order to be compatible with 
on-shell improvement.

Before going into more details, we would like to convey the general idea
of the application of chiral Ward identities. 
For simplicity we assume an isospin doublet of mass-degenerate 
quarks. 
Consider first 
a regularization of QCD which preserves the full
SU($2)_{\rm V} \times $SU($2)_{\rm A}$
flavor symmetry as it is
present in the continuum Lagrangian of massless QCD.
Indeed, such regularizations 
exist \cite{exactchi:neub,exactchi:perfect,exactchi:martin}, see
Peter Hasenfratz' lectures.

We can derive chiral Ward identities 
in the Euclidean formulation of the theory. These 
then provide exact relations between different correlation functions.
Immediate consequences of these relations are that there are  
currents $V_\mu^a,\,A_\mu^a$ which do not get renormalized ($\za=\zv=1$)
and the quark mass does not have an additive renormalization.

In a general discretization, such as the Wilson formulation, 
lattice QCD does {\em not} have the SU($2)_{\rm A}$ flavor symmetry for finite
values of the lattice spacing. Then, the Ward identities 
are not satisfied exactly. From universality, we do, however, expect that the 
correlation functions may be renormalized such that they obey the same Ward identities
as before -- up to $\Oa$ corrections.
Therefore we may impose those Ward identities for
the renormalized currents, to fix their normalizations. 

Furthermore, following Symanzik \cite{impr:Sym1}, it suffices to a add
a few local irrelevant terms to the action and to the currents
in order to obtain an improved lattice theory, where the continuum
limit is approached with corrections of order $a^2$.
The coefficients of these terms can be determined 
by imposing improvement conditions. For example one may 
require certain chiral Ward identities to be valid at finite lattice
spacing $a$. 

\subsection{Chiral Ward identities}

For the moment we do not pay attention to a regularization of
the theory and derive the Ward identities in a formal way. As mentioned
above these identities are exact in a regularization that preserves
chiral symmetry. To derive the Ward identities, one starts from the
path integral representation of a correlation function 
and performs the change of integration variables
\bes
  \psi(x) &\to& \rme^{i \frac{\tau^a}{2}
            [\eps^a_{\rm A}(x)\gamma_5 + \eps^a_{\rm V}(x)]} \psi(x) 
    \nonumber \\
         &=& \psi(x) + i  \eps^a_{\rm A}(x)   \da^a\psi(x)
         + i \eps^a_{\rm V}(x) \dv^a\psi(x)\enspace ,
    \nonumber \\ 
  \psibar(x) &\to& \psibar(x) \rme^{i \frac{\tau^a}{2}
            [\eps^a_{\rm A}(x)\gamma_5 - \eps^a_{\rm V}(x)]}  
    \nonumber \\
         &=& \psibar(x) + i  \eps^a_{\rm A}(x)   \da^a\psibar(x)
         + i \eps^a_{\rm V}(x) \dv^a\psibar(x) \enspace ,
\ees
where we have taken $\eps^a_{\rm A}(x), \eps^a_{\rm V}(x)$ infinitesimal
and introduced the variations
\bes
  \dv^a\psi(x)&=&\frac{1}{2}\tau^a\psi(x),
  \qquad\quad\quad
  \dv^a\psibar(x)=-\psibar(x)\frac{1}{2}\tau^a \enspace , 
  \nonumber \\
  \da^a\psi(x)&=&\frac{1}{2}\tau^a\dirac{5}\psi(x),
  \qquad\phantom{\dirac{5}}
  \da^a\psibar(x)=\psibar(x)\dirac{5}\frac{1}{2}\tau^a \enspace .
  \label{e_var_quarks}
\ees
The Ward identities then follow from the invariance of the
path integral representation of correlation functions with respect to such
changes of integration variables. They obtain contributions from
the variation of the action and the variations of the fields
in the correlation functions. 
The variations of the currents
\bes  
V_\mu^a(x) = \psibar(x)\dirac{\mu}\frac{1}{2}\tau^a\psi(x) \label{e_Vmu}
\ees
and $A_\mu^a$, \eq{e_Amu}, 
is given by
\bes
  \dv^aV^b_{\mu}(x)&=&-i\epsilon^{abc}V^c_{\mu}(x),
  \qquad\,
  \da^aV^b_{\mu}(x)=-i\epsilon^{abc}A^c_{\mu}(x),
  \nonumber \\
  \dv^aA^b_{\mu}(x)&=&-i\epsilon^{abc}A^c_{\mu}(x),
  \qquad
  \da^aA^b_{\mu}(x)=-i\epsilon^{abc}V^c_{\mu}(x) \enspace .
  \label{e_var_currents}
\ees
They form a closed algebra under these variations.

%%%%%%%%%%%%%%%%%%%%%%%%%%%%%%%%%%%%%%%%%%%%%%%%%%%%%%%%%%%%%%%%%%%%%%%%%%%%%%%%%%%%%%%%%%
\begin{minipage}{0.35\textwidth}
\includegraphics*[width=0.9\textwidth]{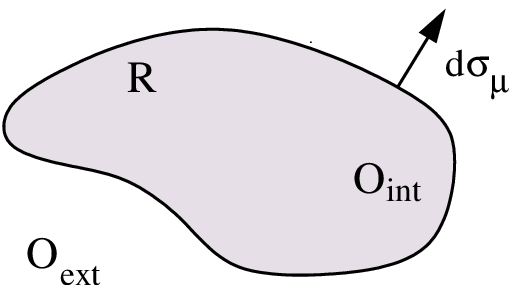}
\end{minipage}
%%%%%%%%%%%%%%%%%%%%%%%%%%%%%%%%%%%%%%%%%%%%%%%%%%%%%%%%%%%%%%%%%%%%%%%%%%%%%%%%%%%%%%%%%%
\hfill
\begin{minipage}{0.57\textwidth}
\vspace{2mm}

Since this is convenient for our applications, we write the
Ward identities in an integrated form. 
Let $R$ be a space-time region with smooth boundary $\partial R$.
Suppose ${\cal O}_{\rm int}$ and ${\cal O}_{\rm ext}$ 
are polynomials in the basic fields localized
in the interior and exterior of $R$ respectively (see left).
The general vector current Ward identity then reads
\vspace{2mm}

\end{minipage}
\bes
  \int_{\partial R}\rmd\sigma_{\mu}(x)\,  
  \left\langle 
  V^a_{\mu}(x) {\cal O}_{\rm int} {\cal O}_{\rm ext} 
  \right\rangle
  =-
  \left\langle 
  \left(\dv^a{\cal O}_{\rm int}\right) {\cal O}_{\rm ext} 
  \right\rangle,
  \label{e_vectorWI}
\ees
while for the axial current one obtains
\bes
  \int_{\partial R}\rmd\sigma_{\mu}(x)\,  
  \left\langle 
  A^a_{\mu}(x) {\cal O}_{\rm int} {\cal O}_{\rm ext} 
  \right\rangle
  &=&-
  \left\langle 
  \left(\da^a{\cal O}_{\rm int}\right) {\cal O}_{\rm ext} 
  \right\rangle  
  \label{e_axialWI} \\
  &&+
  2m\int_R\rmd^4x\,
  \left\langle 
  P^a(x) {\cal O}_{\rm int} {\cal O}_{\rm ext} 
  \right\rangle \enspace .
  \nonumber
\ees
Here volume integrals over for example $\partial_\mu  A^a_{\mu}(x)$
have been changed to surface integrals.
The integration measure $\rmd\sigma_{\mu}(x)$ points 
along the outward normal to the surface $\partial R$
and $P^a(x)$ was defined in \eq{e_density}.

We may also write down  the precise meaning of the PCAC-relation \eq{e_PCAC}. 
It is \eq{e_axialWI} in a differential form, 
\bes
 \left\langle \left[ \partial_{\mu}A_{\mu}^a(x) -2m
  P^a(x)\right]   {\cal O}_{\rm ext} 
  \right\rangle = 0 \enspace , 
  \label{e_PCACnew}
\ees
where now ${\cal O}_{\rm ext}$ may have support everywhere except for at the point
$x$. 

Going through the same derivation in the lattice regularization,
one finds equations of essentially  the same form as the 
ones given above, but with additional terms~\myref{Boch}.
At the classical level these terms are of order $a$. More precisely, 
in \eq{e_PCACnew} the important additional term
originates from the variation of the Wilson term,
$a\,\da^a(\psibar \nabstar{\mu} \nab{\mu} \psi)$, and is a local field of dimension 5.
Such $\Oa$-corrections are present in any observable computed on the lattice
and are no reason for concern. However, as is well known in field theory,
such operators mix with the ones of lower and equal dimensions
when one goes 
beyond the classical approximation.
In the present case, the dimension
five operator mixes among others also with 
$\partial_{\mu}A_{\mu}^a(x)$ and $P^a(x)$.
This means that part of the classical $\Oa$-terms turn into
$\rmO(g_0^2)$ in the quantum theory. The 
essential observation is now that this mixing 
can simply be written in the form of a
renormalization of the terms that are already present
in the Ward identities, since all dimension three and four operators 
with the right quantum number are already there.

We conclude 
that the identities, which we derived above in a formal manner, 
are valid in any proper lattice regularization after 
\begin{itemize}
 \item{replacing the bare fields $A,V,P$ and quark mass $m_0$
 by renormalized ones, where one must allow 
 for the most general
 renormalizations,
 \bes
  (\ar)_{\mu}^a &=& \za A_{\mu}^a \enspace ,\quad
 (\vr)_{\mu}^a = \zv V_{\mu}^a \enspace , \nonumber \\
 (\pr)^a &=& \zp P^a \enspace , \quad
 \mr = \zm \mq\, , \quad \mq=m_0-m_c  \enspace , \nonumber
\ees}
 \item{allowing for the usual $\Oa$ lattice artifacts.}
\end{itemize}
Note that the additive quark mass renormalization $m_c$ 
diverges like $\rmO(g_0^2/a)$  for dimensional
reasons.  

As a result of this discussion, the formal
Ward identities may
be used to determine the normalizations of the currents.
We refer the reader to \cite{reviews:Schlad,impr:pap4} for details and 
explain here
the general idea how one can use the Ward identities to determine
improvement coefficients.

\subsection{On-shell $\Oa$-improvement \label{s:impr}}

\subsubsection{Motivation}

%
%%%%%%%%%%%%%%%%%%%%%%%%%%%%%%%%%%%%%%%%%%%%%%%%%%%%%%%%%%%%%%%%%%%%%%%%%%%%%%%%%%%%%%%%%%
\begin{figure}[tb]
\vspace{0pt}
\centerline{\epsfig{file=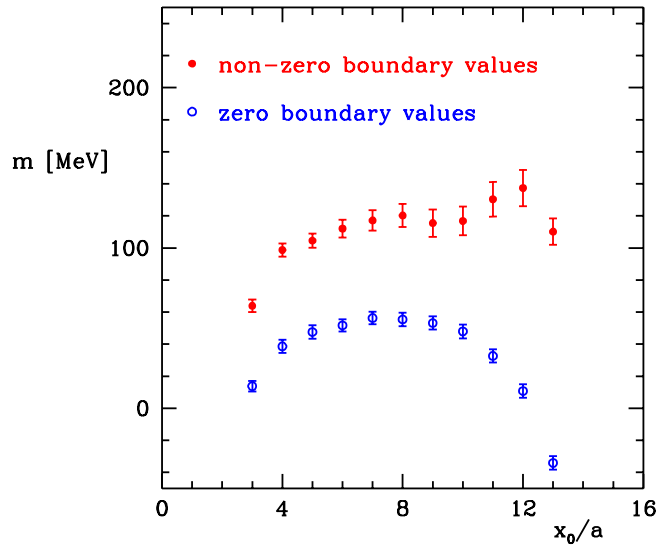,
       width=8.5cm}%7.5 
}
\vspace{0pt}
\caption{Dependence of current quark mass $m$ on the boundary
  condition and the time coordinate~\protect\cite{impr:lett}. 
  The calculation is done in the quenched approximation
  on a $16 \times 8^3$ lattice at $\beta=6.4$, which corresponds to a
  lattice spacing of $a \approx 0.05\,\fm$.
  ``Boundary values'' refer to the
  gauge field boundary conditions in the SF. Their
  values are given in~\protect\cite{impr:lett}.}.
\label{f_letter}
\end{figure}
%%%%%%%%%%%%%%%%%%%%%%%%%%%%%%%%%%%%%%%%%%%%%%%%%%%%%%%%%%%%%%%%%%%%%%%%%%%%%%%%%%%%%%%%%%
%

Let us first recall why one wants to remove lattice spacing effects 
linear in $a$. 
The prime reason is as follows. If linear effects are present, 
one has to vary $a$ in the numerical 
simulations over a large range in 
order to be able to get a reasonable estimate of their magnitude.\footnote{Obviously 
it does not really help to have a large range by considering large values of $a$.
Then one enters the regime where either the higher order terms are significant
or -- more likely -- the whole asymptotic expansion in $a$ breaks down.}  
In contrast if
the cutoff effects are  $\rmO(a^2)$, a range of $0.05\,\fm \leq a \leq 0.1\,\fm$ 
typically allows to check well whether they contribute significantly.
In fact a reasonably well controlled extrapolation to the
continuum can then be done allowing for a term proportional to $a^2$
and also a smaller range in $a$ may be sufficient. Examples can be found
e.g. in \cite{mbar:pap3}.
In addition, it does turn out that linear $a$ effects
can be quite large.
Let us give here just two examples.

The first one is the current quark mass $m$
defined by the PCAC relation. As
detailed below, its value is independent of kinematical variables
such as the boundary conditions. Dependences on such
variables are pure lattice artifacts.
We examined the current quark mass in the valence
approximation by numerical Monte Carlo simulations and found
large lattice artifacts even for quite small lattice spacings~\cite{impr:lett}
(cf. fig.~\ref{f_letter}).

The second example is the mass of the vector meson,
made dimensionless by multiplying with $r_0$. This quantity has large cutoff 
effects in the quenched
approximation~\cite{lat94:rainer}. Depending on the quark 
mass, $a$-effects of around 20\% and more are seen at $a\approx\,0.15\,\fm$;
see for example Fig.~1 of \cite{lat94:rainer},  
Fig.~1 of \cite{reviews:leshouches2}.

\subsubsection{A warning from two dimensions \label{s:unexpect} }

%%%%%%%%%%%%%%%%%%%%%%%%%%%%%%%%%%%%%%%%%%%%%%%%%%%%%%%%%%%%%%%%%%%%%%
\begin{figure}[tb]
\vspace{-0pt}%\vspace{9pt}
\centerline{
\epsfig{file=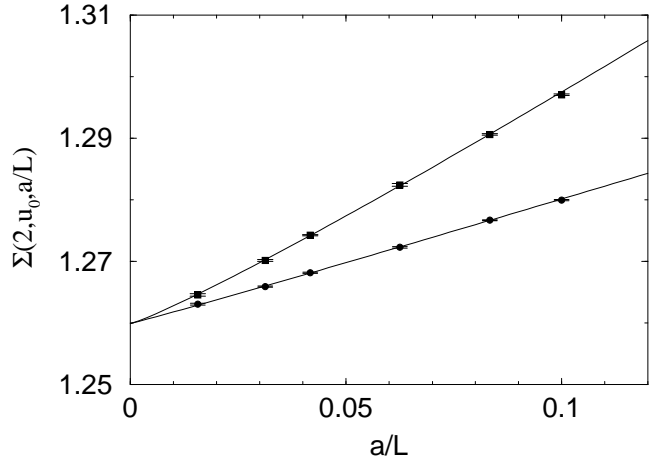,width=8.5cm} %7.5
}
\vspace{-0pt}
\caption{
Lattice spacing dependence of the step scaling function of the 
LWW coupling in the 2-d O(3) sigma model\protect\cite{sigma:nonstandard}
for coupling $u_0=1.0595$. The data points with the smaller 
cutoff effects are for the standard nearest neighbor action.
}
\label{f:lwwo3}
\end{figure}
%%%%%%%%%%%%%%%%%%%%%%%%%%%%%%%%%%%%%%%%%%%%%%%%%%%%%%%%%%%%%%%%%%%%%%

Before entering the discussion of the $\Oa$ improvement programme,
we mention some unexpected results from thorough examinations
of 2-d O($N$) sigma models. 
The theoretical basis for the discussion and removal
of $a$-effects is Symanzik's effective theory, see \sect{s:set}.
O($N$) sigma models were the second class of models
investigated by Symanzik in order to establish this theory. 
For these models the basic statement is that 
(up to logarithmic modifications) the cutoff effects are
quadratic in $a$, when $a$ is small enough. 

It therefore came as a surprise that Hasenfratz and Niedermayer
found in a numerical study of the 
L\"uscher-Weisz-Wolff (LWW) renormalized coupling\cite{alpha:sigma} 
of the 2-d O(3) sigma model that its step scaling function 
shows an $a$-dependence which is roughly linear in $a$ for quite 
small $a$ (large correlation length)\cite{sigma:unexpected}. 
With a further improved algorithm, a Bern-Berlin collaboration
confirmed this behavior with even higher precision
and smaller lattice spacing\cite{sigma:nonstandard}.
We show their result in \fig{f:lwwo3}.
On the other hand it was known that the 
cutoff effects of the step scaling function
are $\rmO(a^2)$ in the large $N$ limit of the 
O($N$) models\cite{alpha:sigma}. Subsequent numerical studies for $N=4,8$ showed
no conclusive results: just like in the $N=3$ case, the cutoff effects
look linear when judged by eye, but they can also be fitted with
$\rmO(a^2)$ functions, in particular when the expected logarithmic modifications
are taken into account\cite{sigma:largeNaeff}. 

%%%%%%%%%%%%%%%%%%%%%%%%%%%%%%%%%%%%%%%%%%%%%%%%%%%%%%%%%%%%%%%%%%%%%%
\begin{figure}[tb]
\vspace{-0pt}%\vspace{9pt}
\centerline{
\epsfig{file=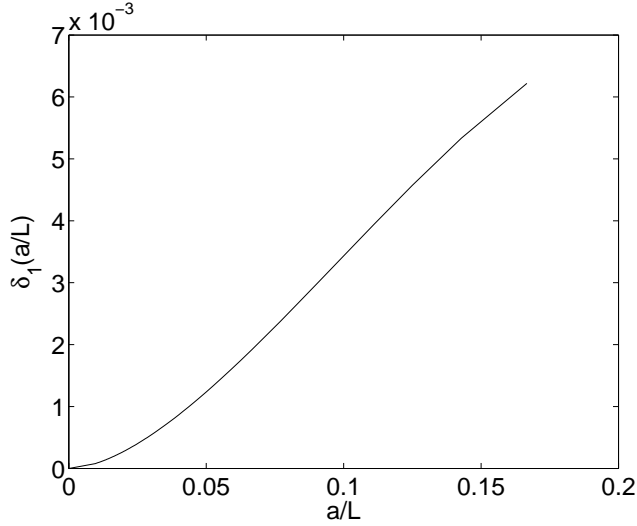,width=8.5cm} %7.5
}
\vspace{-0pt}
\caption{
Coefficient of $1/N$ in the $1/N$-expansion of the 
cutoff effects of the step scaling function of the 
LWW coupling of 2-d O($N$) sigma models. 
Graph prepare by U. Wolff based on \protect\cite{sigma:largeNaeff}.
}
\label{f:largeN}
\end{figure}
%%%%%%%%%%%%%%%%%%%%%%%%%%%%%%%%%%%%%%%%%%%%%%%%%%%%%%%%%%%%%%%%%%%%%%

Later the $1/N$ correction was worked
out at finite lattice spacing\cite{sigma:largeNaeff}. 
Recall that at order $(1/N)^0$ one has $\rmO(a^2)$ effects. 
The cutoff effect proportional to $1/N$ is shown in \fig{f:largeN}. 
Over a large range in $a$ it is almost a linear function
of $a$, but close to $a=0$ it is dominated by an $\rmO(a^2)$
term. Thus our personal conclusion is that there is no
conflict with Symanzik's effective theory in the 
O($N$) models. One should also note that all the 
$a$-effects discussed here are rather small.

However, there is a clear warning that, depending
on model and observable, the lattice spacing may have to be rather 
small for the leading correction term in the effective theory
to dominate. On the more practical side, long continuum
extrapolations with significant slopes may be dangerous, 
since in QCD we do not have much information where the asymptotic
expansion in $a$ is accurate\cite{lat05:ulli}. 
This is one of the reasons why we will
spend much time on understanding the cutoff effects in
the QCD step scaling function of the coupling in \sect{s:finiteaPT}
and \sect{s:finiteaNP}.

\subsubsection{Symanzik's local effective theory (SET)\label{s:set}}

In the following explanation of the theory we follow quite 
closely \cite{impr:pap1}. 
We consider QCD on an infinitely extended lattice with
two degenerate light Wilson quarks of bare mass $m_0$~\cite{Wilson}.
The action is then given as in \sect{s:latt} except that no
boundary conditions or boundary terms are necessary.

Quite some time ago, Symanzik provided arguments
that a lattice theory can be described in terms
of a local effective theory, when the 
lattice spacing is small enough~\cite{impr:Sym1}. The effective action, 
\begin{equation}
 \Seff=S_0+a S_1+a^2 S_2+\ldots\,,
 \label{e_Seff}
\end{equation}
has as a leading order, $S_0$, the action of the continuum theory~\footnote{
If more rigor is desired one may define it on a
lattice with spacing $\varepsilon\ll a$.}.  
The terms $S_k$,
$k=1,2,\ldots$, are space-time integrals of Lagrangians ${\cal
  L}_k(x)$.  These are given as general linear combinations of local
gauge-invariant composite fields which respect the exact symmetries of
the lattice theory and have canonical dimension $4+k$.  We use the
convention that explicit (non-negative) powers of the quark mass $m$
are included in the dimension counting.  A possible basis of fields
for the Lagrangian ${\cal L}_1(x)$ reads
\begin{eqnarray}
{\cal O}_1 &=&\psibar\,\sigma_{\mu\nu}F_{\mu\nu}\psi,\nonumber\\[.2ex]
{\cal O}_2 &=&\psibar\,D_{\mu}D_{\mu}\psi
           +\psibar\,\lvec{D}_{\mu}\lvec{D}_{\mu}\psi,\nonumber\\[.2ex]
{\cal O}_3 &=& m\tr\!\left\{F_{\mu\nu}F_{\mu\nu}\right\},
              \label{e_counterterms}\\[.2ex]
{\cal O}_4 &=& m\left\{\psibar\,\dirac{\mu}D_{\mu}\psi
               -\psibar\,\lvec{D}_{\mu}\dirac{\mu}\psi\right\},
              \nonumber\\[.2ex]
{\cal O}_5 &=& m^2\psibar\psi,\nonumber
\end{eqnarray} 
where $F_{\mu\nu}$ is the gluon field tensor and
$\sigma_{\mu\nu}=\frac{i}2[\dirac\mu,\dirac\nu]$.

When considering correlation functions of local gauge invariant fields
the action is not the only source of cutoff effects.  If $\phi(x)$
denotes such a lattice field (e.g.~the axial density or the isospin
currents), one expects the connected $n$-point function
\begin{equation}
  G_n(x_1,\ldots,x_n)=(\zphi)^n
  \left\langle\phi(x_1)\ldots\phi(x_n)\right\rangle_{\rm con}
\end{equation}
to have a well-defined continuum limit, provided the renormalization
constant $\zphi$ is correctly tuned and the space-time arguments
$x_1,\ldots,x_n$ are kept at a physical distance from each other.

In the effective theory the renormalized lattice field $\zphi\phi(x)$
is represented by an effective field,
\begin{equation}
  \phieff(x)=\phi_0(x)+a\phi_1(x)+a^2\phi_2(x)+\ldots,
\end{equation} 
where the $\phi_k(x)$ are linear combinations of composite, local
fields with the appropriate dimension and symmetries.  For example, in
the case of the axial current~(\ref{e_Amu}), $\phi_1$ is
given as a linear combination of the terms
\begin{eqnarray}
  (\op{6})_{\mu}^a &=&
  \psibar\,\dirac{5}\frac{1}{2}\tau^a\sigma_{\mu\nu}
      \bigl[{D}_{\nu}-\lvec{D}_{\nu}\bigr]\psi,
  \nonumber\\[.5ex]
  (\op{7})_{\mu}^a&=&\psibar\,\frac{1}{2}\tau^a\dirac{5}
                  \bigl[{D}_{\mu}+\lvec{D}_{\mu}\bigr]\psi,
  \label{e_impr_current}\\[.5ex]
  (\op{8})_{\mu}^a&=&m\psibar\,\dirac{\mu}\dirac{5}\frac{1}{2}\tau^a\psi.
  \nonumber
\end{eqnarray}
The convergence of $G_n(x_1,\ldots,x_n)$
to its continuum limit can now be studied in the 
effective theory,
\bes
  G_n(x_1,\ldots,x_n) &=& \left\langle\phi_0(x_1)\ldots\phi_0(x_n)\right\rangle_{\rm con} 
  \nonumber\\ && \label{e_continuum_approach}
  -a\int\rmd^4y\,\left\langle\phi_0(x_1)\ldots\phi_0(x_n)
  {\cal L}_1(y)\right\rangle_{\rm con}
   \\  && \nonumber
  \,+a\,\sum_{k=1}^n
  \left\langle\phi_0(x_1)\ldots\phi_1(x_k)\ldots\phi_0(x_n)
  \right\rangle_{\rm con}
  \,+\,\rmO(a^2), \nonumber
\ees
where the expectation values on the right-hand side 
are to be taken in the continuum theory with action $S_0$.

\subsubsubsection{Using the field equations \label{s:fe}}

For most applications, it is sufficient to
compute on-shell quantities such as particle
masses, S-matrix elements and correlation functions
at space-time arguments, which are separated by
a physical distance. It is then possible to make use of the
field equations to reduce first the number of 
basis fields in the effective Lagrangian ${\cal L}_1$
and, in a second step, also in the O($a$) counter-term
$\phi_1$ of the effective composite fields.

Of course, in the quantum theory, the field equations have to
be used with care. 
Performing changes in the integration variables
of the path integral shows immediately that the 
field equations are only valid up to additional terms.
In particular, 
if one uses the field equations in the Lagrangian ${\cal L}_1$ under
the space-time integral in \eq{e_continuum_approach}, the
errors made are contact terms that arise when $y$ comes close to one
of the arguments $x_1,\ldots,x_n$.  Using the operator product 
expansion and the dimensions
and symmetries, one easily verifies that these contact terms must have
the same structure as the insertions of $\phi_1$ in the last term of
\eq{e_continuum_approach}. The use of the field equations in 
${\cal L}_1$ therefore just means a redefinition of the coefficients
in the counter-term $\phi_1$ (apart from contributions
that are absorbed in the renormalization factor $Z_\phi$,
which originate from the mixing of $\phi_1$ with $\phi$).

It turns out that one may eliminate two of the terms in
\eq{e_counterterms} by using the field equations. 
A possible choice is to stay with the
terms ${\cal O}_1$, ${\cal O}_3$ and ${\cal O}_5$, which yields the
effective continuum action for on-shell quantities to order $a$.
Having made this choice one may apply the field equations once again
to simplify the term $\phi_1$ in the effective field as well.  In the
example of the axial current it is then possible to eliminate the
term~$\op{6}$ in \eq{e_impr_current}.

\subsubsection{Improved lattice action and fields \label{s:simpr}}

The on-shell O($a$) improved lattice action is obtained by
adding a counter-term to the unimproved lattice action
such that the effects of the action $S_1$ in the effective
theory are canceled in on-shell amplitudes. 
This can be achieved by adding lattice representatives
of the terms ${\cal O}_1$, ${\cal O}_3$ and ${\cal O}_5$
to the unimproved lattice Lagrangian, with coefficients that
are functions of the bare coupling $g_0$ only.
Leaving the discussion of suitable improvement conditions 
for later, we here note that the fields 
${\cal O}_3$ and ${\cal O}_5$ already appear in the unimproved
theory and thus merely lead to a re-parametrization of 
the bare parameters $g_0$ and $m_0$. In the following, 
we will not consider these terms any further. 
Their relevance in connection with
massless renormalization
schemes is discussed in detail in \Ref{impr:pap1}.

We choose the standard discretization $\widehat{F}_{\mu\nu}$
of the field tensor\,\cite{impr:pap1} and add the improvement term
to the Wilson-Dirac operator, \eq{e_Dlat},
\begin{equation}
 D_{\rm impr}=D+\csw\,{{ia}\over{4}}\sigma_{\mu\nu}\widehat{F}_{\mu\nu}.
 \label{e_dimpr}
\end{equation} 
With a properly chosen coefficient $\csw(g_0)$, 
this yields the on-shell O($a$) improved lattice action which 
has first been proposed by Sheikholeslami and Wohlert~\cite{impr:SW}.

The perturbative expansion of $\csw$ reads  
$\csw=1+\csw^{(1)}g_0^2+\rmO(g_0^4)$, with~\cite{impr:pap2,impr:Wohlert}
$ \csw^{(1)} = 0.26590(7)$ where the Wilson plaquette gauge action
is assumed. 

The O($a$) improved isospin currents and the axial density can
be parametrized as follows,
\begin{eqnarray}
 (\ar)_\mu^a&=&
 \za(1+\ba a\mq)\bigl\{A_\mu^a+a\ca\tilde{\partial}_\mu
       P^a\bigr\},\nonumber\\[.5ex]
 (\vr)_\mu^a&=&
 \zv(1+\bv a\mq)\bigl\{V_\mu^a+a\cv\tilde{\partial}_\nu
       T_{\mu\nu}^a\bigr\},\nonumber\\
  [.5ex]
 (\pr)^a&=&
 \zp(1+\bp a\mq)P^a\,, \label{e_renfields}
\end{eqnarray}
\vspace{-2mm}
where\vspace{-2mm}
$$
T^a_{\mu\nu}=i\psibar\sigma_{\mu\nu}\frac12\tau^a\psi \,.
$$
We have
included the normalization constants $Z_{\rm A,V,P}$, which have to be
fixed by appropriate normalization conditions.  
Again,
the improvement coefficients $b_{\rm A,V,P}$ and $c_{\rm A,V}$ are
functions of $g_0$ only. At tree level of perturbation theory, they are
given by
$\ba=\bp=\bv=1$ and $\ca=\cv=0$~\cite{pert:heatlie,impr:pap1} and to 
1-loop accuracy and with the plaquette gauge action we have
\cite{impr:pap2,impr:pap5} 
\begin{eqnarray}
\ca &=&-0.005680(2)\times g_0^2\CF+\rmO(g_0^4),\quad \nonumber\\
\cv &=&-0.01225(1)\times g_0^2\CF+\rmO(g_0^4),\quad \nonumber\\
\ba &=& 1 + 0.11414(4)\times g_0^2\CF+\rmO(g_0^4),\quad \nonumber\\
\bv &=& 1 + 0.11492(4)\times g_0^2\CF+\rmO(g_0^4),\quad \nonumber\\
\bp &=& 1 + 0.11484(2)\times g_0^2\CF+\rmO(g_0^4).\quad 
 \label{e_cApert}
\end{eqnarray}
Here $\CF=4/3$.
Non-perturbative
determinations will be mentioned below.

\subsection{The PCAC relation}

We assume for the moment that on-shell O($a$) improvement has been
fully implemented, i.e.~the improvement coefficients are assigned
their correct values. If ${\cal O}$ denotes a renormalized on-shell
O($a$) improved field localized in a region not containing $x$, we
thus expect that the PCAC relation,
\bes
%    \frac12(\drv\mu+\drvstar\mu)
    \tilde{\partial}_\mu
    \langle(\ar)_\mu^a(x)\,{\cal O}\rangle&=&
     2\mr\langle(\pr)^a(x)\,{\cal O}\rangle
 \label{e_PCAC_impr} \\
  \tilde{\partial}_\mu &=& \frac{1}{2}(\drvstar{\mu}+\drv{\mu})
\ees
holds up to corrections of order $a^2$. 
At this point we note that the field ${\cal O}$ need not
be improved for this statement to be true. To see this
we use again Symanzik's local effective theory and
denote the O($a$) correction term
in ${\cal O}_{\rm eff}$ by $\phi_1$.
Eq.\,(\ref{e_PCAC_impr}) then receives an order $a$ contribution
\begin{equation} 
   a\bigl\langle
  \left\{\partial_\mu(\ar)^a_{\mu}(x)-2\mr(\pr)^a(x)\right\}
  \phi_1\bigr\rangle,
 \label{e_PCACcorrection}
\end{equation}
which is to be evaluated in the continuum theory.
The PCAC relation holds exactly in the continuum and
the extra term~(\ref{e_PCACcorrection}) thus
vanishes, 
a conclusion that holds also
in the Schr\"odinger functional.

\subsection{Non-perturbative improvement \label{s_NPI}}

The coefficients of the different improvement terms need to be fixed by suitable
improvement conditions. One considers pure lattice artifacts,
i.e. combinations
of observables that are known to vanish in the continuum limit
of the theory. Improvement conditions require
these lattice artifacts to vanish,
thus defining the values of the improvement 
coefficients as a function of the lattice spacing.

In perturbation theory, lattice artifacts 
can be obtained from any (renormalized) quantity by subtracting its
value in the continuum limit. The improvement coefficients are 
unique and some of them have been cited above.

Beyond perturbation theory, one wants to determine the improvement coefficients by
MC calculations and it requires significant effort
to take the continuum limit.
It is therefore advantageous to use lattice artifacts that derive from
a symmetry of the continuum field theory that is not respected by the
lattice regularization. One may require rotational invariance
of the static potential $V({\bf r})$, e.g.
$$
 V({\bf r}=(2,2,1)r/3) - V({\bf r}=(r,0,0)) =0 \, ,
$$
or 
Lorentz invariance,
$$
 [E({\bf p})]^2 - [E({\bf 0})]^2  - {\bf p}^2 =0 \, ,
$$
for the momentum dependence of a one-particle energy $E$.

For $\Oa$ improvement of QCD it is advantageous
to use violations of the PCAC relation~(\ref{e_PCACnew}), instead.
PCAC can be used in the context of the 
Schr\"odinger functional (SF), where one has a large flexibility to
choose appropriate improvement conditions and can compute the improvement
coefficients also for small values of the bare coupling $g_0$, making contact
with their perturbative expansions. A further -- and maybe 
the most significant --  advantage of the SF in
this context is the following. As we have seen
earlier, in the SF we may choose 
boundary conditions such that the induced
background field, has non-vanishing components $F_{\mu\nu}$. 
Remembering eq.~(\ref{e_dimpr}), we observe that 
correlation functions are then sensitive to the improvement coefficient 
$\csw$ already
at tree level of perturbation theory. With a vanishing
background field this would be the case 
at higher orders only. Also in the non-perturbative
regime this is the basis for a good sensitivity of
the improvement conditions to $\csw$. 
 
As a consequence of the freedom to choose improvement conditions, 
the resulting values of 
improvement coefficients such as $\csw,\,\ca$
depend on the exact choices made.  The corresponding variation of
$\csw,\,\ca$ is of order $a$. It changes the effects of
order $a^2$ in physical observables computed after improvement.
We will come back to this point, but first we 
proceed to sketch the non-perturbative calculation of 
the coefficient $\csw$~\cite{impr:pap1,impr:pap2}.

We define a bare current quark mass, $m$, viz.
\bes
 m \equiv {  \left\langle \left[ \partial_{\mu}(\aimpr)_{\mu}^a(x) \right]
              {\cal O}^{a} \right\rangle 
           \over
             2 \left\langle  P^a(x)   {\cal O}^{a} 
                 \right\rangle }
                 \, , \quad 
   (\aimpr)_\mu^a=A_\mu^a+a\ca \frac12({\partial}_\mu +{\partial}^*_\mu)
                  P^a \enspace 
  \label{e_PCACnn}
\ees
with the pseudo scalar boundary operator 
${\cal O}^{a}$ from \eq{e_boundops}.
When all improvement coefficients have their proper values, 
the renormalized quark mass, defined by
the renormalized PCAC-relation, is related
to $m$ by
\bes
  \mr = {{\za (1+\ba a \mq)}\over{\zp (1+\bp a \mq)}} m + \rmO(a^2)\enspace .
  \label{e_mrn}
\ees
It now suffices to choose 3 different versions of \eq{e_PCACnn} by different
choices of the time coordinate $x_0$ and/or boundary conditions and
obtain 3 different values of $m$, denoted by $m_1,m_2,m_3$.
Since the prefactor in front of $m$ in \eq{e_mrn} is just a numerical 
factor, we may conclude that all $m_i$ have to be equal
in the improved theory up to errors of order $a^2$.
$\csw$ and $\ca$ may therefore be computed by requiring~\footnote{
In the practical calculations actually four different
masses $m_i$ were used to compute $\csw$ and two more 
to extract $\ca$.}
\bes
  m_1 = m_2 = m_3 \label{e_imprcond}\enspace .
\ees
This simple idea has first been used to compute 
$\csw$ and $\ca$ as a function of $g_0$ in the quenched 
approximation~\myref{impr:pap1}. 
A good accuracy has been reached in these determinations
for $a\leq0.1\,\fm$. 
We comment on more recent 
determinations below, after showing some tests of the effectiveness
of $\Oa$-improvement, namely the test whether cutoff effects are 
indeed significantly reduced in practical calculations. 

\subsubsubsection{Verification of $\Oa$-improvement}

%%%%%%%%%%%%%%%%%%%%%%%%%%%%%%%%%%%%%%%%%%%%%%%%%%%%%%%%%%%%%%%%%%%%%%
\begin{figure}[tb]
\vspace{-0pt}%\vspace{9pt}
\centerline{
\epsfig{file=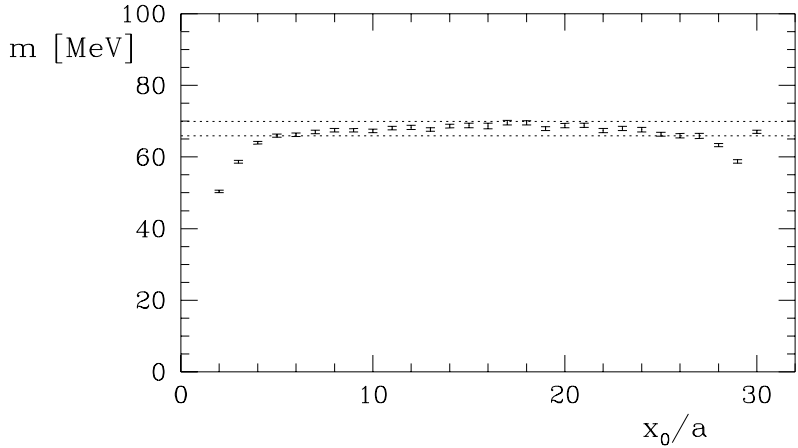,
       width=10.5cm} % 9.0
}
\vspace{-0pt}
\caption{
Unrenormalized current quark mass $\m$ 
in the improved theory, with non-perturbatively determined $\csw$
and $\ca$, as a function 
of the time $x_0$ on a $32\times16^3$ lattice
at $\beta=6.2$ ($a\approx0.07\,\fm$)and $\hop=0.1350$. 
The width of the corridor bounded by the 
dotted horizontal lines is $4\,\MeV$.
}
\label{f_mimpr}
\end{figure}
%%%%%%%%%%%%%%%%%%%%%%%%%%%%%%%%%%%%%%%%%%%%%%%%%%%%%%%%%%%%%%%%%%%%%%

The first test of the size of residual $\rmO(a^2)$ effects is provided again
by the PCAC relation. To set the scale, remember that the cutoff
effects in the PCAC mass $m$ were as large as several tens of $\MeV$
before improvement (cf. fig.~\ref{f_letter}). The situation 
after improvement, and for a somewhat larger value of the lattice spacing,
is illustrated in fig.~\ref{f_mimpr}. Away from the boundaries, where
the effect of states with energies of the order of the cutoff induces
noticeable effects, $m$ is independent of time to within $\pm 2\,\MeV$.
Compared to the situation before $\Oa$ improvement, this is a change
by more than an order of magnitude. 

The second test is in the scaling of the vector meson mass, 
which is improved dramatically in the quenched approximation.
In the range $0.01 \leq a^2/r_0^2 \leq 0.035$ ($a\leq0.1\,\fm$)\footnote{
We use the 
reference scale $r_0$ defined in terms of the static quark
potential.\cite{pot:r0} It has a phenomenological value of
$r_0\approx0.5\,\fm$.}
and with statistical errors of 1-3\%, no $a$ effect was found
(see Fig.~3, \Ref{mbar:pap3}). This is a significant 
test because i) the $a$-effects were large before improvement,
ii) the statistical precision is good and iii) $a^2$ varies by
more than a factor 3.

{\em In the quenched approximation} several other tests of cutoff 
effects in the
improved theory have been carried out
\cite{impr:jochen,mbar:pap3} revealing that $\Oa$-improvement works.
More precisely we mean by this that for $a\leq 0.1\,\fm$ cutoff effects found after 
improvement are generally reasonably small ($\rmO(5\,\%)$) and
well compatible with an $\rmO(a^2)$ behavior. Still, no miracles have been
achieved: at $a\approx0.1\,\fm$ also cutoff effects of about 15\% have been found
\cite{mbar:pap3}.

We now proceed to discuss a rather relevant detail in the non-perturbative
determination of improvement coefficients.

\subsubsection{The constant physics condition \label{s:cpc}}

We have mentioned before that beyond perturbation theory 
the improvement coefficients are not unique; they depend
on the improvement condition. This ambiguity is of $\Oa$
in the coefficients and then of $\rmO(a^2)$ in the physical
observables after improvement. So everything is correct,
but if these ambiguities are large, one has to take extra care.

\begin{figure}[tb]
\vspace{-0pt}%\vspace{9pt}
\centerline{
\epsfig{file=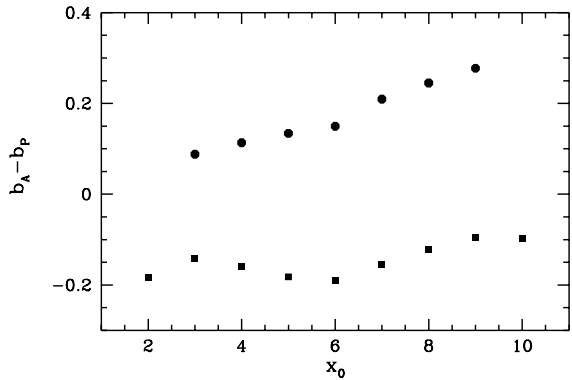, width=7.5cm} %7.0
}
\vspace{-0pt}
\caption{The improvement coefficient $\ba-\bp$ extracted on a $12 \times 8^3$
lattice and a lattice spacing of $a\approx0.1\,\fm$. Axial current and density
are inserted at distance $x_0$. Squares are obtained with standard 
lattice representatives of the derivatives $\partial_\mu$ and
$\partial^*_\mu\partial_\mu$, while circles come from improved
(and therefore less local) derivatives~\protect\cite{impr:babp}.
}
\label{f:babp1}
\end{figure}

To give an explicit example, consider the quenched approximation
but with non-degenerate quark masses, $m_{\mrm{R},i}\neq m_{\mrm{R},j}$.
In this theory the renormalized, improved currents and axial densities
are given by a straight forward generalization of what we wrote down 
earlier\cite{impr:roma2_1},
\bes
 (\ar)_\mu^{ij}&=&
 \za(1+\ba \frac12(a\mqi+a\mqj))\bigl\{A_\mu^{ij}+a\ca\tilde{\partial}_\mu
       P^{ij}\bigr\},\nonumber\\[.5ex]
 (\pr)^{ij}&=&
 \zp(1+\bp \frac12(a\mqi+a\mqj))P^{ij}\,, \label{e_renfields_nd} \\[.5ex]
 A_\mu^{ij} &=& \psibar_i \gamma_\mu\gamma_5 \psi_j\,,\quad
 P^{ij} = \psibar_i \gamma_5 \psi_j \,, \nonumber
\ees 
and the bare quark masses, $m_{0,i}$, of the Lagrangian are related to
the renormalized ones via~\footnote{
In the improved action, $\bm$ is the coefficient of the field $\op{5}$.}
\bes
  m_{\mrm{R},i} = {\za \over \zp}\, Z\, \mqi(1+a\,\bm \mqi)\,,
  \quad \mqi = m_{0,i} - \mcrit\,
 . \label{e:mri}
\ees
One then derives the (unrenormalized) PCAC relation  \cite{impr:roma2_1}
\bes
  &&{\partial_\mu A_\mu^{ij}+a\partial^*_\mu\partial_\mu  P^{ij}
                  \over P^{ij}}
                [1+a\frac12 (\ba-\bp)\,(\mqi+ \mqj)]\nonumber \\
 && \;=\;
               Z \, [\mqi+ \mqj + a\,\bm(\mqi^2+ \mqj^2)]
                +\rmO(a^2) \,.
\ees
This operator identity has to be understood as in \eq{e_PCAC_impr}. 
Applying it with a few external operators/kinematical
conditions, one can extract
$\ba-\bp$, $\bm$ and $Z$~\cite{impr:roma2_1,impr:ca_gupta}.  
Suitable correlation functions have been implemented in the \SF.
As shown in \fig{f:babp1}, the result for $\ba-\bp$ at $a\approx0.1\,\fm$
does depend a lot on the details, namely here $x_0$ and the exact
lattice representation of 
derivatives. This is an extreme example of an improvement coefficient that 
is difficult to determine. The reasons are surely that the $\Oa$-effects
are not dominating over the $\rmO(a^2)$ effects in the considered
correlation functions. Presumably the $\Oa$-effects are just not
very large.  We proceed to discuss what one should do in these cases.

%%%%%%%%%%%%%%%%%%%%%%%%%%%%%%%%%%%%%%%%%%%%%%%%%%%%%%%%%%%%%%%%%%%%%%
%%%%%%%%%%%%%%%%%%%%%%%%%%%%%%%%%%%%%%%%%%%%%%%%%%%%%%%%%%%%%%%%%%%%%%
 
When the improvement conditions are formulated through \SF correlation
functions, we can also study them in perturbation theory. 
One chooses kinematical variables such as $T/L\,,x_0/L$ exactly as in 
the non-perturbative set up. One then computes
the expansion of an
improvement coefficient, here denoted generically by $b$, as a series
\bes
  b(g_0,a/L) \sim b^{(0)}(a/L) + g_0^2\, b^{(1)}(a/L) +\rmO(g_0^4)\,. 
\ees
Taking the case of $\ba-\bp$ and $L/a=8, T/a=12$ as in \fig{f:babp1},
the perturbative coefficients 
$b^{(0)},\,b^{(1)}$ show a similar dependence on the kinematics as
the non-perturbative results~\cite{impr:babp}. 

While generically the functions $b^{(0)}(a/L),\,b^{(1)}(a/L)$ depend on the kinematical
choices made in the improvement coefficients, the values
$b^{(0)}(0),\,b^{(1)}(0)$ are unique. This is the precise meaning of 
our earlier statement that improvement coefficients are unique in 
perturbation theory. 

%%%%%%%%%%%%%%%%%%%%%%%%%%%%%%%%%%%%%%%%%%%%%%%%%%%%%%%%%%%%%%%%%%%%%%%%%%%%%%%%%%%%%%%%%%
\begin{minipage}{0.46\textwidth}
\includegraphics*[width=1.0\textwidth]{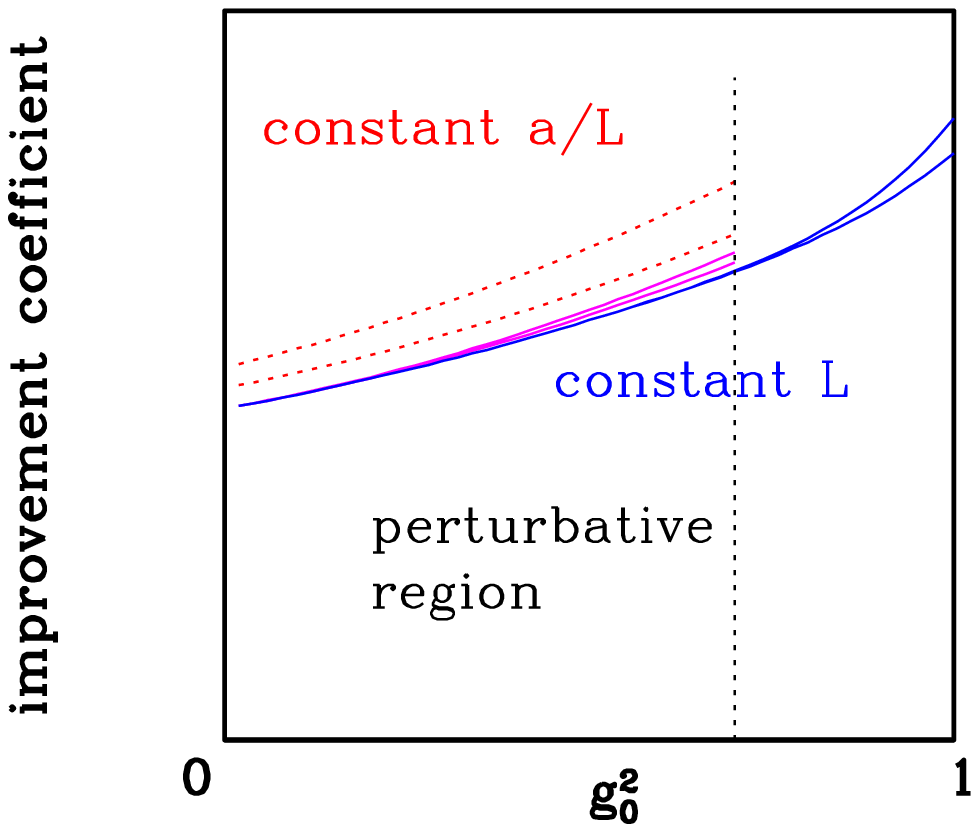}
%\vspace{5mm}
\footnotesize
Generic behavior of improvement coefficients. Dotted lines
correspond to two different improvement conditions at fixed $a/L$,
while the full lines are for fixed $L$ in physical units.
\end{minipage}
%%%%%%%%%%%%%%%%%%%%%%%%%%%%%%%%%%%%%%%%%%%%%%%%%%%%%%%%%%%%%%%%%%%%%%%%%%%%%%%%%%%%%%%%%%
\hfill
\begin{minipage}{0.40\textwidth}
\vspace{2mm}

On the left we 
illustrates what happens generically. 
The continuous lines correspond to two different improvement
conditions at {\em constant physics}. By this we mean that
all length scales in the correlation functions which define
the improvement condition are kept fixed in physical units, 
for example in units of $r_0$. When $g_0$ is varied, only 
the lattice spacing $a$ 
changes and this is the situation to which the SET can be applied. 
The two different improvement coefficients then very rapidly
go to an almost  unique function of $g_0$ 
as $g_0\to0$. 
\vspace{2mm}

\end{minipage}

On the other hand, 
one may want to set improvement conditions at fixed $L/a$
for practical reasons\footnote{
In fact this was done originally for 
$\csw,\,\ca$~\cite{impr:pap3}. Note, however,
that the small tree-level $a$-effect was subtracted from the 
non-perturbative ones to insure the improvement coefficients go 
to their tree-level values exactly and further the
conditions were chosen such that also $|b^{(1)}(a/L) - b^{(1)}(0)|$ 
is very small.}.
In this setting, the $\Oa$ 
ambiguity in the improvement coefficients does not go to zero
as $g_0 \to 0$ (and $a\to 0$) (dotted lines).

If the dependence of $b^{(i)}(a/L)$ on $a/L$ is rather weak,
say 
\bes
   |b^{(i)}(a/L) - b^{(i)}(0)|  \ll 10^{-1}\,, 
   \label{e:bratio}
\ees
for the relevant $a/L$,
one may also choose fixed $a/L$. Still it is clear from our discussion
that fixed $L/r_0$ is to be preferred whenever possible.
In cases where it is advantageous 
to work at finite quark masses, also the combination
$r_0\,m_{\mrm{R},i}$ should be kept constant.
Note that these {\em constant physics} conditions do
not have to be satisfied very precisely since we are talking about
a correction to an $\Oa$ term. 

%%%%%%%%%%%%%%%%%%%%%%%%%%%%%%%%%%%%%%%%%%%%%%%%%%%%%%%%%%%%%%%%%%%%%%
\begin{figure}[tb]
%\centerline{
\epsfig{file=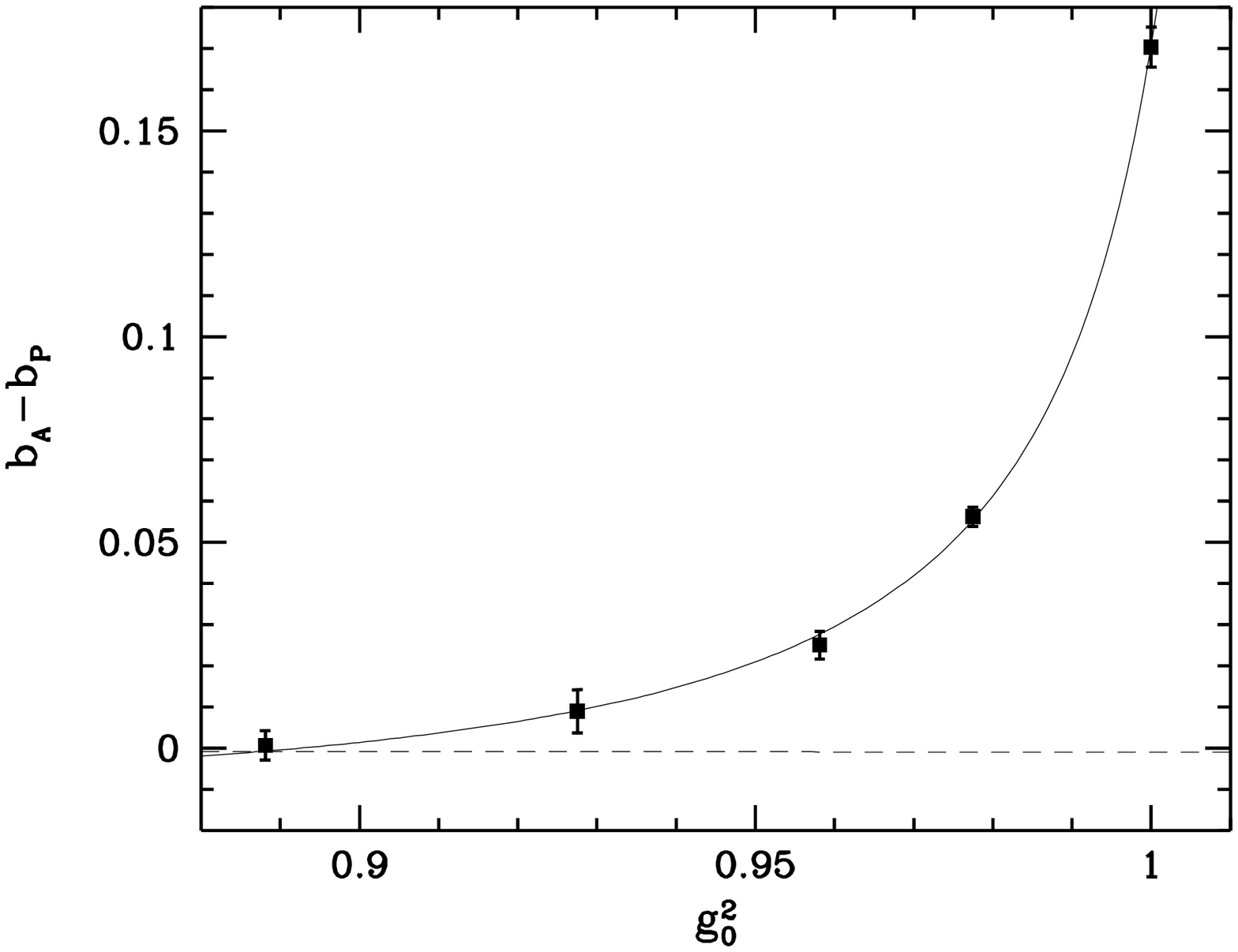, width=0.49\textwidth}\hfill
\epsfig{file=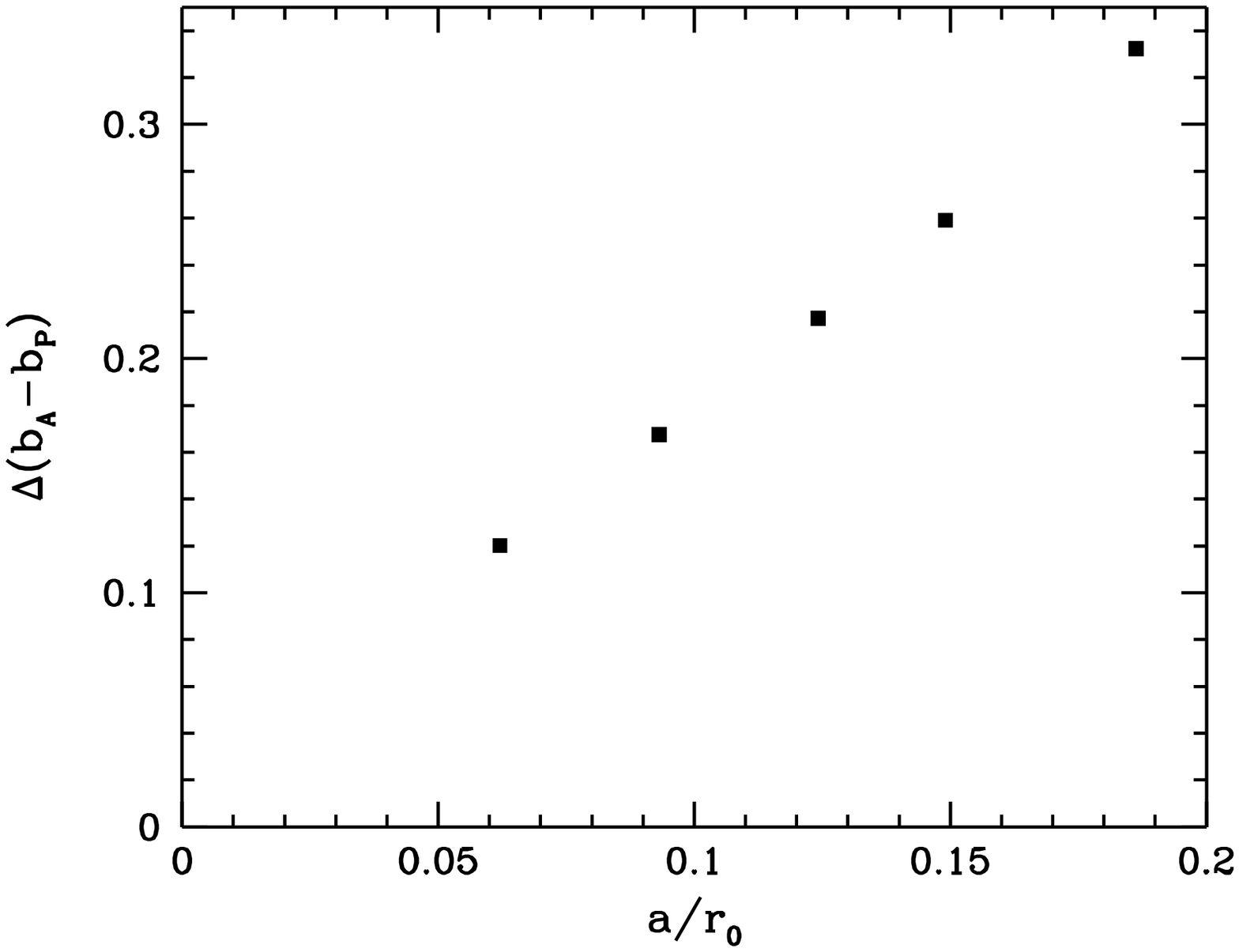, width=0.49\textwidth}
%}
\caption{Non-perturbative $\ba-\bp$ for $\nf=0$ and plaquette 
gauge action (left) as well as its ambiguity $\Delta(\ba-\bp)$.
The latter is the difference of $\ba-\bp$ for one improvement 
condition and $\ba-\bp$ for another one.
}
\label{f:babp3}
\end{figure}
%%%%%%%%%%%%%%%%%%%%%%%%%%%%%%%%%%%%%%%%%%%%%%%%%%%%%%%%%%%%%%%%%%%%%%

Such conditions have first been imposed
in \Ref{impr:babp} -- exactly because of \fig{f:babp1}.
We show the result for $\ba-\bp$, with the constant physics 
condition implemented, in \fig{f:babp3}.
In this extreme case the ambiguity is of the same 
order as the improvement coefficient itself. In such a situation 
it is rather tempting
to just put the coefficient to zero. However, it should be obvious that
it is then not guaranteed that linear $a$-effects are absent after improvement.
An extrapolation to the continuum using an $\rmO(a^2)$ model
function for the cutoff effects might then give significantly
wrong results. We repeat this relevant fact in different
words. While an unfortunately chosen improvement condition, but implemented
at constant physics, may even enlarge the cutoff effects for intermediate $a$, 
it guarantees\footnote{Here we use a strong wording. Remember
of course that everything is based on the SET, which has not been 
proved. Remember also \sect{s:unexpect}.} that linear $a$-effects are absent. 
Putting the improvement coefficient to zero or some
perturbative approximation does not guarantee the latter and the 
linear effects should at least be estimated in some way.

Before we review what is known in $\Oa$-improvement,
we return once more to the improvement coefficients $\csw,\ca$.
With plaquette gauge action, $a/L=1/8$ and the chosen
improvement condition, \eq{e:bratio} is very well 
satisfied for $i=0,1$~\cite{impr:pap2}.
It appears justified to subtract these effects
perturbatively as it was done in \cite{impr:pap3,impr:csw_nf2}. 
For $\ca$ there is evidence for a significant $\Oa$ 
ambiguity~\cite{impr:ca_gupta,impr:ca_ukqcd,lat03:stephan}.
As a consequence, the computation with $\nf=2$ has been carried out 
with an improvement condition at constant physics~\cite{impr:ca_nf2}.

\subsubsection{Summary of results \label{s:summary} }

\begin{table}[t] 
{\begin{center} \small
\begin{tabular}{lllcl}
\hline \\[-1.5ex]
improvement coefficient  & order & $\nf$ &  gauge action  & reference\\
or improved fields        & in PT &       &                \\
\hline\\[-1.5ex]
 $\csw,\ca$              & 1-loop & -- & Wilson          &
 \cite{impr:pap2,impr:Wohlert} \\
 $\bg$              & 1-loop & -- & Wilson          &
 \cite{impr:pap1,pert:1loop} \\
 $\bm,\ba,\bp,\bv,\cv$ \\
 $V_\mu^a, A_\mu^a$  & 1-loop & -- & Wilson & 
 \cite{impr:pap5,impr:curr_iwa_pert} \\
 $\csw$              & 1-loop & -- & various         &
 \cite{impr:csw_1loop_pbc} \\
 $\csw,\ca$ & NP & 0 & Wilson & \cite{impr:pap3} \\
 $\bm,\ba-\bp,\bv,\ca,\cv$ \\
 $V_\mu^a, A_\mu^a$   & NP & 0 &  Wilson & \cite{impr:ca_gupta} \\
 $\ca$                & NP & 0 &  Wilson & \cite{impr:ca_ukqcd} \\
 $\ba-\bp,\bv-\bs,\bm$  & NP & 0 & Wilson & \cite{impr:roma2_1} \\
 $\ba-\bp,\bm$  & NP at CP & 0 & Wilson & \cite{impr:babp} \\
 $\ba-\bp,\bm$  & NP at CP & 0 & Wilson & \cite{hqet:pap2} \\
 $\csw$ & NP & 2 & Wilson & \cite{impr:csw_nf2} \\
 $\csw$ & NP & 3 & Wilson & \cite{impr:csw_nf3} \\
 $\ca$  & NP at CP & 2 & Wilson & \cite{impr:ca_nf2} \\
 $\bm,\ba,\bp,\bv,\cv,b_\mrm{T}$ &       &     \\
 $V_\mu^a, A_\mu^a, T_{\mu\nu}^a$  & 1-loop & -- & $1\times1$ \& $1\times2$ 
 & \cite{impr:curr_iwa_pert} \\
 $\csw$              & 1-loop & -- & $1\times1$ \& $1\times2$          &
 \cite{impr:csw_iwa_pert} \\
 $\csw$ & NP at CP & 0,2,3 & Iwasaki & \cite{impr:csw_iwa_allnf} \\
 $\ca$  & NP at CP & 3   & Iwasaki & CP-PACS \& JLQCD \\
 \\[0.5ex]     \hline
\end{tabular}
\end{center}
}
\caption{Literature on $\Oa$ improvement coefficients. 
NP stands for non-perturbative, CP for constant physics condition, 
Wilson for the Wilson plaquette action,
``$1\times1$ \& $1\times2$'' for the gauge action including 
a $1\times2$ loop and Iwasaki for the action where the
coefficient of that term is set to the value proposed by Y. Iwasaki.
}
\label{t:imprlit}
\end{table}

%%% Local Variables: 
%%% mode: latex
%%% TeX-master: "sect_Lambda"
%%% End: 

A summary of the available results for $\Oa$ improvement
coefficients is listed in \tab{t:imprlit}.
Many investigations have been carried out accumulating
a quite advanced knowledge. It is impossible 
for me to review it all, rather I add only a few comments. 
\bi
\item   In the quenched approximation there is a rather complete
        knowledge for the Wilson gauge action, but $\ca$ has not
        been determined with a constant physics condition.
        Conservative continuum extrapolations in the improved theory
        usually start only with values of $a$ which are a bit below 
        $0.1\,\fm$. \\
        The problem with ``exceptional configurations''\cite{exceptional},
        which is specific to the quenched approximation,
        is enhanced by $\csw>0$\cite{exceptional:boulder}. 
        Thus small quark masses
        can only be reached near the continuum limit.  
\item   For $\nf=2$ and Wilson gauge action, the determination 
        of improvement coefficients is quite advanced.
        Interesting computations of physical observables
        can be done in the improved theory.
\item   For $\nf=3$ and Wilson gauge action, evidence for the
        existence of a first order phase transition in the 
        $(\beta,\kappa)$-plane has been reported\cite{impr:nf3_phasetransition}.
        The value of $\csw$ was fixed to typical values found for $\nf=0,2$.
        The authors conclude that one has to remain
        with lattice spacings significantly below $a=0.1\,\fm$ for this
        action. Recent investigations\cite{algo:stability} 
        have shown that simulations
        with Wilson fermions at small quark masses are 
        only algorithmically stable
        when the physical volume is sufficiently large. \footnote{The 
        lattice spacing should also be sufficiently small, but this is
        less relevant here. See also \cite{algo:cutoff}.} 
        It is possible that the simulations 
        that lead to the evidence for a phase 
        transition\cite{impr:nf3_phasetransition} suffer from this problem. 
        We therefore consider the existence (and even more its position)
        of this phase transition {\em not} as settled. Unfortunately
        it appears that only large volume simulations with good statistics 
        can tell. Note that this statement may be  controversial.
\item   With Iwasaki gauge action, $\csw$ is now known\cite{impr:csw_iwa_allnf}
        non-perturbatively
        for $\nf=0,2,3$. Non-perturbative computations of $\ca$ are in 
        progress and the relevant other improvement coefficients are known
        to 1-loop order\cite{impr:curr_iwa_pert}. A small grain of
        salt is the following.
        The perturbative expansion seems less trustworthy for this
        action, 
        since the bare coupling is significantly larger than for the Wilson gauge
        action at the same lattice spacing, while the perturbative
        coefficients are of a similar size. 
\item   In \sect{s:more} 
        we will come to the question whether more information is needed
        to do interesting computations in QCD.
\ei

\subsubsection{$\rmO(a^2)$ effects after improvement for $\nf=2$ \label{s:asq}}

The ALPHA collaboration has performed a large series of simulations
with the \SF in small volumes of a linear extent of at most
$L \approx 0.5\,\fm$ \cite{alpha:letter,alpha:nf2,mbar:nf2}. 
The purpose of these simulations was the
non-perturbative determination of the running of scale dependent
quantities (see \sect{s_p}). Extracting the continuum limit
of the step scaling functions would have been impossible  
without $\Oa$-improvement. It is fair to say that 
$\Oa$-improvement worked marvelously well in this situation.
The interested reader may in particular study the discussion
in~\cite{alpha:letter}.
The emerging picture for QCD in the Wilson
formulation is that the Symanzik expansion and improvement works very
well for lattice spacings around $0.05\,\fm$ and below. 

At larger lattice spacings, the situation is unclear at present.
In particular indirect evidence has been shown that 
$\rmO(a^2)$ effects are rather large at  
$a\approx0.1\,\fm$\cite{lat03:rainer,impr:za_nf2}. 
Large scale simulations in large volume
will tell, whether the remaining $\rmO(a^2)$ effects are 
a problem in practice. We emphasize that there is no evidence
anywhere that the Symanzik theory of $a$-effects is invalid. 
The remaining issue is how small $a$ has to be made for it 
to be precise when truncated to the first one or two correction
terms. It is also expected that smaller lattice spacings
are needed for smaller quark masses. This issue is discussed 
in Steve Sharpe's lectures.\cite{nara:steve}

\subsubsection{Do we need more? \label{s:more}}

The question arises, whether the knowledge described above 
is sufficient to do QCD computations.

First, we recall that we assumed that one is interested in 
on-shell quantities. In the popular MOM-scheme\cite{RIMOM}
used for the renormalization of composite operators,
one does, however, consider off-shell correlation functions.
While improvement in that situation has been investigated, 
too\cite{impr:offshell1,impr:offshell2,lat01:sharpe},
we continue under the assumption that off-shell 
correlation functions are avoided. So far we have 
restricted our discussion to the case of QCD with 
mass-degenerate quarks. This does of course not correspond 
to real QCD. General quark masses are no complication
for quenched quarks; we already referred to that situation 
in \sect{s:cpc}. However, dynamical quarks of different 
masses allow for a whole series of new $\rmO(a)$ counter-terms
which are proportional to the difference of quark masses \cite{impr:nondeg}.
Determining them non-perturbatively would be a considerable challenge.
However, let us look at their influence in practice\footnote{
We thank M. L\"uscher for emphasizing this point.}. For $\nf=3$,
the biggest mass difference is of order $\mstrange$. For $a\leq 0.1\,\fm$,
we have 
\bes
   a m_\mrm{q,s} \leq 1/30 \,.
\ees
Thus we are talking about very small effects even at the largest 
typical lattice spacing. It is then sufficient to know the coefficient
with, an accuracy of 0.25 or so which is surely 
possible by perturbation theory. 
The same argument applies to $\bg$\cite{impr:pap1}. 
Note, however, that for non-degenerate 
masses, \eq{e:mri} is modified already at the level
$a^0$ and this has to be taken into account\cite{impr:nondeg},
or better it is avoided
by relying on the PCAC masses only.

Obviously, for a dynamical charm quark, the question should
be reconsidered carefully.

Another issue is the improvement of more complicated
composite operators, such as 4-fermion operators.
Many improvement terms may be necessary. For these difficult cases
one may consider a mixed action approach, where (some part)
of chiral symmetry forbids the $\Oa$ terms\cite{mixedact:brs}.
Also twisted mass lattice QCD offers an interesting strategy
for these problems. We refer to Stefan Sint's lectures.

In conclusion, a very large class of interesting problems do not 
seem to need more beyond what is presently known or is being
determined. One may therefore take advantage of the
simplicity of the formulation and work on the algorithms
\cite{algo:L1,algo:L1a,algo:L2,algo:GHMC,algo:urbach,algo:RHMC} 
to be able to work at small 
lattice spacings and quark masses; see also the lectures of 
Tony Kennedy\cite{nara:tony}.

%%% Local Variables: 
%%% mode: latex
%%% TeX-master: "sect_Lambda"
%%% End: 

\lecture{II}{Fundamental parameters of QCD from the lattice}
\renewcommand\thesection      {II.\arabic{section}}

\section{The problem of scale dependent renormalization\label{s_p}}
Let us investigate the extraction of short distance parameters 
(\sect{s_sdr}a) in some detail.
After some brief comments on the conventional way of 
obtaining $\alpha$ from experiments
we explain how one can compute it at large energy 
scales using lattice QCD. This paves the road for a computation of
the $\Lambda$-parameter and then also for the renormalization group
invariant quark masses. We will finally comment briefly on other scale-dependent
renormalization.

\subsection{The extraction of $\alpha$ from experiments}
One considers experimental observables $O_i$ depending on an 
overall energy scale $q$ and possibly some additional kinematical 
variables denoted by $y$. 
The observables can be computed in a perturbative series 
which is usually written in terms of the $\MSbar$ coupling $\alphaMSbar$,
\footnote{
%%%%%%%%%%
We can always arrange the definition of the observables such that 
they start with a term $\alpha$. For simplicity we neglect all quark
mass dependences; they are irrelevant for the main points of the
present discussion.
}
\bes
 O_i(q,y) &=& \alphaMSbar(\mu) + A_i(y)\alphaMSbar^2(\mu)+\ldots\,, \quad
 \mu=q \enspace . \label{e_O_i}
\ees
For example $O_i$ may be constructed from jet cross sections and
$y$ may be related to the details of the definition of a jet.

\begin{figure}[t!]
\begin{center}
\psfig{file=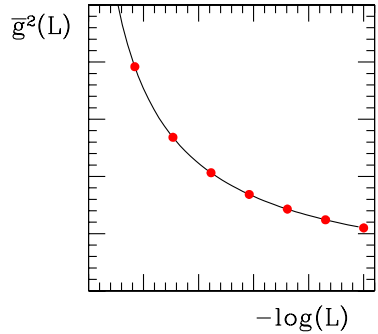, width=5cm}
\end{center}
\caption{Running of $\gbar^2$ and
  its reconstruction from the step scaling function (points).
 \label{f_running1}}
\end{figure}

The renormalization group describes the energy dependence
of $\alpha$ in a general scheme ($\alpha \equiv \gbar^2/(4\pi)$),
\bes
  \mu {\partial \bar g \over \partial \mu} &=& \beta(\bar g) \enspace ,
     \label{e_RG}
\ees
where the $\beta$-function has an asymptotic expansion     
\bes     
 \beta(\bar g) & \buildrel {\bar g}\rightarrow0\over\sim &
 -{\bar g}^3 \left\{ b_0 + {\bar g}^{2}  b_1 + \ldots \right\}
                      \enspace ,  \label{e_RGpert} \\ \nonumber
 &&b_0=\frac{1}{(4\pi)^2}\bigl(11-\frac{2}{3}\nf\bigr) 
                      \enspace ,\quad
   b_1=\frac{1}{(4\pi)^4}\bigl(102-\frac{38}{3}\nf\bigr) \enspace , 
\ees
with higher order coefficients $b_i, \, i>1 $ that depend on the scheme.
\Eq{e_RGpert} entails asymptotic freedom:
at energies that are high enough for \eq{e_RGpert} to be applicable and
for a number of quark flavors, $\nf$, that is not too large, 
$\alpha$ decreases with increasing energy as indicated in \fig{f_running1}.
The solution of \eq{e_RG} contains an integration constant,
the renormalization group invariant parameter $\Lambda$. It is (exactly)
given by
\be
 \Lambda =\mu \left(b_0\gbar^2\right)^{-b1/(2b_0^2)} \rme^{-1/(2b_0\gbar^2)}%%\\         %% && \times
           \exp \left\{-\int_0^{\gbar} \rmd x
          \left[\frac{1}{ \beta(x)}+\frac{1}{b_0x^3}-\frac{b_1}{b_0^2x}
          \right]
          \right\} \enspace , \label{e_lambdapar}
\ee
where $\gbar\equiv\gbar(\mu)$
Note that $\Lambda$ is different
in each scheme. If a coupling $\alpha_X$ is related to another one $\alpha_Y$
at the same energy scale 
via 
\be
 \alpha_Y(\mu) = \alpha_X(\mu) + c_1^{XY}\, [\alpha_X(\mu)]^2 +
       c_2^{XY}\, [\alpha_X(\mu)]^3 + \ldots \enspace ,
       \label{e_match}
\ee
their $\Lambda$-parameters are converted via
\be
     \Lambda_X/\Lambda_Y  =  \exp\{-c_1^{XY}/(8\pi b_0)\}\,.
     \label{e_s0}
\ee
From the above equations it is easy to show that \eq{e_s0} is exact.
For large $\mu$ one reads off the asymptotics
\be
\gbar^2  \buildrel {\mu}\rightarrow \infty \over\sim 
   {1 \over b_0 \ln(\mu^2/\Lambda^2)}
  -{b_1 \ln[\ln(\mu^2/\Lambda^2)] \over b_0^3 [\ln(\mu^2/\Lambda^2)]^2}
  + \rmO\left( {\{\ln[\ln(\mu^2/\Lambda^2)]\}^2 \over [\ln(\mu^2/\Lambda^2)]^{3}} \right)\,.
\ee

%%%%%%%%%%%%%%%%%%%%%%%%%%%%%FIGURE%%%%%%%%%%%%%%%%%%%%%%%%%%%%%%%%%%%
%%%%%%%%%%%%%%%%%%%%%%%%%%%%%%%%%%%%%%%%%%%%%%%%%%%%%%%%%%%%%%%%%%%%%%

\begin{figure}[tb]
\begin{center}
\psfig{file=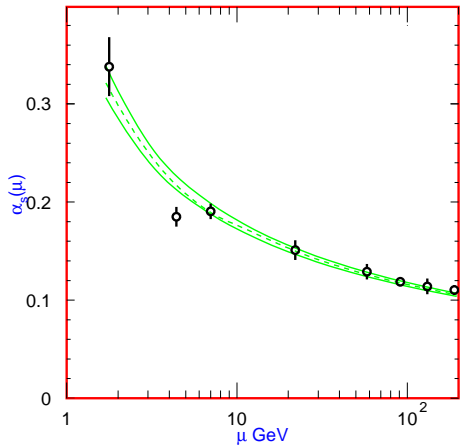, width=6cm}
\end{center}
\caption{The running coupling in the $\msbar$ scheme extracted from 
	various scattering experiments compared to the perturbative
	scale dependence. Graph of the particle data group 
	\protect\cite{PDBook}.
 \label{f:pdgalpha}}
\vspace{0.5cm}
\end{figure}

We note that -- neglecting experimental uncertainties -- $\alphaMSbar$ 
extracted in this way is obtained with a precision given by the terms
that are left out in \eq{e_O_i}. In addition to $\alpha^3$-terms, there
are non-perturbative contributions which may originate
from ``renormalons'', ``condensates'' (the two possibly being related),
``instantons'' or -- most importantly -- may have an origin that no physicist
has yet uncovered. Empirically, one observes that values
of $\alphaMSbar$ determined
at different energies and evolved to a common reference point 
using the renormalization group equation \eq{e_RG} including $b_2$
agree rather well with each other; the aforementioned uncertainties are 
apparently not very large (\fig{f:pdgalpha}).
Nevertheless, determinations of $\alpha$ are limited in precision
because of these uncertainties and in particular if there was a significant
discrepancy between $\alpha$ determined at different energies it would
be difficult to say whether this was due to the terms left out in \eq{e_O_i}
or was due to terms missing in the Standard Model Lagrangian.

%%%%%%%%%%%%%%%%%%%%%%%%%%%%%FIGURE%%%%%%%%%%%%%%%%%%%%%%%%%%%%%%%%%%%
%%%%%%%%%%%%%%%%%%%%%%%%%%%%%%%%%%%%%%%%%%%%%%%%%%%%%%%%%%%%%%%%%%%%%%

It is an obvious possibility and at the same time a challenge for 
lattice QCD to achieve a determination of $\alpha$ in one (non-perturbatively) 
well defined scheme and evolve this coupling to high energies. There one may use 
\eq{e_lambdapar} with a perturbative approximation for $\beta(\gbar)$. For a 
good precision $b_2$ should be known.

The $\Lambda$-parameter can then serve as an input for perturbative predictions
of jet cross sections or the hadronic width of the Z-boson 
and compare to high energy experiments
to test the agreement between theory and experiment. Since in the lattice
regularization, QCD is naturally renormalized through the hadron spectrum,
such a calculation provides the connection between low energies
and high energies, verifying that one and the same theory describes 
both the hadron spectrum and the properties of jets. 

\paragraph{Note.} A dis-satisfying property of $\alphaMSbar$ is that it is 
{\it only} defined in a perturbative framework; strictly speaking there is
no meaning of phrases like ``non-perturbative corrections'' in the extraction 
of $\alphaMSbar$ from experiments. The way that I have written \eq{e_O_i} 
suggests immediately what should be done instead. An observable $O_i$ itself 
may be taken as a definition of $\alpha$ -- of course with due care. 
Such  schemes called {\it physical schemes} are defined
without ambiguities. 
This is what will be done below
for observables that are easily handled in MC-simulations of QCD.
For an additional example see \cite{alpha:Grunberg_lett,alpha:Grunberg_pap}.

\subsection{Reaching large scales in lattice QCD \label{s_Rls}}
Let us simplify the discussion and restrict ourselves to
 the pure Yang-Mills theory
without matter fields in this section.
A natural candidate for a non-perturbative definition of $\alpha$
is the following. Consider a quark and an anti-quark separated by a distance
$r$ and in the limit of infinite mass. They feel a force $F(r)$, 
the derivative of the static potential $V(r)$, which can be 
computed from Wilson loops~(see e.g. \cite{books:MM}). A physical coupling is defined as
\bes
 \alpha_{\rm q \bar q} (\mu) &\equiv& { \frac{1}{C_F} r^2 F(r) \, , \quad \mu=1/r,} 
 { \quad C_F = 4/3} \enspace . \label{e_alphaqq}
\ees
It is related to the $\MSbar$ coupling by \eq{e_match} with a certain constant
$c_1^{\MSbar \,\rm q \bar q}$, which also determines the ratio  of the $\Lambda$-parameters
(\eq{e_s0}).
Note that $\alpha_{\rm q \bar q}$ is a renormalized coupling defined 
in continuum QCD.

\paragraph{Problem.}
If we want to achieve what was proposed in the previous subsection,
the following criteria must be met.
\begin{itemize}
 \item{Compute $\alpha_{\rm q \bar q} (\mu)$ at energy scales of 
       $\mu\sim 10\,\GeV$ or higher in order to be able to make the connection 
       to
       other schemes with controlled perturbative errors.}
 \item{Keep the energy scale $\mu$ removed from the cutoff $a^{-1}$ to
       avoid large discretization effects and to be able to 
       extrapolate to the continuum limit.} 
 \item{Of course, only a finite system can be simulated by MC.
       To avoid finite
       size effects one must keep the box size $L$ large
       compared to both the mass of the lightest physical state
       (the pion) as well as to a typical QCD scale, say
       the potential scale $r_0$ \cite{pot:r0}.
       }       
\end{itemize}
These conditions are summarized by
\be
  L \quad \gg \,\, r_0, {1\over \mpi} \,\sim\,{ 1 \over 0.14\GeV} \,\, \gg \,\,
  {1 \over \mu} \, \sim \, {1 \over 10\GeV}  \,\,  \gg a \enspace ,
   \label{e_conditions}
\ee      
which means that one must perform a MC-computation of an $N^4$ lattice
with $N \equiv L/a \gg 70$. In the near future 
it is impossible to perform such a computation. 
The origin of this problem 
is simply that the extraction of short distance parameters
requires that one has control over physical scales that are quite disparate.
To cover these scales in one simulation requires a very fine resolution,
which is too demanding for a MC-calculation.  
  
Of course, one may attempt to compromise in various ways. E.g. one may 
perform phenomenological corrections for lattice artifacts,
keep $1/\mu \sim a$ and at the same time reduce the value of $\mu$ compared to 
what I quoted in \eq{e_conditions}.  Calculations of
$\alpha_{\rm q \bar q}$ along these lines have been 
performed in the Yang-Mills theory  \myref{pot:michael_SU2,pot:UKQCDSU2,pot:bali_last}. 
It is difficult
to estimate the uncertainties due to the approximations that are necessary 
in this approach. More recently, results in the continuum limit could be obtained
up to $\mu = 4\,\GeV$ by simulating very large 
lattices\cite{pot:intermed,pot:lett}, and still it is not
obvious that one has reached the perturbative region.
We will come back to this.

\paragraph{Solution.}
Fortunately these compromises can be avoided altogether 
\myref{alpha:sigma}. 
The solution is to identify the two physical scales, above,
\bes
 \mu =1/L \enspace  .
\ees 
In other words, one
takes a finite size effect as the physical observable. The evolution of the coupling
with $\mu$ can then be computed in several steps, changing $\mu$ by factors of
order $2$ in each step. In this way, no large scale ratios appear
and discretization errors are  small for $L/a \gg 1$.

\begin{figure}[ht]
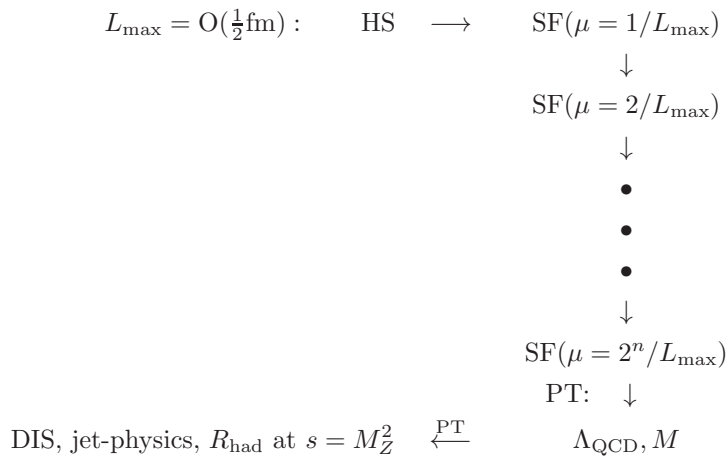

{\small
\bes
 { L_{\rm max}}=\rmO(\frac{1}{2}\fm): \qquad 
 {\rm HS} \quad \longrightarrow \quad
      &{\rm SF} (\mu=1/{ L_{\rm max}})& \quad 
               \nonumber \\
      &\downarrow&  \nonumber \\
      &{\rm SF} (\mu=2/{ L_{\rm max}})&  \nonumber \\ 
      &\downarrow&  \nonumber \\
      &\bullet&  \nonumber \\
      &\bullet&  \nonumber \\
      &\bullet&  \nonumber \\
      &\downarrow&  \nonumber \\
      &{\rm SF} (\mu=2^n/{ L_{\rm max}})& \nonumber \\
   &\mbox{ \small PT:} \quad  \downarrow \qquad \quad &  \nonumber \\
\hbox{DIS, jet-physics, }R_\mrm{had}\mbox{ at } s=M_Z^2 \quad \stackrel{\rm  PT}{\longleftarrow} 
     \quad   &\Lambda_{\rm QCD}, M & \nonumber
\ees
}
\vspace{-0.4cm}
\caption{The strategy for a non-perturbative computation of 
         short distance parameters.
         SF refers to the \SF scheme introduced in \sect{s:alphasf}
\label{f_strategy}}
\end{figure}

The complete strategy to compute short distance parameters 
is summarized in \fig{f_strategy}.
%%%%%%%%%%%%%%%%%%%%%%%%%%%%%FIGURE%%%%%%%%%%%%%%%%%%%%%%%%%%%%%%%%%%%
%%%%%%%%%%%%%%%%%%%%%%%%%%%%%%%%%%%%%%%%%%%%%%%%%%%%%%%%%%%%%%%%%%%%
One first renormalizes QCD replacing the bare parameters by hadronic
observables. This defines the hadronic scheme (HS) as explained in 
Sect.~\ref{s_hs}. At a low energy scale $\mu=1/L_{\rm max}$ this scheme 
can be related to the finite volume scheme denoted by SF in the graph. 
Within this scheme one then computes the scale evolution
up to a desired energy $\mu=2^n/{ L_{\rm max}}$. As we will see it 
is no problem to choose the number of steps $n$ large enough to be
sure that one is in the perturbative regime. There 
perturbation theory (PT) is used to evolve further to infinite energy and
compute the $\Lambda$-parameter and the renormalization group invariant
quark masses. Inserted into perturbative expressions these provide
predictions for jet cross sections or other high energy observables. 
In the graph all
arrows correspond to relations in the continuum;
the whole strategy is designed such that lattice calculations
for these relations can be extrapolated to the continuum limit.  

For the practical success of the approach, the finite volume coupling
(as well as the corresponding quark mass) must satisfy a number of criteria.
\begin{itemize}
     \item{They should have an easy perturbative expansion, such that
           the $\beta$-function (and $\tau$-function, which describes the 
           evolution of the running masses) can be computed to 
           sufficient
           order.} 
     \item{They should be easy to calculate in MC (small variance!).}
     \item{Discretization errors must be small to allow
           for safe extrapolations to the continuum limit.}   
\end{itemize}
Careful consideration of the above points led to the introduction
of renormalized coupling and quark mass through the 
Schr\"odinger functional (SF) of QCD 
\myref{SF:LNWW,alpha:SU2,alpha:SU3,SF:stefan1,impr:lett},
introduced in \sect{s:SF}. 
In the Yang-Mills theory, an alternative 
finite volume coupling was 
studied in detail in \cite{deDivitiis:1995yp,alpha:SU2impr}.

The criteria \eq{e_conditions} apply quite generally to any
scale dependent renormalization, e.g. the one of
4-fermion operators of the effective weak Hamiltonian at 
scales $\mu \ll M_\mrm{W}$.
Indeed, details have 
been worked out for several 
cases\cite{mbar:pap1,zastat:pap1,zastat:pap3,4ferm:pert,4ferm:nf0,stat:zbb_pert,lat06:filippo}.
 
A frequently applied alternative is to search for a ``window'' where
$\mu$ is high enough to apply PT but not too close to $a^{-1}$, 
the cutoff~\myref{renorm_mom:paper1}. 
An essential advantage of the details of the 
approach of \Ref{renorm_mom:paper1} as applied to the
renormalization of composite quark operators 
is its simplicity: formulating the renormalization conditions
in a {\it MOM}-scheme~\footnote{
We do not use the prefix {\it RI} (regularization independent), 
since also the SF-scheme is  regularization independent.
}, 
one may use results from perturbation theory in infinite volume
in the perturbative part of the matching. Since, however, it is 
difficult to reach
high energies $\mu$ in this approach, we will not discuss it further and refer to
\cite{renorm_mom:paper1,renorm_mom:paper2,renorm_mom:paper3,renorm_mom:paper4} 
for an idea of the present 
status and further references, instead.

%%% Local Variables: 
%%% mode: latex
%%% TeX-master: "latticen.bib"
%%% End: 

\section{The computation of $\alpha(\mu)$ and $\Lambda$}

We are now in the position to explain the details of \fig{f_strategy}
\myref{alpha:sigma,mbar:pap1}. The problem has been solved ``completely'' in the 
SU($N$) Yang-Mills theories for $N=2,3$ 
\cite{alpha:SU2,alpha:SU2impr,alpha:SU3,mbar:pap1,pot:r0_SU3,pot:intermed}.
In QCD with only two dynamical quarks the strategy has been 
carried out well
\cite{alpha:letter,alpha:nf2}, except for the last line in the graph, which needs
more work. 
In the present context, the Yang-Mills theory is of course equivalent to
the quenched approximation of QCD or the limit of zero flavors.
We will therefore also refer to results in quenched QCD.

Our central observable is the step scaling function that describes
the scale-evolution of the coupling, i.e. moving vertically in
\fig{f_strategy}. The analogous function for
the running quark mass will be discussed in the following section.

\subsection{The step scaling function}

We start from a given value of the coupling, $u=\gbar^2(L)$.
When we change the length scale by a factor $s$, the coupling
has a value $\gbar^2(sL)=u'$.\footnote{\Fig{f_strategy} is for the most
frequently used case $s=2$.}
The step scaling function, $\sigma$ 
is then defined as
\bes
 \sigma(s,u)&=&u' \enspace .
\ees
The interpretation is obvious. $\sigma(s,u)$ is a 
discrete $\beta$-function. Its knowledge allows for the 
recursive construction
of the running coupling at discrete values of the
length scale,
\bes
  u_k = \gbar^2(s^{-k} L) \enspace , \label{e_uk}
\ees
once a starting value $u_0 = \gbar^2(L)$ is specified
(cf. the points in \fig{f_running1}). The step scaling function,
$\sigma$, which is readily expressed as an integral 
of the $\beta$-function, has a perturbative expansion 
\bes
 \sigma(s,u)&=& u + 2 b_0 \ln(s) u^2 +\ldots  \enspace .
\ees
%%%%%%%%%%%%%%%%%%%%%%%%%%%%%FIGURE%%%%%%%%%%%%%%%%%%%%%%%%%%%%%%%%%%%
\begin{figure}
\vspace{0.0cm}
\centerline{\psfig{file=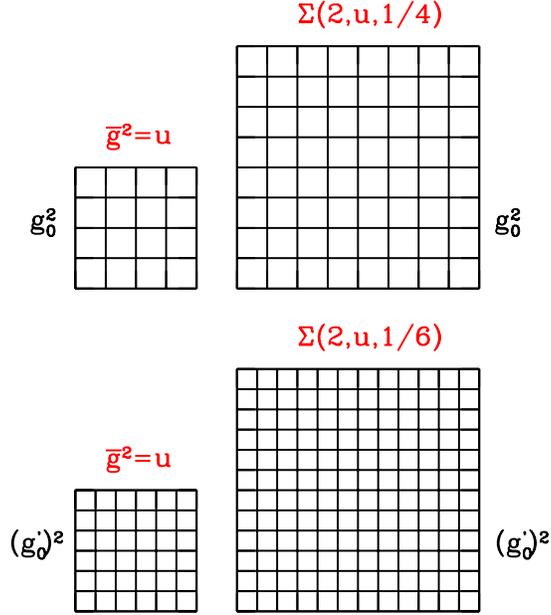,width=0.5\textwidth}}
\vspace{0.0cm}
\caption{The computation of a lattice step scaling function.
       \label{f:ssf_latt} }
\end{figure}
%%%%%%%%%%%%%%%%%%%%%%%%%%%%%%%%%%%%%%%%%%%%%%%%%%%%%%%%%%%%%%%%%%%%%%

On a lattice with finite spacing, $a$, the step scaling function 
will have an additional dependence on 
the resolution $a/L$. We define
\bes
 \Sigma(s,u,a/L)&=&u'\enspace ,
\ees
with   
\bes
 \gbar^2(L)&=&u, \quad  \gbar^2(sL)=u'\, , \quad 
 \mbox{$g_0$ fixed, $L/a$ fixed} \enspace .
\ees
The continuum limit $\sigma(s,u) = \Sigma(s,u,0)$ is then 
reached by performing calculations for several
different resolutions and extrapolation $a/L \to 0$. 
The computation of $\sigma(2,u)$ is illustrated
in \fig{f:ssf_latt}.  
In detail, one performs the following steps:
\begin{itemize}
 \item[1.] Choose a lattice with $L/a$ points in each direction.
 \item[2.] Tune the bare coupling $g_0$ such that the renormalized
           coupling $\gbar^2(L)$ has the value $u$ and tune the bare quark mass, $\mbare$,
           such that the PCAC-mass, defined at fixed physical kinematics~\cite{alpha:letter},
           vanishes.\footnote{The details of the definition of
             the kinematics affects only the cutoff effects, see \sect{s:impr}.} 
 \item[3.] At the same value of $g_0$, simulate a lattice 
           with twice the linear size; compute 
           $u'=\gbar^2(2L)$. This determines the lattice step scaling function
           $\Sigma(2,u,a/L)$.
 \item[4.] Repeat steps 1.--3. with different resolutions $L/a$ and extrapolate
           $a/L \to 0$.       
\end{itemize}
Note that step 2. takes care of the renormalization and 3. determines
the evolution of the {\it renormalized} coupling.

%%%%%%%%%%%%%%%%%%%%%%%%%%%%%FIGURE%%%%%%%%%%%%%%%%%%%%%%%%%%%%%%%%%%%
\begin{figure}
\vspace{0.0cm}
\centerline{\psfig{file=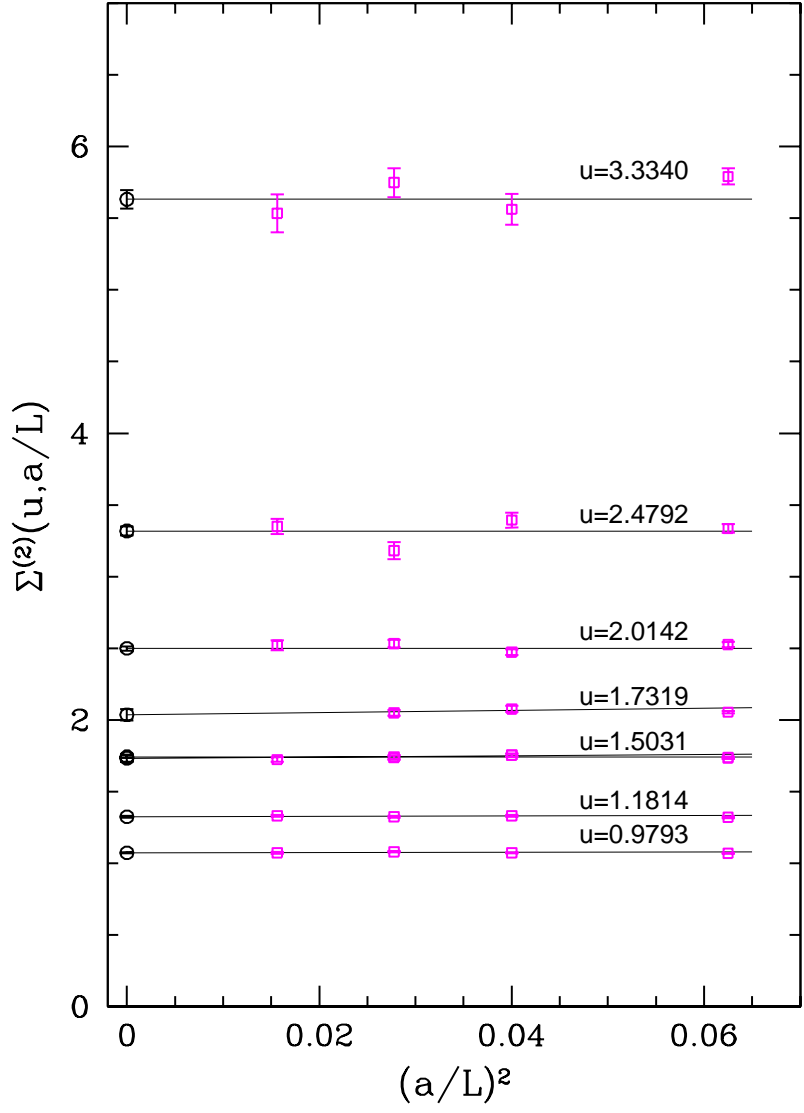,width=0.8\textwidth}}
\vspace{0.0cm}
\caption{The lattice step scaling function
         after 2-loop observable improvement for $\nf=2$.
       \label{f:Sigma_nf2} }
\end{figure}
%%%%%%%%%%%%%%%%%%%%%%%%%%%%%%%%%%%%%%%%%%%%%%%%%%%%%%%%%%%%%%%%%%%%%%

The presently most advanced numerical results are displayed in \fig{f:Sigma_nf2}. 
The coupling used is exactly the one defined in \sect{s:alphasf}
and the calculation is done in the theory with $\nf=2$ flavors of $\Oa$-improved fermions.
One observes that the dependence on the resolution is very weak, 
in fact it is not observable within the precision of the data in 
\fig{f:Sigma_nf2}. We now investigate in more detail how the
continuum limit  of $\Sigma$ is reached.  As a first step,
we turn to perturbation theory.

\subsection{Lattice spacing effects in perturbation theory \label{s:finiteaPT}}

Symanzik has investigated the cutoff dependence of field theories
in perturbation theory~\cite{Symanzik:1982}. Generalizing
his discussion to the present
case, one concludes that the lattice spacing effects
have the expansion
\bes
 {\Sigma(2,u,a/L)-\sigma(2,u) \over \sigma(2,u)} &=& \delta_1(a/L) \, u +
 \delta_2(a/L) \, u^2+ \ldots 
  \label{e_delta} \\ 
 \delta_n(a/L)  &{\buildrel {a/L}\rightarrow0\over\sim } &
 \sum_{k=0}^n e_{k,n} [\ln(\frac aL)]^k { \left(\frac aL\right)} +
              d_{k,n} [\ln(\frac aL)]^k { \left(\frac aL\right)^2} + \ldots
              \enspace .
 \nonumber
\ees
We expect that the continuum limit 
is reached with corrections $\rmO(a/L)$ also beyond perturbation
theory. In this context $\rmO(a/L)$
summarizes terms that contain at least one power of
$a/L$ and may be modified by logarithmic corrections as it is the case
in \eq{e_delta}.
To motivate this expectation recall \sect{s_Io}, where we explained
that lattice artifacts correspond to irrelevant
operators %\footnote{For a more precise meaning of this terms
%one must discuss Symanzik's effective theory. We refer the reader 
%to \cite{paper1} for such a discussion.}, 
which carry explicit factors of the lattice spacing. Of course,
an additional $a$-dependence comes from their anomalous dimension,
but in an asymptotically free theory such as QCD, this just is 
a logarithmic (in $a$) modification.

%%%%%%%%%%%%%%%%%%%%%%%%%%%%%FIGURE%%%%%%%%%%%%%%%%%%%%%%%%%%%%%%%%%%%
\begin{figure}[t]
\vspace{0cm}

\centerline{
\psfig{file=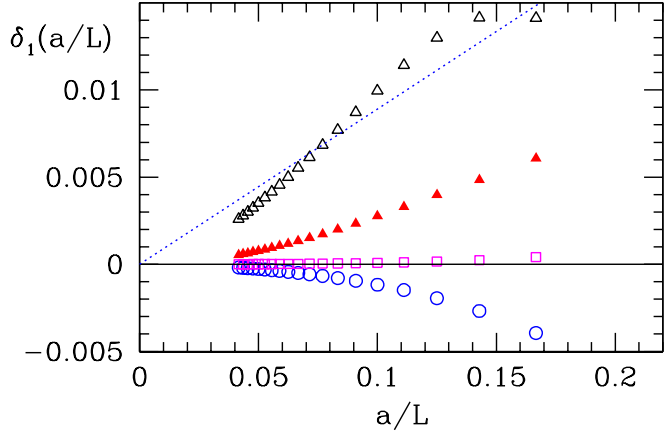,width=0.6\textwidth}
}
\vspace{0cm}
\caption{Lattice artifacts at 1-loop order. Points show
         $\delta_1(a/L) $ 
         for the SU(3) Yang-Mills theory with 1-loop improvement.
	 Wilson (circles), Symanzik (squares), Iwasaki (filled triangles)
	 and DBW2 (empty triangles) are shown. 
         The dotted line corresponds to the linear piece in $a$,
         when only tree-level improvement is used, instead (only for Wilson).
         \label{f_delta1}}
\end{figure}
%%%%%%%%%%%%%%%%%%%%%%%%%%%%%%%%%%%%%%%%%%%%%%%%%%%%%%%%%%%%%%%%%%%%%%

As mentioned in \sect{s:latt}, the lattice artifacts may be
reduced to $\rmO((a/L)^2)$ by canceling the 
leading irrelevant operators. In the case at hand, this is achieved by  a 
proper choice of $\ct(g_0)$.
It is interesting to note, that by using the perturbative
approximation 
\bes
 \ct(g_0) = 1+ \ct^{(1)}g_0^2
\ees
one does not only eliminate $e_{k,n}$ for $ n=0,1$ but also the logarithmic 
terms generated at higher orders ($n>1$)are reduced, 
\bes
e_{n,n}=0, \quad e_{n-1,n}=0\enspace .
\ees
For tree-level improvement, $\ct(g_0) = 1$, the corresponding
statement is $e_{n,n}=0$. Heuristically, the latter is easy to understand.
Tree-level improvement means that the propagators and vertexes
agree with the continuum ones up to corrections
of order $\rmO(a^2)$. Terms proportional to $a$ can then arise 
only through a linear divergence of the loop integrals. Once this happens, 
one cannot have the maximum number of 
logarithmic divergences any more; consequently 
$e_{n,n}$ vanishes.

To demonstrate further that the abelian field introduced in the previous 
section induces small lattice artifacts, we show $\delta_1(a/L)$ for
the one loop improved case in \fig{f_delta1} (circles). 
The term that is canceled by the proper choice 
$\ct^{(1)}=-0.089$ is shown as a dashed line. 
The left over 
$\rmO((a/L)^2)$-terms are  below the 1\% level
for couplings $u\leq2$ and lattice sizes $L/a\geq 6$. 
For not too large coefficient of the $1\times2$ loop in the action,
they are close to quadratic in $a/L$ in the range of interest. 
Also the fermion contribution and the 2-loop cutoff effects
are known~\cite{pert:1loop,pert:2loop_fin}. When the $\Oa$-improved
theory is used (and the coupling has reasonable values), 
they turn out to be smaller than the 1-loop terms discussed here.

Of course the cutoff effects also depend on the gauge action. In one
class of such actions, one adds a $1\times2$ rectangular loop. Its
coefficient has originally been determined to have 
tree-level Symanzik $\Oasq$ improvement in the pure gauge 
theory~\cite{impr:onshell}.
Later other choices have been proposed by Y. Iwasaki \cite{Iwasakiact}
and the QCD-TARO collaboration \cite{DBW2coeff}, based on renormalization
group considerations. We refer to these as ``Symanzik,Iwasaki,DBW2''.
The choice of the action close to 
the \SF boundaries does of course involve an additional
freedom discussed in \cite{SF:Iwasaki,pert:1loopshinji}. For the choice 
favored in \cite{pert:1loopshinji}, we included the corresponding $\delta_1(a/L)$ 
in \fig{f_delta1}. For
the tree-level Symanzik improved theory the cutoff effects are remarkably
small. One might think that this is so by construction, but note that here we 
are discussing 1-loop effects and in addition also boundary operators contribute to
the SET expansion of the cutoff effects. 

Altogether, we now understand better
why the $a/L$-dependence is so small in \fig{f:Sigma_nf2}. Still,
because the continuum limit is so important, and we have seen
that unexpected difficulties may be present (\sect{s:unexpect}),
we continue its discussion in the pure gauge theory.
There, numerical simulations with better resolutions have been carried 
out.

%%%%%%%%%%%%%%%%%%%%%%%%%%%%%FIGURE%%%%%%%%%%%%%%%%%%%%%%%%%%%%%%%%%%%
\begin{figure}[t]
\vspace{0.0cm}
\centerline{\psfig{file=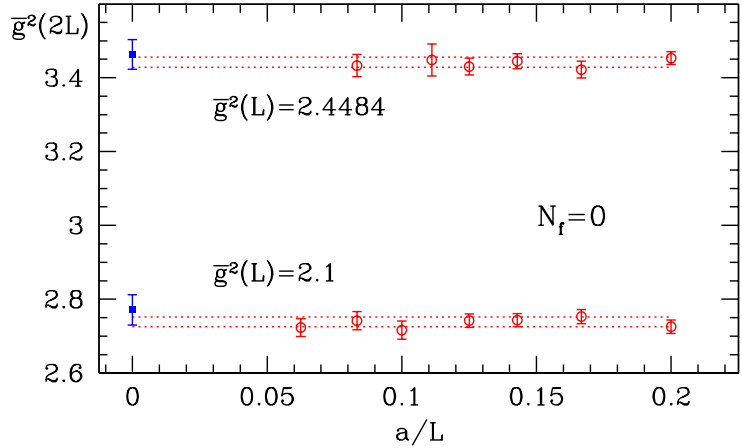,width=0.7\textwidth}}
\vspace{0.0cm}
\caption{The lattice step scaling function
         after 2-loop observable improvement for $\nf=0$. Data from
	\protect\cite{mbar:pap1,lat01:jochen}.
	The point at $a=0$ is the continuum limit of \protect\cite{mbar:pap1}
	taken with a subset of the data.
       \label{f_Sigma_nf0}}
\end{figure}
%%%%%%%%%%%%%%%%%%%%%%%%%%%%%%%%%%%%%%%%%%%%%%%%%%%%%%%%%%%%%%%%%%%%%%

\subsection{The continuum limit -- universality  \label{s:finiteaNP}}

In fact, it is not only of interest to investigate 
how exactly the continuum
limit is approached.  Its very existence and its universality, i.e. 
the independence of the renormalized continuum observable from
the lattice action, is worth testing 
by precise MC computation. 

Above we have discussed how one can add boundary terms to the 
{\em action} to systematically reduce the cutoff effects linear in $a$.
Given the knowledge of the observables in perturbation theory at
finite lattice spacing, we can go further and remove the cutoff-effects
for arbitrary $a/L$, i.e. for instance including $\rmO((a/L)^2)$, {\em but up to a
fixed order in $u$}. This is achieved by the {\em improvement of the
observable} \cite{alpha:SU2impr,alpha:letter,alpha:nf2}
\be
   \Sigma^{(k)}(2,u,a/L) = {\Sigma(2,u,a/L) \over 1 + \sum_{i=1}^k\delta_i(a/L) \, u^i} \,.
\ee

The approach has been carried out with $k=2$ for the $\nf=2$ data (\fig{f:Sigma_nf2})
and is tested thoroughly for  $\nf=0$ in \fig{f_Sigma_nf0}. 

\begin{figure}[t]
%\vspace{-1.8cm}
%
\centerline{\psfig{file=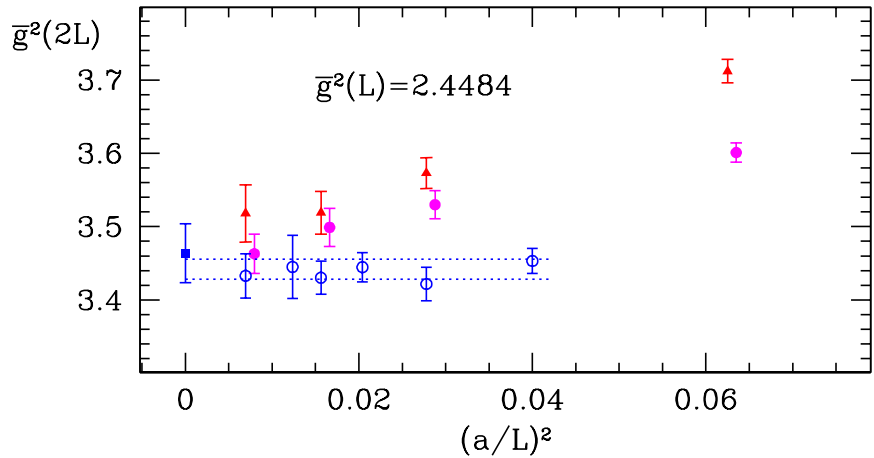,width=0.8\textwidth}}
%\vspace{-0.8cm}
\caption{Universality test in the SU(3) Yang Mills theory. The data from 
top (triangles) to 
bottom (open circles) are for the Iwasaki, Symanzik  
and Wilson gauge action. Both the 
boundary improvement of the action and the improvement of the observables
are carried out. At present this is possible at the  2-loop level for the 
Wilson gauge action (same as \protect\fig{f_Sigma_nf0})
and at the 1-loop level otherwise. Data are from 
\protect\cite{alpha:Iwasaki1,alpha:letter}.
         \label{f:universal} }
\end{figure}

Recently,
a nice universality test has been carried out by the CP-PACS collaboration 
\cite{alpha:Iwasaki1}. Obviously the results in \fig{f:universal} are
well compatible with an $a\to0$ limit which is independent of the
action. Similar results exist for the SU(2) theory \cite{alpha:SU2impr}.
These results leave little doubt
that the continuum limit of the \SF exists and is independent 
of the lattice action. In turn this also supports the statement
that the \SF is renormalized after the renormalization of the coupling 
constant.

%%%%%%%%%%%%%%%%%%%%%%%%%%%%%FIGURE%%%%%%%%%%%%%%%%%%%%%%%%%%%%%%%%%%%
%%%%%%%%%%%%%%%%%%%%%%%%%%%%%%%%%%%%%%%%%%%%%%%%%%%%%%%%%%%%%%%%%

We return to the extraction of $\sigma$ in $\nf=2$ QCD. 
Even though $a$-effects are not statistically significant,
one has to be careful how one extracts the continuum limit. 
The worry does not so much concern the central values, but the
correct estimate of their uncertainties.
For example just averaging data at all values of $a/L$ produces an
unrealistically small statistical error, because one has then 
assumed that $a$-effects are entirely absent, although the data
tell only that they are smaller than statistical uncertainties.
One possible strategy for the continuum extrapolation
is thus a fit to a constant that uses the lattices with $L/a=6,8$ only.
As a check of this procedure, different variants of 
a combined continuum extrapolation of all the data sets, but excluding $L/a=4$
were carried out.
For example the ansatz
\begin{eqnarray}
\label{eq:rho}
  \Sigma^{(2)}(2,u,a/L) = \sigma(2,u) + \rho \,u^4\, (a/L)^2 \nonumber
\end{eqnarray}
with a constant $\rho$ was fitted to the data. The final conclusion
was that the simple fit to a constant (for $L/a=6,8$)
yields realistic error estimates for the existing data set \cite{alpha:nf2}.
A further check of this procedure is \fig{f_Sigma_nf0} where the dotted lines
represent the error band obtained in this way and the four points at smaller
lattice spacing are in perfect agreement with this band.

After this long -- but important -- discussion of cutoff effects,
we are convinced that we have continuum results for the 
step scaling function with realistic uncertainties. They are 
ready to be used to construct the running coupling and the 
$\Lambda$-parameter.

\subsection{The running of the coupling}

%%%%%%%%%%%%%%%%%%%%%%%%%%%%%%%%%%%%%%%%%%%%%%%%%
\begin{figure}[t]
\vspace{9pt}
\centerline{\psfig{file=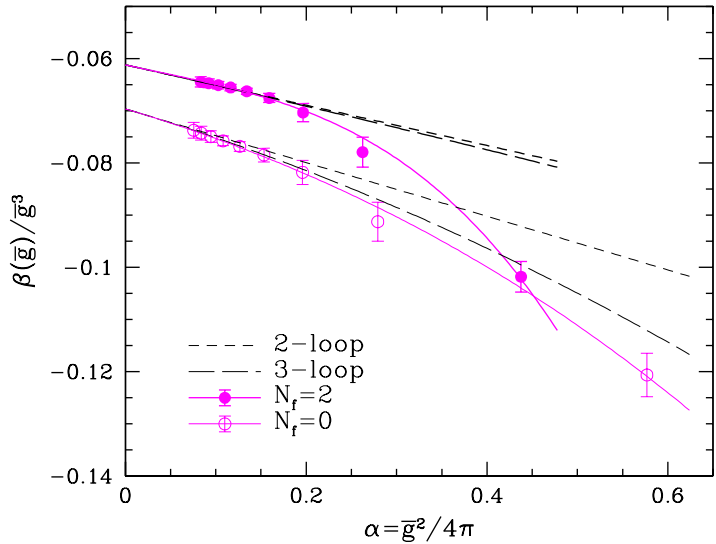,width=9.5cm}}
\vspace{-0mm}
\caption{The QCD $\beta$--function in the SF scheme.}
\label{f:beta}
\end{figure}
%%%%%%%%%%%%%%%%%%%%%%%%%%%%%%%%%%%%%%%%%%%%%%%%%

\begin{figure}[!b]
%\hspace{-2mm}
\psfig{file=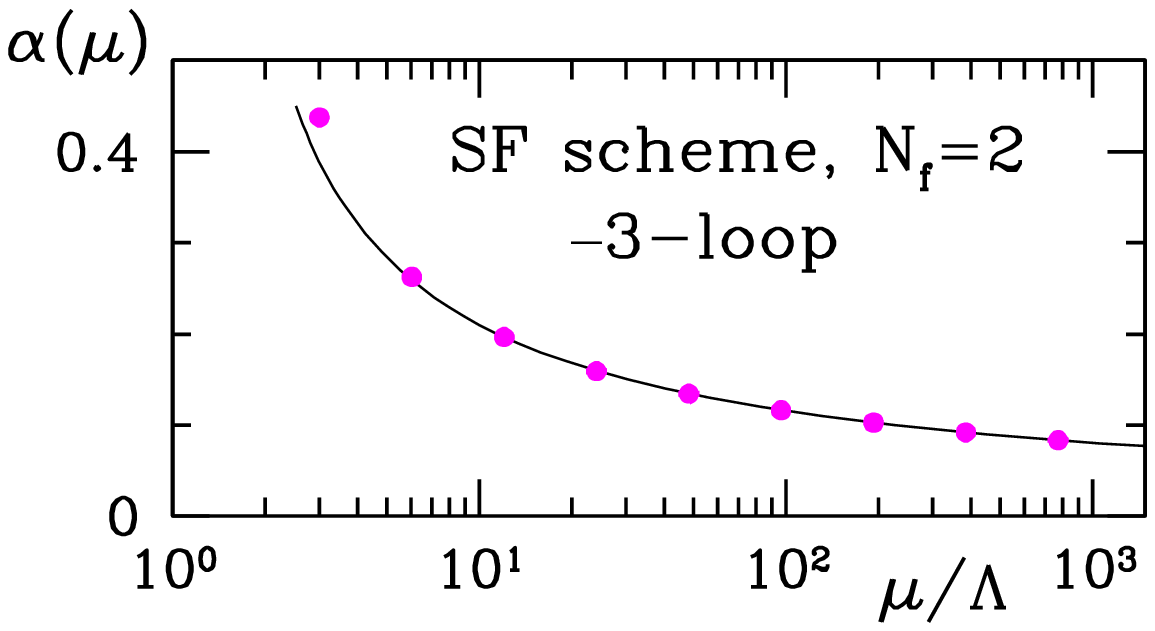,width=0.49\textwidth}\hfill
\psfig{file=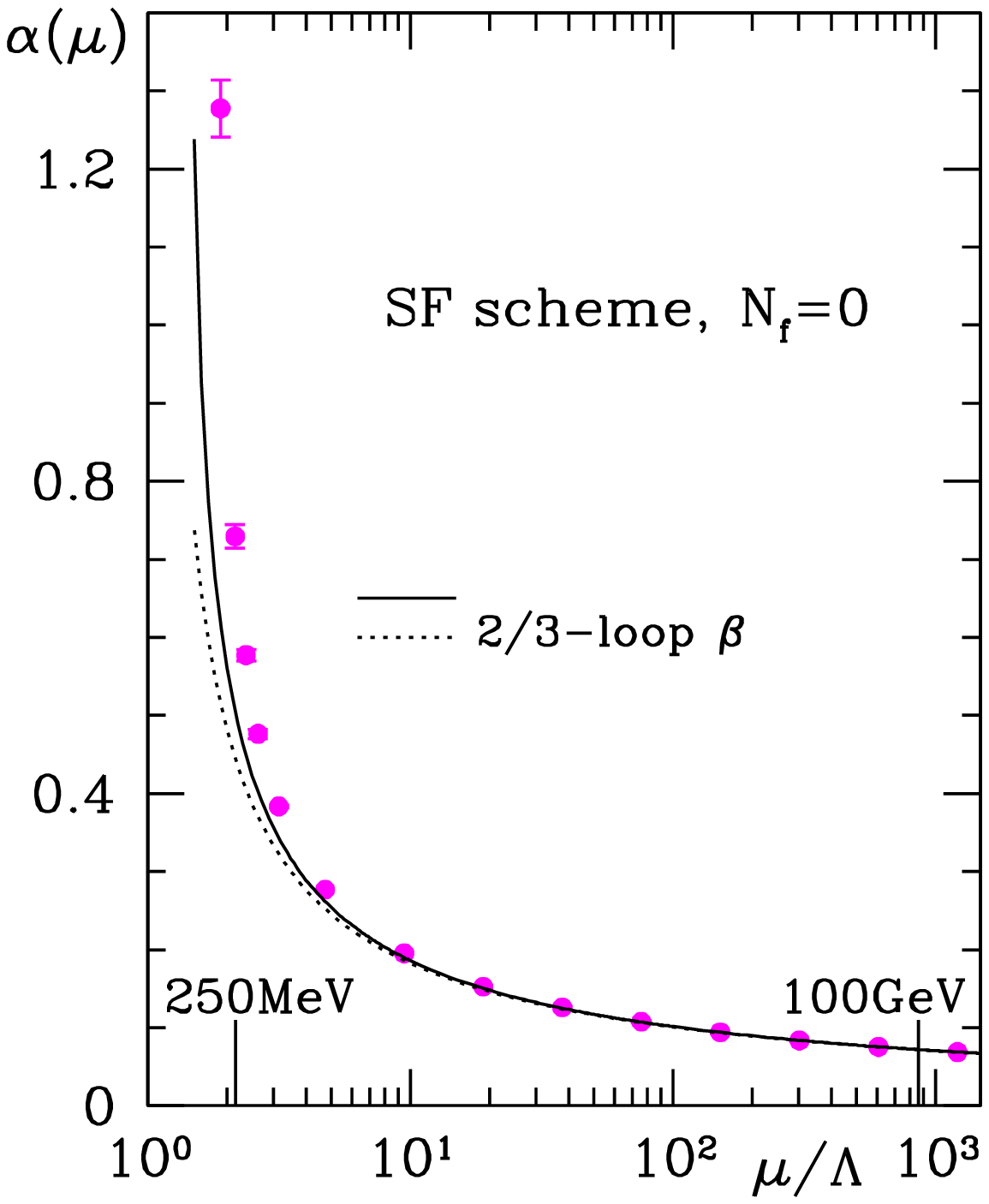,width=0.49\textwidth}
%\hspace{-2mm}
\vspace{-0mm}
\caption{Running coupling for $\nf=2$ (left) and
 $\nf=0$ (right).}
\label{f:runn}
\end{figure}

The numerical values of $\sigma(u)$ 
are next represented by 
a smooth interpolating function (a polynomial in $u$). 
With this function the running coupling $\gbar^2(2^{-i}\Lmax)\equiv u_i$ 
can be constructed
from the recursion
\bes
 u_\mrm{max}\equiv u_0 =\gbar^2(\Lmax)\,,\quad \sigma(u_{i+1}) = u_i\,, \; i=0 \ldots n\,;  
\ees
the result is shown in \fig{f:runn} for the (arbitrary) choice 
$u_\mrm{max}=5.5$. One can also 
set up a recursion for the $\beta$-function itself \cite{alpha:nf2},
\bes
  \beta(\sqrt{u_{i}}) = \sqrt{u_{i+1} / u_{i} }\, \sigma'(u_{i+1})\,\beta(\sqrt{u_{i+1}})\,.
\ees
Together with a
start value for the $\beta$-function taken from perturbation 
theory (3-loop) at the weakest coupling ($\alpha \approx 0.08$)
this yields the numerical results \fig{f:beta}.
Their agreement with perturbation theory 
is excellent at weak couplings $\alpha<0.2$, while at
the largest couplings significant deviations
from perturbation theory are present for $\nf=2$. 
Indeed the difference between non-perturbative
points and 3-loop can't be described by an
effective 4-loop term {\em with a reasonable coefficient}.
At the same time the perturbative series just by itself 
does not show signs of its failure at, say, $\alpha\approx0.3$:
instead successive orders yield smaller and smaller 
corrections.

We return to the running couplings shown in 
 \fig{f:runn}. In the zero flavor case,
also the region of $\mu$ of around $250\,\MeV$ was 
investigated with a specifically adapted strategy~\cite{lat01:jochen}. 
In this region, the SF coupling 
shows the rapid growth expected from a strong coupling 
expansion. 

Initially, the graphs  \fig{f:runn}
are obtained for $\mu$ in units of $\mu_\mrm{min}=1/\Lmax$.
One chooses 
 $u_\mrm{max}$ relatively large, but within the range covered
by the non-perturbative computation of $\sigma(u)$.
The artificial scale $\Lmax$ has been replaced by the  
$\Lambda$ parameter by use of \eq{e_lambdapar}. We proceed
to explain this step.

%%%%%%%%%%%%%%%%%%%%%%%%%%%%%%%%%%%%%%%%%%%%%%%%%%
%%%%%%%%%%%%%%%%%%%%%%%%%%%%%%%%%%%%%%%%%%%%%%%%%

\subsection{The $\Lambda$ parameter}

We may evaluate \eq{e_lambdapar} for the last few data points
in \fig{f:runn} using the 3-loop approximation to the 
$\beta$-function in the SF-scheme. 
The resulting $\Lambda$-values are essentially independent 
of the starting 
point, since the data follow the perturbative running very accurately
at large $\mu$.
This excludes a sizable contribution to the $\beta$-function beyond
3-loops in this region. Indeed, a typical estimate of a 4-loop term 
in the $\beta$-function would change the value of $\Lambda$ 
by a tiny amount. The corresponding uncertainty can be neglected compared
to the statistical errors.

Changing $\Lambda$ from the SF scheme to the $\msbar$ scheme we then
have 
\bes
  \nf=0\,,&&u_\mrm{max}=3.48\,:\; \ln(\Lambda_{\msbar} \Lmax)= -0.84(8)\,,
  \label{e:lll0}\\
  \nf=2\,,&&u_\mrm{max}=4.61\,:\; \ln(\Lambda_{\msbar} \Lmax)= -0.40(7)\,.  
  \label{e:lll2}
\ees
It remains to connect the artificially defined
length scale $\Lmax$ to an experimentally measurable
low energy scale of QCD such as the proton mass or the
kaon decay constant, $\Fk$. 

So far it has been
convenient to first evaluate $\Lmax$ in units of the 
low energy scale $r_0$. This reference scale is precisely 
defined through the QCD static
quark potential \cite{pot:r0} but related to experiments
only through potential models: $r_0\approx0.5\,\fm$. 
For $\nf=0$ a detailed investigation resulted 
in the continuum limit \cite{pot:r0_SU3,pot:intermed} 
\be
   \nf=0\,,\; u_\mrm{max}=3.48\,:\;\Lmax/r_0=0.738(16) 
\ee 
and thus  
\be
  \Lambda_{\MSbar}^{(0)} r_0 = 0.60(5)\,. \label{e:lambdar0_nf0}
\ee 

In the $\nf=2$ theory, the situation is illustrated in
\tab{t:lamr0} which relies on results for $r_0/a$
from \cite{JLQCD:nf2b52,QCDSF:nf2mstrange}. On the one hand,
all the numbers in italic are consistent, indicating
that lattice spacing effects are small, on the other
hand the first column shows that $a$ is not yet varied 
very much. At the moment 
$$
 \Lambda_{\MSbar}^{(2)} r_0 = 0.62(4)(4) 
$$
is quoted, where the second error generously
covers the range of numbers in italic in \tab{t:lamr0}
and the first one comes from \eq{e:lll2}.

\begin{table}[tbh] 
{\begin{center} \small
\begin{tabular}{cccccc}
\hline \\[-1.5ex]
              &  & \multicolumn{2}{c}{$u_{\rm max}=3.65$} 
                  & \multicolumn{2}{c}{$u_{\rm max}=4.61$} \\[0.5ex]
 $\beta=6/g_0^2$ & $r_0/a$ & $L_{\rm max}/a$ & $\Lambda_{\MSbar}\, r_0$ &
                        $L_{\rm max}/a$ & $\Lambda_{\MSbar}\, r_0$ \\[0.5ex]
\hline\\[-1.5ex]
5.00  & 5.45(5)(20) &  4.00(6) & 0.655(27) &  6.00(8)  & {\em 0.610(25)} \\           
5.29  & 6.01(4)(22) &  4.67(6) & {\em 0.619(25)} &  6.57(6)  & {\em 0.614(24)} \\ 
5.40  & 7.01(5)(15) &  5.43(9) & {\em 0.621(17)} &  7.73(10) & {\em 0.609(16)} \\[0.5ex]     
\hline
\end{tabular}
\end{center}
}
\caption{$\Lambda$--parameter in units of $r_0$ for $\nf=2$ for
        different resolutions in the low energy part 
        of the calculation. Two different values of 
      $u_{\rm max}$ are considered, but $\Lambda_{\MSbar}\, r_0$ 
    should be independent of these when cutoff effects are small.} 
\label{t:lamr0}
\end{table}

%%% Local Variables: 
%%% mode: latex
%%% TeX-master: "sect_Lambda"
%%% End: 

\subsection{Discussion}

The scale dependence of the SF coupling is
close to perturbative below $\alpha_\mrm{SF} =0.2$ and becomes
non-perturbative above $\alpha_\mrm{SF} =0.25$. In fact 
a strong coupling expansion suggests that {\em this} 
coupling grows exponentially for large $L$.
In the $\nf=0$ theory it was possible to verify this
behavior for $L$ close to $1 \,\fm$ \cite{lat01:jochen} (\fig{f:runn}).
Apart from the determination of $\Lambda$, 
an achievement of this investigation is the confirmation that the 
transition between this non-perturbative region and the perturbative one
is very smooth.

%%%%%%%%%%%%%%%%%%%%%%%%%%%%%%%%%%%%%%%%%%%%%%%%%%
\begin{figure}[tbh]
\centerline{\psfig{file=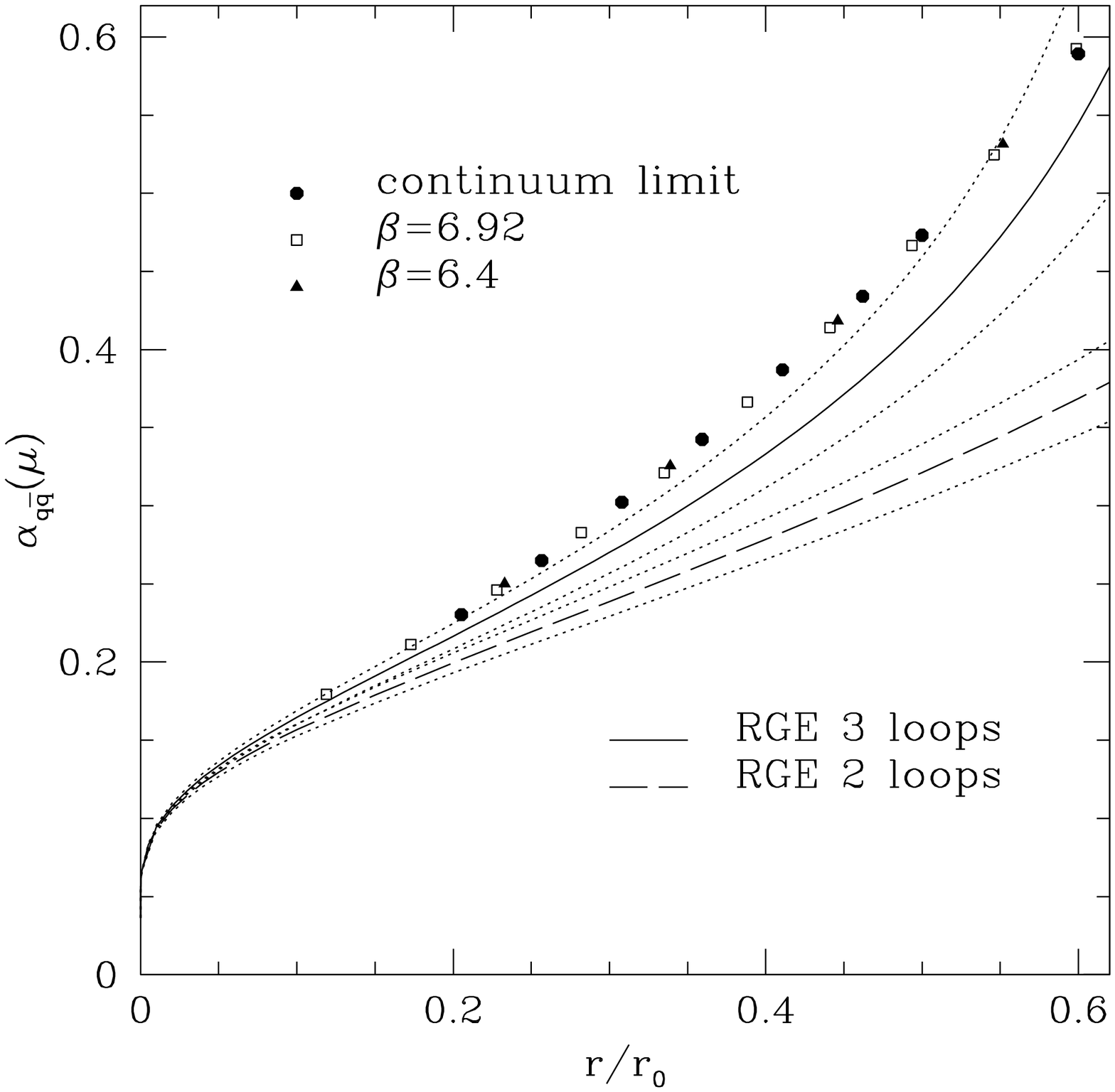,width=8.0cm}}
\vspace{-0mm}
\caption{Running coupling in the $\mrm{q\bar q}$-scheme (\eq{e_alphaqq}).
        The perturbative prediction is given by the 
        relation \eq{e_lambdapar} between the running
        coupling and $\Lambda_\mrm{q\bar q}^{(0)}$. The dotted
        lines show the uncertainty of $\Lambda$ \eq{e:lambdar0_nf0}. }
\label{f:alphaqq}
\end{figure}
%%%%%%%%%%%%%%%%%%%%%%%%%%%%%%%%%%%%%%%%%%%%%%%%%

With some effort this transition region was also bridged for 
$\alphaqq$ in the SU(3) pure
gauge theory. As shown in \fig{f:alphaqq}, the non-perturbative
continuum results agree reasonably well with the perturbative prediction at
the largest $\mu$ (small $r$). The prediction is parameter free, since
the $\Lambda$-parameter, \eq{e:lambdar0_nf0}, can be changed to
the $\rm q\bar q$-scheme, \eq{e_alphaqq}, where also the 3-loop $\beta$-function
is known~\cite{pot:2loop1,pot:2loop2,pot:lett}. The figure also 
illustrates that a reliable determination of $\Lambda$ is
not possible from $\alphaqq$ because for the accessible $\mu$ 
the 3-loop contribution to the $\beta$-function is still
very significant.

\begin{table}[htb] 
{\small
\begin{center}\begin{tabular}{lcccc}
\hline \\[-1.5ex]
         source & \multicolumn{4}{c}{$\Lambda_{\MSbar}\, r_0$} \\[0.5ex]
          &   $\nf=$0  &  $\nf=$2  &  $\nf=$4  &  $\nf=$5 \\[0.5ex]
\hline\\[-1.5ex]
        ALPHA\cite{alpha:nf2,alpha:SU3}  & 0.60(5) & 0.62(6) & & \\[0.5ex]
        world average\cite{Bethke:2004uy} & & & 0.74(10) &0.54(8) \\[0.5ex]
        DIS\@NNLO\cite{alpha:blum1}  &&& 0.57(8) &  \\[0.5ex]     
\hline
\end{tabular}\end{center}
}
\caption{Results for  $\Lambda_{\MSbar} r_0$ for different number of flavors.
    } 
\label{t:lamr01}
\end{table}

%%% Local Variables: 
%%% mode: latex
%%% TeX-master: "sect_Lambda"
%%% End: 

In \tab{t:lamr01} we compare our
results for  $\Lambda_{\MSbar} r_0$ to selected
phenomenological ones, where we set $r_0=0.5\,\fm$. 
There appears to be an irregular $\nf$-dependence, but we note that\\
1.~~the errors are not very small yet,\\
2.~~the 4-flavor  $\Lambda$ is obtained from the 5-flavor one
by perturbation theory \cite{alpha:bernwetz}. 
For this to be accurate, perturbation theory has to 
be accurate for  $\mu \ll m_\mrm{beauty}$, which is not completely
obvious.

\subsection{Improvements are necessary}

The most urgently needed improvement
of the present results is to eliminate
the model dependence which is intrinsic in the 
use of $r_0$. So one should replace $\Lmax/\r_0$ by
$\Fk\times\Lmax$ (computed at small enough 
light quark masses and small $a$)
and an effort is presently being made.
Also the strange quark sea has to be included (``2+1'') and one should 
estimate the effects of the charm quark. Such 2+1 simulations are
for example being carried out by JLQCD and CP-PACS, who have also 
studied the computation of the SF coupling with the gauge actions they 
are using~\cite{alpha:Iwasaki1}.

These improvements will come and I am convinced that  lattice results 
will yield the best
controlled and most precise results for $\Lambda$ in the long run.
The reason is simple: 
I essentially described {\em all} the sources of systematic errors.
The kind of assumptions one has to make are minimal.

Let me also mention that there is a large number of other results
for $\alpha$ from lattice gauge theory, 
where $\alpha(\mu \sim a^{-1})$ is extracted from quantities related
to the cutoff. Some of them cite a very small error
\cite{alpha:lepagemilk}. As discussed in \cite{reviews:Schlad},
it is not easy to estimate the systematic errors in these computations,
mainly because one cannot separately discuss higher order 
perturbative corrections and discretization errors.
We thus think it is very desirable to carry out the program of 
\fig{f_strategy} with good precision and the relevant number of flavors.

%%% Local Variables: 
%%% mode: latex
%%% TeX-master: "sect_Lambda"
%%% End: 

% quark mass
\section{Renormalization group invariant quark masses}

The computation of running quark masses and the 
renormalization group invariant (RGI) quark mass proceeds in
analogy to the computation of $\alpha(\mu)$.
Since we are using a mass-independent renormalization scheme (cf. \sect{s_Rm}),
the renormalization
(and thus the scale dependence) is independent of the flavor of
the quark.  When we consider ``the'' running mass below, any one flavor 
can be envisaged; the scale dependence is the same for all of them.

The renormalization group equation for the coupling \eq{e_RG}
is now accompanied
by one describing the scale dependence of the mass,
\bes
  \mu {\partial \mbar \over \partial \mu} &=& \tau(\bar g) \enspace ,
     \label{e_RG_m}
\ees
where $\tau$ has an asymptotic expansion
\bes     
 \tau(\bar g) & \buildrel {\bar g}\rightarrow0\over\sim &
 -{\bar g}^2 \left\{ d_0 + {\bar g}^{2}  d_1 + \ldots \right\}
                      \, , \qquad
 d_0={8}/{(4\pi)^2} 
 \enspace ,  \label{e_RGpert_m}
\ees
with higher order coefficients $d_i, \, i>0 $ which depend on the scheme.

Similarly to the $\Lambda$-parameter, we may define a 
renormalization group invariant quark mass, $M$, by
the asymptotic behavior of $\mbar$,
\bes
 M &=& \lim_{\mu \to \infty} \mbar(\mu) [2 b_0\gbar(\mu)^2]^{-d_0/2b_0}\enspace .
\ees
It is an easy exercise to show that $M$ does not depend 
on the renormalization scheme. 
It can be computed in the SF-scheme and used
afterward to obtain the running mass in any other scheme by inserting
the proper $\beta$- and $\tau$-functions in the renormalization
group equations.

%%%%%%%%%%%%%%%%%%%%%%%%%%%%%FIGURE%%%%%%%%%%%%%%%%%%%%%%%%%%%%%%%%%%%
\begin{figure}[ht]
\centerline{
\psfig{file=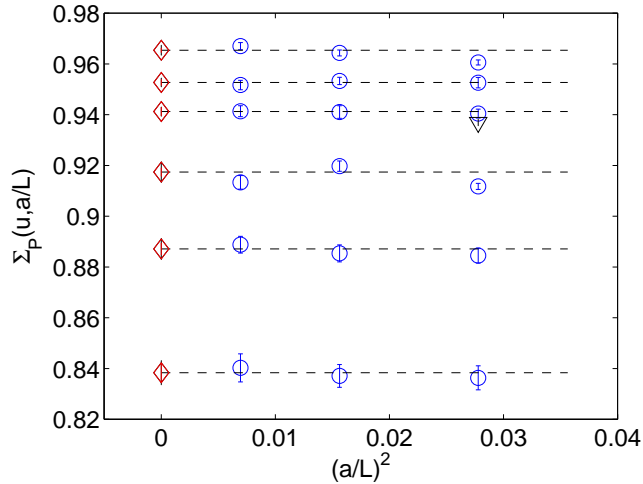,%
width=9cm}}
\caption{Lattice spacing dependence of
the step scaling function for the quark mass for $\nf=2$.
The coupling $u$ ranges from $u=0.979$ to $u=3.33$. 
\label{f_sigmap}}
\end{figure}
%%%%%%%%%%%%%%%%%%%%%%%%%%%%%%%%%%%%%%%%%%%%%%%%%%%%%%%%%%%%%%%%%%%%
To compute the scale evolution of the mass non-perturbatively, 
we introduce a new step scaling function,
\bes
 \Sigmap(u,a/L) = \left. \zp(2L) \over \zp(L) \right|_{\gbar^2(L)=u} \enspace ,
\ees
with $\zp$ of \eq{e_zp}.
Results for $\Sigmap$ at finite lattice spacing and the 
extracted continuum limit 
are displayed in \fig{f_sigmap}. Details may be found
in  \cite{mbar:pap1,mbar:nf2}.
 
\begin{figure}[ht]
\centerline{
\psfig{file=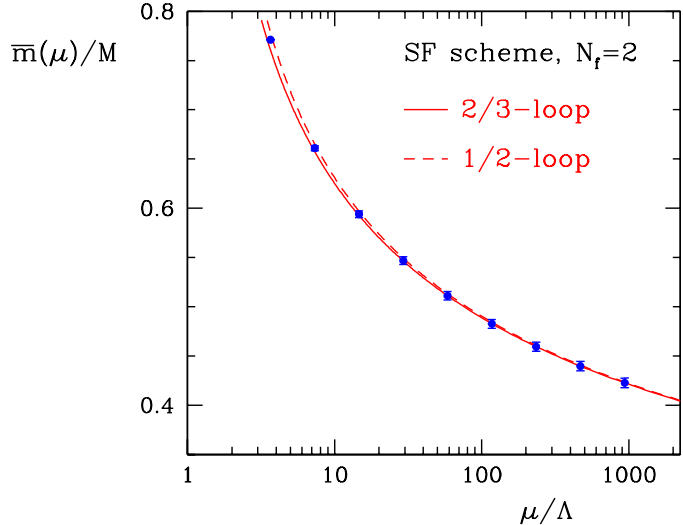,%
width=9.0cm}}
\caption{The running quark mass as a function of $\mu\equiv1/L$ for $\nf=2$.
\label{f_mbar}}
\end{figure}

Applying $\sigmap$ and $\sigma$ recursively one then obtains
the series, 
\bes
 \mbar(2^{-k} \Lmax) / \mbar(2\Lmax)\, , \,\,k=0,1,\ldots \enspace ,
\ees  
up to a largest value of $k$, which corresponds to the smallest
$\gbar$ that was considered in \fig{f_sigmap}.  
From there on,
the perturbative
2-loop approximation to the $\tau$-function and
3-loop approximation to the $\beta$-function
(in the SF-scheme) 
may be used to integrate the renormalization group equations to infinite
energy, or equivalently  to $\gbar=0$. The result is the
renormalization group invariant mass,
\bes
  M = \mbar\,(2 b_0\gbar^2)^{-d_0/2b_0} 
   \exp \left\{- \int_0^{\gbar} \rmd g \left[{\tau(g) \over \beta(g)}
     - {d_0 \over b_0 g} \right] \right\}  \enspace .
\ees
In this way, one is finally able to express the
running mass $\mbar$ in units of the renormalization group invariant mass, 
$M$, as shown in \fig{f_mbar}. Remember that $M$ has the same value in all
renormalization schemes, in contrast to the running mass $\mbar$.  

%%%%%%%%%%%%%%%%%%%%%%%%%%%%%FIGURE%%%%%%%%%%%%%%%%%%%%%%%%%%%%%%%%%%%
%%%%%%%%%%%%%%%%%%%%%%%%%%%%%%%%%%%%%%%%%%%%%%%%%%%%%%%%%%%%%%%%%%%%

The {\em perturbative} evolution of the quark masses follows 
very accurately the non-perturbative results down to rather
low energy scales. Of course, this result may not be generalized 
to running masses in other schemes. 

The point at lowest scale $\mu$ in \fig{f_mbar} corresponds to
\bes
 M/\mbar = 1.296(16) \quad \mbox{at} \quad L=2\Lmax \enspace .
\ees
Remembering the very definition of the renormalized mass \eq{e_mbar},
one can use this result to relate the renormalization group invariant mass
and the bare current quark mass $m$ on the lattice
through
\bes
  M= m \times 1.296(16) \times \za(g_0) / \zp(g_0,2\Lmax/a)=\zM(g_0)\,m \enspace . 
  \label{e_M_final}
\ees
In this last step, one then inserts the bare current quark mass, e.g. of
the strange quark, and extrapolates the result to the continuum limit. 
The bare current quark masses themselves are the ones for which
the appropriate pseudo scalar masses are fixed to their experimental
values. The presently available results from this strategy
are listed in \tab{t:masses}. Note that the computation of
the b-quark mass required a detour through Heavy Quark Effective Theory,
the subject of the following lecture.

\begin{table}[b!]
{\begin{center} \small
\begin{tabular}{lcccl}
\hline \\[-1.5ex]
  $i$     & $\nf$ &  input & $M_i/\GeV$ & ref. \\[0.5ex]
\hline\\[-1.5ex]
 strange & 0 & $\mk,r_0$ & 0.137(05) & \cite{mbar:pap3} \\
 strange & 2 & $\mk,r_0$ & 0.137(27) & \cite{mbar:nf2} \\
 charm & 0 & $\md,r_0$  & 1.654(45) & \cite{mbar:charm1} \\
 beauty & 0 & $\mbs,\mbsstar,r_0$ & 6.771(99) & \cite{lat05:nicolas} \\[0.5ex] 
\hline
\end{tabular}
\end{center}
}
\caption{Quark masses determined with full NP
        renormalization and continuum limit. We use
      $r_0=0.5\,\fm$.} 
\label{t:masses}
\end{table}

%%% Local Variables: 
%%% mode: latex
%%% TeX-master: "latticen.bib"
%%% End: 

For the charm and the beauty quark masses determinations with $\nf>0$
and NP renormalization are still missing. However, in our
opinion it is even quite early concerning the determinations of the
light quark masses. Although some $\nf$-dependence of the strange
quark mass has been reported in the literature, one can presently
not exclude that this is due to perturbative uncertainties
or discretization errors.
Note that the quark masses computed in the quenched approximation
were in a similar stage in 1996 but very soon
afterward the uncertainties shrunk by an order
of magnitude due to NP renormalization and continuum extrapolations.
This remains to be achieved for the real theory with $\nf>0$!

\section{Renormalization scale dependence of other composite operators}

\begin{figure}[htb]
\centerline{
\psfig{file=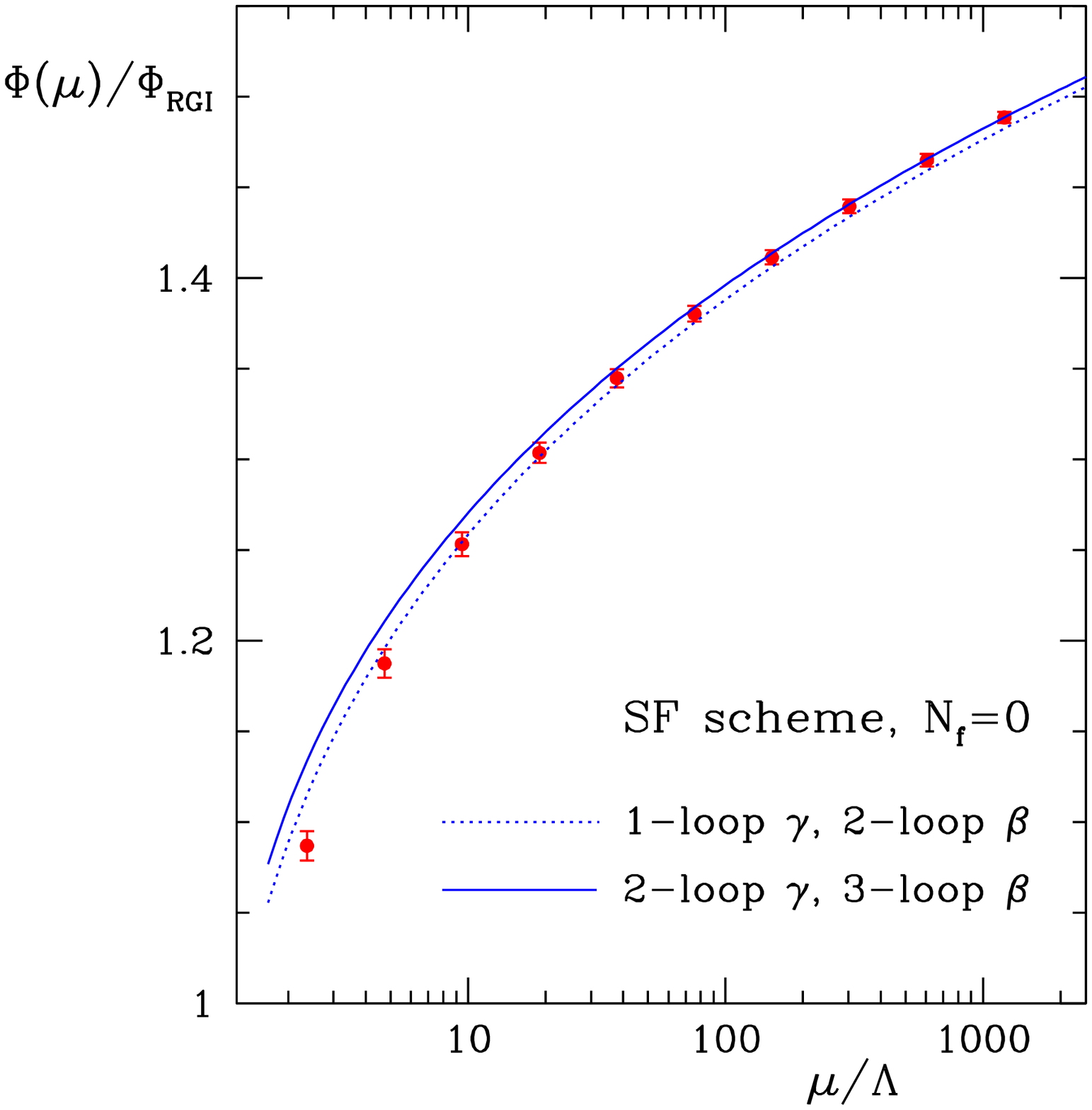,%
width=0.48\textwidth} \hfill
\psfig{file=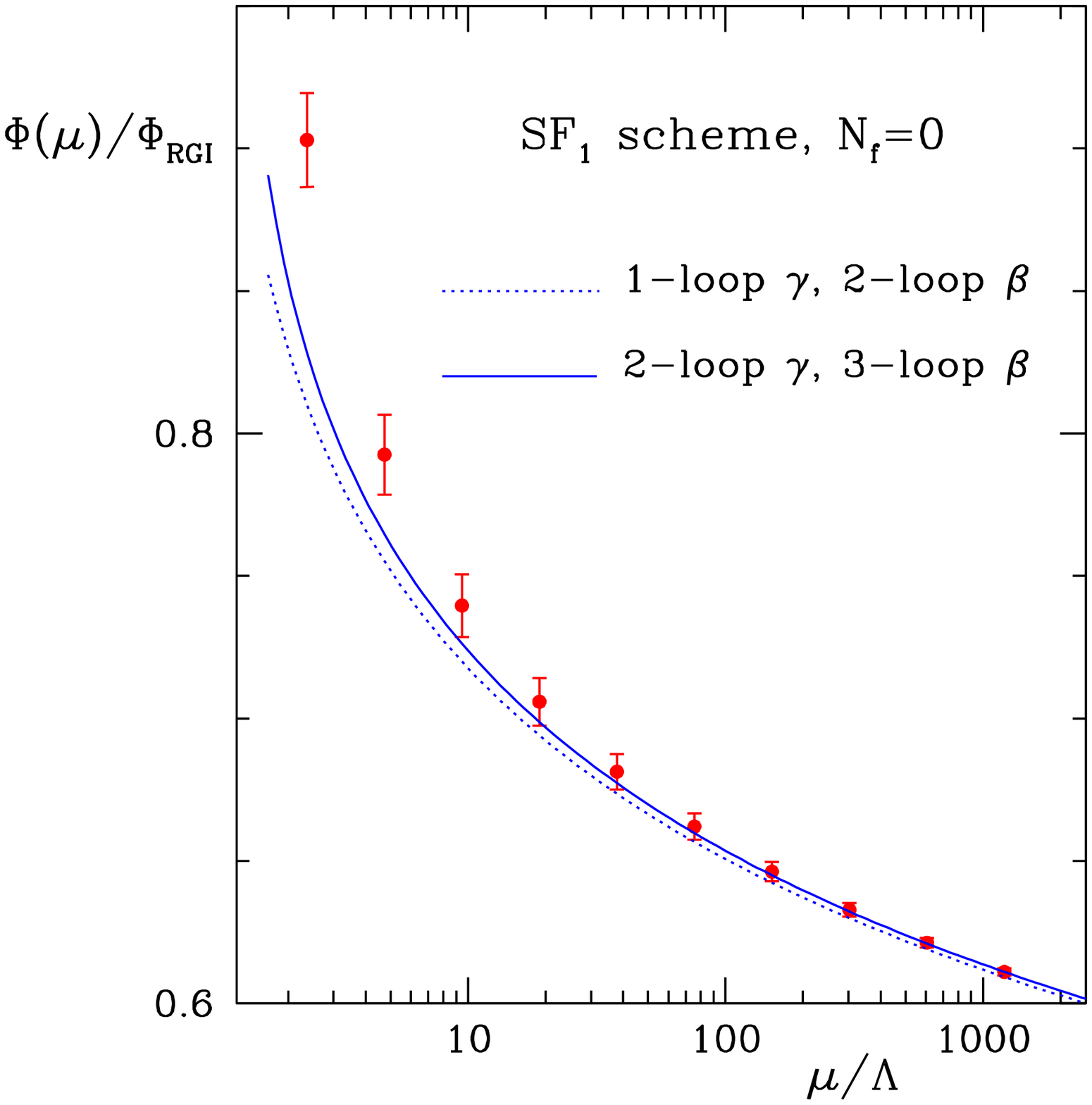,%
width=0.48\textwidth}
}
\caption{The running of the static-light axial current $\mu\equiv1/L$ 
	for $\nf=0$ (left)\protect\cite{zastat:pap3}
	and a left-left 4-fermion operator (right)\protect\cite{4ferm:nf0}.
\label{f_phibar}}
\end{figure}

Due to our definition of the renormalized quark mass, 
its scale dependence is given by the one of the composite operator $P^a(x)$. Other 
composite operators can be considered and indeed the strategy described
here has been applied to 4-fermion operators in the 
weak effective Hamiltonian~\cite{4ferm:nf0,4ferm:pert,stat:zbb_pert,lat06:filippo}, the
HQET axial current~\cite{zastat:pap1,zastat:pap3} and in
the operator which yields  $ \langle x \rangle $ of
the non-singlet structure functions~\cite{struct:nonsingpert,struct:singpert,struct:run_o44}.
It is worth pointing out that the non-perturbative
scale dependence disagrees more  (and significantly) from
the perturbative one in some of these cases. They provide examples
which emphasize that a fully non-perturbative renormalization
is necessary to control the associated uncertainties. We show two 
examples in \fig{f_phibar}. For details we have to refer to the cited 
papers. 

%%%%%%%%%%%%%%%%%%%%%%%%%%%%%FIGURE%%%%%%%%%%%%%%%%%%%%%%%%%%%%%%%%%%%
%%%%%%%%%%%%%%%%%%%%%%%%%%%%%%%%%%%%%%%%%%%%%%%%%%%%%%%%%%%%%%%%%%%%

However, we want to emphasize again one general feature, which was
noted already for the $\nf=2$ running coupling. Significant differences
between the perturbative and the non-perturbative result are present
in cases, where the subsequent order of perturbation theory 
are very close to each other. The difference of 2/3-loop order
running to the non-perturbative results can't be parameterized
by the next order perturbative term with a reasonable coefficient. 

%%% Local Variables: 
%%% mode: latex
%%% TeX-master: "sect_running"
%%% End: 

\lecture{III}{Non-perturbative Heavy Quark Effective Theory}
\renewcommand\thesection      {III.\arabic{section}}
\section{Introduction  \label{s:hqeti}}

Similarly to the scale dependent renormalization covered in
the previous lecture, the inclusion of heavy quarks in a 
lattice gauge theory simulation is a multi-scale problem.
Such problems are always difficult and require the development of 
new techniques. By ``heavy quark'' we mean a quark whose
mass is large compared to the intrinsic scale of QCD, 
$\Lambda_\mrm{QCD}$. If we take $\Lambda_\mrm{QCD}\approx500\,\MeV$,
the charm quark mass is about a factor 2 higher, but the
bottom quark mass has
\bes
   \mbeauty \approx 5\,\GeV \sim 10 \Lambda_\mrm{QCD}\,. \label{e:bscaleratio}
\ees
Here $\mbeauty$ is a quark mass defined at the scale 
$\mu=\mbeauty$, but the scheme is irrelevant 
at this point; we will continue to 
use the symbol $\mbeauty$ when this is the case.
Since  practical large volume simulations do not (yet) 
reach lattice spacings as small as $a = 1/(5\,\GeV)$, we are faced with
\bes
  a \mbeauty > 1  \label{e:amb}\,,
\ees
and a b-quark does not propagate properly on the lattice, at least 
when it is discretized with a standard relativistic QCD Lagrangian.

%%%%%%%%%%%%%%%%%%%%%%%%%%%%%FIGURE%%%%%%%%%%%%%%%%%%%%%%%%%%%%%%%%%%%
\begin{figure}[htb]
\centerline{
\psfig{file=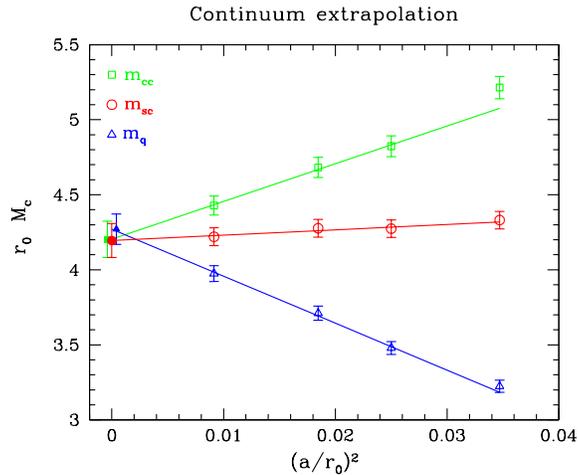,width=7.5cm}}
\caption{The RGI mass of the charm quark in units of $r_0$,
  evaluated with the $\Oa$-improved formulation, as a function 
  of the lattice spacing~\protect\cite{mbar:charm1}. The physical input
  is the mass of the $\mrm{D}_\mrm{s}$ meson (and $\mk,r_0$). 
  The RGI-mass at finite lattice spacing is defined through
  the PCAC-mass, \protect\eq{e_mrn}, of a charm-charm correlator ($m_\mrm{cc}$)
  or a charm-strange
  correlator ($m_\mrm{sc}$) and through the bare quark mass, $\mq$, and
  \protect\eq{e:mri}.
\label{f:Mc}}
\end{figure}
%%%%%%%%%%%%%%%%%%%%%%%%%%%%%%%%%%%%%%%%%%%%%%%%%%%%%%%%%%%%%%%%%%%%
 
For the charm quark, values $ a \mcharm \ll 1$ are achievable, but still
care has to be taken. In \fig{f:Mc} we show 
the RGI mass of the charm quark in the quenched approximation, 
computed for three definitions, which differ at finite lattice spacing. 
While an extrapolation to a common continuum value is convincing,
it is already clear from this figure, that the four times
heavier bottom quark
can't be treated this way. In fact, the main point is that 
the cutoff-effects become entirely non-linear in $a^2$, when 
$am$ is too large. This breakdown of the Symanzik expansion 
can be seen explicitly in perturbation theory~\cite{zastat:pap2}.
In a 1-loop calculation, it
has been estimated 
to happen around~\cite{zastat:pap2} $a \mbeauty \approx 1/2$ or $a \Mbeauty \approx 0.7$.

Various ways of coping with this problem have been proposed and investigated.
Referring the reader to reviews for 
other approaches\cite{lat03:kronfeld,reviews:hashimogi,lat06:onogi},
we directly turn to HQET. Already in 1987  Estia Eichten suggested
that to describe the non-perturbative dynamics of hadrons with 
a single heavy quark, it is a good approximation to consider
this quark to be static, i.e. it propagates only in time (in the rest frame of
the hadron) \cite{stat:eichten}. This static approximation
describes the correct asymptotics of bound state properties 
as $\mbeauty\to\infty$ and corrections of order
$\minv$ can be included systematically. The 
expansion in $\minv$ is then given by an effective field
theory~\cite{stat:eichhill1,hqet:cont3,hqet:cont4}. It has been extended 
to transition form factors, e.g. between B- and D-mesons, 
assuming that also the charm quark can be described by the effective
field theory\cite{hqet:cont1,hqet:cont2,hqet:cont3}. 
This is done by considering heavy quark fields with
finite velocities in the limit of large mass. We will here ignore this
phenomenologically very interesting possibility and restrict ourselves
to HQET at zero velocity~(The formulation of HQET in Euclidean space
and at non-zero velocity is more subtle~\cite{hqet:agliettif}).

In this theory considerable progress has been made recently, which we want 
to explain. In particular, in order to compute $\minv$ corrections, one
has to know the leading order (including its renormalization) non-perturbatively
(\sect{s:need}).
This necessitates a so-called non-perturbative matching of effective theory
and QCD. A strategy for this \cite{hqet:pap1,hqet:pap4}
will be described in \sect{s:hqetstrat} including results for $\minv$ corrections
in a test case. 
The strategy solves at the same time the problem of a proper definition
and computation of $\minv$-corrections, which is present in any regularization
of the theory,  and the problem of power divergences ($\sim a^{-n}$) in 
a theory with a hard cutoff such as the lattice regularization. 
After a more general introduction 
we will concentrate on the inclusion of $\minv$ corrections. Because of limitations of space
we mention the also very relevant developments in the static approximation
only rather briefly. 

%%% Local Variables: 
%%% mode: latex
%%% TeX-master: "Nara"
%%% End: 

\subsection{% Classical HQET: 
            Derivation of the classical theory\label{s:hqetc}}

\def\lag#1{{\cal L}_{\rm #1}}
\def\psibar{\overline{\psi}}
\def\rme{{\rm e}}
\def\LD{\lag{}}
\def\Dop{{\cal D}}
\def\Lh{\lag{h}}
\def\Lhb{\lag{\bar h}}
\def\Lhhb{\lag{h \bar h}}
\def\nab#1{{\nabla_{#1}}}
\def\lnabstar#1{\overleftarrow{\nabla}\kern-0.5pt\smash
             {\raise 4.5pt\hbox{$\ast$}}\kern-4.5pt_{#1}}
\def\nabstar#1{\nabla\kern-0.5pt\smash{\raise 4.5pt\hbox{$\ast$}}
               \kern-4.5pt_{#1}}
\def\vecD{{\bf D}}
\def\vecB{{\bf B}}
\def\vecsig{{\mathbf\sigma}}
\newcommand{\vecg}{{\boldsymbol \gamma}}
\newcommand{\sigb}{\vecsig\!\cdot\!\vecB}
\def\Dg{D_k\gamma_k}

We here go through some steps to derive the effective 
theory at the classical level. The main point is to see what 
assumptions have to be made and to present the explicit form
of the Lagrangian. We follow the idea of \cite{hqet:cont5}, but 
work in Euclidean space since we are ultimately interested in the 
lattice theory. 

In this section, we keep the dependence of the fields on the space-time coordinates
implicit and also drop the label b on the quark field and its mass.
We start from the 
Dirac-Lagrangian of a b-quark with a large mass, $m$, in the continuum,
\bes
  \LD &=& \psibar(D_\mu\gamma_\mu + m)\psi \\
      &=& \psi^\dagger \Dop \psi\,,\quad \Dop=m\gamma_0 + D_0+\gamma_0\Dg \,.
\ees
The light quark fields and gauge fields are not touched by
our considerations.
We write $\psi^\dagger$, but it is just another independent 
Grassmann integration variable in the path integral.  
Since we are considering the classical theory, we can assume
that the fields are smooth. We can therefore perform an expansion
in $D_\mu$. More precisely, we have to refer to a special 
kinematical situation. We want to describe the dynamics of a
hadron containing one heavy quark, where the hadron is at rest.
For infinite mass, the heavy quark propagates only in time. 
Denoting the expansion parameter by $\varepsilon$, the dynamics
thus dictates
\bes
  D_0/m =\rmO(1) \,, \quad D_k/m = \rmO(\varepsilon)  \label{e:countclass}\,,
\ees
when these derivatives act on the heavy quark fields. This is often called
a power counting scheme. In the
quantum theory we will have $\varepsilon=\Lambda_\mrm{QCD}/m$. 
Obviously quantities such as $F_{\mu\nu}=\rmO(1)$ are 
not touched by this consideration.
At the lowest order in this expansion the (``large components'') quark field 
 ($P_{\pm}= \frac{1\pm\gamma_0}{ 2}$)
\bes
   \heavy = P_+\psi\,,\quad  \heavyb=\psibar P_+,  \quad
    \,,
\ees 
propagates forward in time, while the anti-quark field,
\bes
   \aheavy = P_-\psi\,,\quad  \aheavyb = \psibar P_-\,,  
\ees 
propagates backward. In a somewhat sloppy notation we will
often write $\rmO(1/m)$ instead of $\rmO(\varepsilon)$.
The $\rmO(1/m)$ terms in the Lagrangian
\bes
  \LD   
        &=& \Lh^{\rm stat} + \Lhb^{\rm stat}  \label{e:lstat1}
       +\rmO(\frac{1}{ m}) \\[1ex] \label{e:lstat1a}
 \Lh^{\rm stat} &=&   \heavyb(D_0+m)\heavy\,,\quad  
\Lhb^{\rm stat} =  \aheavyb(-D_0+m)\aheavy\,,\quad 
\ees
connect quark and anti-quark fields. They 
can be decoupled through a Foldy-Wouthuysen rotation,
\bes
  \LD &=& \phi^\dagger \Dop' \phi\,,\quad \phi=\rme^{S}\psi\,,\;
                                         \phi^\dagger=\psi^\dagger\rme^{-S}\\
      && \Dop'=\rme^{S} \Dop \rme^{-S}\,,\quad
        S= \frac{1}{2 m}\Dg = - S^\dagger =
       \rmO(\frac{1}{m})\,,
\ees
which yields explicitly
\begin{eqnarray}
  \Dop' &=& \Dop + \frac{1}{2 m} [\Dg,\Dop]+\frac{1}{8 m^2}
                                      [D_l\gamma_l,[\Dg,\Dop]] 
                                    + \rmO(\frac{1}{ m^2}) \\
        &=& \Dop + \frac{1}{2 m} [\Dg,\Dop]-\frac{1}{4 m}
                                      [D_l\gamma_l,\gamma_0\Dg ] 
                                    + \rmO(\frac{1}{ m^2}) \nonumber\\
        &=& \gamma_0 \left\{
                     \gamma_0 D_0 + m + \frac{1}{2 m}
                       (-D_kD_k - {1\over 2i} F_{kl} \sigma_{kl})
                     + \frac{1}{2 m}F_{k0}\gamma_0\gamma_k
                     \right\} + \rmO(\frac{1}{ m^2})\,. \nonumber
\ees
The Lagrangian then reads
\bes
      \LD  &=& \Lh^{\rm stat} + \Lhb^{\rm stat}  \label{e:leff}
       + \left\{ \Lh^{(1)}+\Lhb^{(1)}+\Lhhb^{(1)} \right\}
       +\rmO(\frac{1}{ m^2}) \\[1ex]
\Lh^{(1)} &=& \frac{1}{2m} \heavyb(-D_kD_k -{1\over 2i} F_{kl}\,\sigma_{kl})\heavy \,, \\
        && \sigma_{\mu\nu}\!=\!{i\over2}[\gamma_\mu,\gamma_\nu] \,, 
           \quad F_{kl}=[D_k,D_l] \,.
\end{eqnarray}
For hadrons (or correlation functions) with a single b-quark (or anti-b-quark) 
only double insertions of  $\Lhhb^{(1)}$ contribute. These are of order 
$1/m^2$ and may be dropped at the order written explicitly. 

For later convenience we introduce the short hand 
\bes
  \Lh^{(1)} &=& -\frac{1}{ 2m}(\Okin + \Ospin) \,, \label{e:Lclass}\\
        \Okin &=& \heavyb\, D_kD_k\, \heavy = \heavyb \,\vecD^2\, \heavy\,, 
	\label{e:okin} \\
        \Ospin &=& \heavyb \,{1\over 2i} F_{kl}\,\sigma_{kl}\, \heavy = 
                   \heavyb\,\sigb\, \heavy\,.       \label{e:ospin}     
\ees
We note that $\LD$, \eq{e:leff} is a low energy
effective Lagrangian \cite{effft:weinb,effft:polch,effft:georg}. 
It describes the long wave length modes of the fields
accurately and makes truncation errors, which are of increasing 
relevance for shorter wave lengths. This becomes particularly apparent
when we remove the mass terms from the static Lagrangian 
and define 
\bes
 \Lh^{\rm stat} =   \heavyb (D_0+\eps)\heavy\,,\quad  
\Lhb^{\rm stat} =  \aheavyb (-D_0+\eps)\aheavy\,,\quad  \label{e:lstat2}
\ees
where the limit $\eps\to0_+$ is to be understood in order to
select the proper propagation in time. 
Replacing \eq{e:lstat1a} by   \eq{e:lstat2}
corresponds {\em exactly} to an energy shift by an amount $m$
of all states containing a single heavy quark or anti-quark.
For Euclidean correlation functions it just leads
to an additional factor of $\exp(-m\,(y_0-x_0))$ 
 for correlation functions
where a quark propagates from $x_0$ to $y_0\geq x_0$. (For the
anti-quark there is a factor $\exp(+m\,(y_0-x_0))$ with
$y_0\leq x_0$).

We note again that the essential assumption is \eq{e:countclass},
namely the spatial covariant
derivatives are counted as small compared to the mass term and 
the time derivative. This is the correct physical situation
in a frame where the hadron is at rest and therefore at 
lowest order also the quark is at rest. 

Instead of carrying out the 
expansion of the action, one could also expand the heavy 
quark propagator in terms of $1/m$. 

Quantum fluctuations are not smooth and invalidate the above 
``derivation''. 
However, one expects that they do not modify the structure 
of the effective Lagrangian, but rather only modify the coefficients
of the various terms by non-trivial renormalizations due to these
{\em short distance} fluctuations. After all, arguing heuristically,
long wavelength terms have been identified correctly 
and are described by local interaction terms. In
local quantum field theory, also effective local quantum field theory, 
such terms are renormalized
by a renormalization of the coefficients of the local fields.
Below, we will discuss 
this in some detail.  

%%% Local Variables: 
%%% mode: latex
%%% TeX-master: "Nara"
%%% End: 

\section{The effective quantum field theory \label{s:hqetq}}

\subsection{The static approximation and its symmetries \label{s:stat}}
We start with the lowest order and just consider the heavy
quark; the antiquark action and propagator is completely
analogous. In continuum Euclidean space, the classic
Lagrangian density $\Lh^{\rm stat}$, \eq{e:lstat2}, contains
local fields of a mass dimension $D\leq4$. 
It is power counting renormalizable. The
static effective field theory is thus expected to be renormalizable
in the usual sense, i.e. by a finite number of counter-terms.
The possible counter-terms are restricted by the symmetries.
Apart from the usual ones (parity, gauge symmetry \ldots), 
there is a well known invariance under spin rotations
\cite{Shifman:1987sm,hqet:cont1,hqet:cont2,hqet:cont3}. 
The infinitesimal variations
of these transformations can be written as 
\bes
  \label{e:spin}
	\delsig^k \heavy = \sigma_k \heavy\,,\qquad 
        \delsig^k \heavyb = -\heavyb \sigma_k \,, \qquad
        \sigma_{k} \equiv - \frac12 \epsilon_{ijk}\sigma_{ij}\,,
\ees
with $[\sigma_k,\sigma_l]=i \epsilon_{klm}\sigma_m$.
In addition, the action is invariant under 
phase transformations,
\be
  \label{e_quarknumber}
   \heavy\longrightarrow \rme^{i\eta(\vecx)}\,\heavy,
  \qquad \heavyb\longrightarrow\heavyb \rme^{-i\eta(\vecx)},
\ee
with an arbitrary space- (but not time-) dependent parameter $\eta(\vecx)$.
This invariance corresponds to the local conservation of b-quark number,
ensuring that the quark propagates only in time. 
The only counter-term of dimension $D\leq4$, which involves $\heavy,\heavyb$ 
and is invariant under these symmetries is $\heavyb \heavy$. 
Denoting its coefficient by $\dmstat$,
the formal continuum quantum Lagrangian is thus simply
\be  \label{e:lstat3}
 \Lh^{\rm stat} =   \heavyb (D_0+\dmstat)\heavy\,.\quad  
\ee
In order to discuss the equivalence of the effective theory
and QCD for correlation functions, we introduce also the 
time component of the 
axial current in the effective theory,
\be
  \label{e_StatAxial}
  \Astat(x)=\lightb(x)\gamma_0 \gamma_5\heavy(x)\,,
\ee 
as a prototype for a composite field.
There is no other operator with its quantum numbers and
with dimension $D \leq 3$. Hence $\Astat(x)$ renormalizes multiplicatively.

\subsubsection{Lattice formulation \label{s:statlat}}

Just like the continuum theory we formulate the 
lattice theory in terms of formally 4-component fields, 
satisfying 
\be
  \label{e_constraint}
  P_{+}\heavy=\heavy\,,\quad \heavyb P_{+}=\heavyb\,.\quad
  % P_{+}=\frac12(1+\gamma_0)\,
\ee
The time-doubler is removed by choosing the lattice backward derivative 
\be \label{e:D0lat}
D_0^{\rm W} \heavy(x) = {{1}\over{a}} \left[ \heavy(x) -W^{\dagger} (x-a\hat{0},0)
 \heavy (x-a\hat{0}) \right] \; ,
\ee
in the action
\be
\label{e:stat_action}
S_{\rm h}^{\rm W} =a^4 {{1}\over{1+a\; \delta m_{\rm W} }} \sum_x
\heavyb(x) (D_0^{\rm W} + \delta m_{\rm W} ) \heavy(x) \;.
\ee
With  $W(x,0)=U(x,0)$, the standard time-like links, this is the 
Eichten-Hill action\cite{stat:eichhill1}. For the Monte Carlo evaluation 
it is however of a considerable advantage to define the theory
with more general parallel transporters $W(x,0)$, equivalent
to  $U(x,0)$ up to $\rmO(a^2)$. In this way statistical errors 
of correlation functions at large separation $x_0$
can be reduced exponentially (in $x_0$) and
at the same time discretization errors have been found to be somewhat 
smaller~\cite{stat:letter,stat:actpaper}. Concentrating on conceptual
issues, we refer to the cited papers for details on this more practical issue.

It is an easy exercise to show that the static propagator in the 
presence of gauge fields is ($a\dmstathat=\ln(1+a\dmstat)$)
\bes
G_{\rm h}^{\rm W}(x,y) &=& \theta(x_0-y_0) \;
 \delta({\bf{x-y}}) \;\exp\big(- \dmstathat_{\rm W}\,(x_0-y_0)\big)\;
{\cal{P}}^{\rm W}(y,x)^{\dagger} \; P_+ \; , 
 \nonumber \\ \label{e:stat_prop} \\[-1ex] \nonumber
{\cal{P}}^{\rm W}(x,x)&=&1 \; ,\quad
{\cal{P}}^{\rm W}(x,y+a\hat\mu) = {\cal{P}}^{\rm W}(x,y)W(y,\mu)\,,
\ees
when the fields are normalized as in \eq{e:stat_action}. Here $\theta(x_0), \delta(\vecx)$
are straight forward lattice transcriptions of the continuum $\theta$- and $\delta$-functions.
While in the continuum case, an $\eps$-prescription is necessary to select the
forward propagation $\theta(x_0-y_0)$, the propagator of the lattice action with 
backward derivative, \eq{e:D0lat}, has this property automatically. 
The symmetries \eq{e:spin} and \eq{e_quarknumber} are preserved
by the lattice regularization \eq{e:stat_action}.

\subsubsubsection{Improvement}

With these symmetries one easily goes through
the steps introduced in \sect{s:impr} to find the structure
of Symanzik's effective action. It turns 
out that the only allowed $\Oa$ terms are proportional to
the light quark masses\cite{zastat:pap1}. They are therefore numerically 
not very relevant and in addition their coefficients vanish for 
$\nf=0$. It is interesting to note that through the static effective
theory one can also give a convincing argument that the force
between static quarks is free of linear $a$-effects, if the
light quark action is $\Oa$-improved~\cite{pot:intermed}. 
% We have not seen a different argument for this property.

For ease of notation we drop the sub- and superscript W from now on.

\subsubsection{Renormalization \label{s:statren}}

The only term needed for the renormalization
of the action, $\dmstat$, has been included above. The explicit form of the
propagator shows that $\dmstat$ enters in a purely kinematical way, 
just as an energy shift by an amount  $\dmstathat$
compared to the unrenormalized, $\dmstat=0$ case. Thus all
masses of hadrons (we consider only those  with a single heavy quark) 
have the same shift and
their splittings can be predicted in static approximation up to
$\Lambda_\mrm{QCD}^2/\mbeauty$ corrections without adjusting any parameter except for those
in the ``light part of QCD''. Obviously, the splittings are also
the same if the b-quark is replaced by the charm quark, but now
up to $\Lambda_\mrm{QCD}^2/\mcharm$, if $\mcharm$ is large enough for the 
$1/\mcharm$ expansion to make sense. 
 
But also in the static approximation non-trivial renormalizations occur
as soon as hadronic matrix elements are considered.
The most prominent example is the computation of the B-meson decay constant,
$\fb$. In QCD it can be obtained from the correlation
function~\footnote{It is technically of advantage
to consider so-called smeared-smeared and local-smeared correlation functions,
but this is irrelevant in the present discussion.}
\bes \label{e:caa}
  \caa(x_0) = \za^2 a^3\sum_{\vecx} \Big\langle A_0(x)  (A_0)^{\dagger}(0) 
              \Big\rangle
\ees
with the heavy-light axial current in QCD,
$A_\mu=\lightb\gamma_\mu\gamma_5\psi_\beauty$, and  
$A_\mu^{\dagger}=\psibar_\beauty\gamma_\mu\gamma_5\light$.
In QCD, $\za(g_0)$ is fixed by chiral Ward identities~\cite{Boch,impr:pap4},
which shows that it does not depend on a renormalization scale.
The decay constant may e.g. be obtained from (we use the finite volume 
normalization $\langle B | B \rangle = 1$ for the zero momentum state 
$|B\rangle$)
\bes \label{e:fb}
   \big[\Phiqcd\big]^2 &\equiv& \fb^2\,\mB  \\ \nonumber
        &=& 2L^3\,\big[\langle B | \za A_0 |0 \rangle\big]^2 = 
         2\lim_{x_0\to\infty} \exp(x_0\,\Gamma_\mrm{AA}(x_0)) \caa(x_0)\,,
\ees
where the effective mass
($\dzero    f(x_0) =  {1 \over 2a} [f(x_0+a) - f(x_0-a)]$)
\be \label{e:meff}
    \Gamma_\mrm{AA}(x_0) = -\dzero\, \ln(\caa(x_0))\, 
\ee
appears. In the static approximation, the
QCD current is represented by 
\be
  \Aren(x) = \zastat(g_0,\mu a) \Astat(x)\,,
\ee
where we have allowed for a $\mu$-dependence of the renormalization. 
This is expected because chiral symmetry is not just broken softly,
but in the effective theory it is not present at all.
We will come back to it in \sect{s:fb}. 

%%%%%%%%%%%%%%%%%%%%%%%%%%%%%%%%%%%%%%%%%%%%%%%%%%%%%%%%%%%%%%%%%%%%%%%%%%%%%%%%%%%%%%%%%
\subsection{Including $\minv$ corrections \label{s:1om}}
We work directly in lattice regularization. 
The continuum formulae are completely analogous. 
The expressions for $\Okin,\Ospin$, \eq{e:okin}, \eq{e:ospin}, 
are discretized in a straight forward way, 
$D_kD_k \to \nabstar{k}\nab{k}$ and
$ F_{kl} \to \widehat{F}_{kl}$ with the latter
defined in \cite{impr:pap1}. Of course other discretizations of these
composite fields are possible.

Apart from the terms in  the classical Lagrangian,
renormalization can in principle introduce new local fields  
compatible with the symmetries (but not  \eq{e:spin} 
and \eq{e_quarknumber}, which are broken by $\Ospin,\Okin$)
and with dimension  $D\leq5$. Also the field equations
can be used to eliminate terms. With these rules one easily finds 
that no new terms
are needed and it suffices to treat the coefficients of
$\Ospin,\Okin$ as free parameters which depend on the bare coupling of the 
theory and on $\mbeauty$.

The $\minv$ Lagrangian then reads
\be
  \Lh^{(1)}(x) = -(\omegakin\,\Okin(x) + \omegaspin\,\Ospin(x)) \,.
\ee
Since these terms are fields of dimension five, the theory defined
with a path integral weight
$P\propto\exp(-a^4\sum_x[\lag{light}(x)+\Lh^\mrm{stat}(x)+\Lh^{(1)}(x)])$ 
is {\em not}
renormalizable. In perturbation theory,
new divergences will occur at each order in the loop expansion, 
which necessitate to introduce
new counter-terms. The continuum limit of the lattice theory will
then not exist. However, that effective theory is NRQCD not 
HQET. Since 
the effective theory is ``only'' 
supposed to reproduce the $\minv$ expansion of the observables
order by order in $\minv$, 
we expand the weight $P$ in $\minv$, counting 
$\omegakin=\rmO(\minv)=\omegaspin$. 
This defines HQET. The same step has already
been used in Symanzik's effective theory.

Up to and including $\rmO(\minv)$, 
expectation values in HQET are defined as 
\bes
  \langle \op{} \rangle &=& 
         \langle  \op{}  \rangle_\mrm{stat} 
        + \omegakin  a^4\sum_x \langle  \op{} \Okin(x) \rangle_\mrm{stat}
        + \omegaspin a^4\sum_x \langle \op{} \Ospin(x) \rangle_\mrm{stat} 
        \nonumber \\
  &\equiv&  \langle  \op{}  \rangle_\mrm{stat} 
        + \omegakin\langle  \op{}  \rangle_\mrm{kin} 
        + \omegaspin\langle  \op{}  \rangle_\mrm{spin} \,,
        \label{e:exp}
\ees
where 
\be
  \label{e:expval}
         \langle  \op{}  \rangle_\mrm{stat} = {1 \over \cal Z} \int_\mrm{fields} 
          \op{} \exp(-a^4\sum_x[\lag{light}(x)+\Lh^\mrm{stat}(x)])\, 
\ee
is defined  with respect to the lowest order action, which is
power counting renormalizable. The path integral defining
the average extends over 
all fields and the normalization $\cal Z$ is
fixed by $\langle 1  \rangle_\mrm{stat} = 1$. 

In order to compute matrix elements or correlation functions in the 
effective theory, we also need the effective composite fields. 
At the classical level they can again be obtained from the 
Foldy-Wouthuysen rotation. In the quantum theory one
adds all local fields with the proper quantum numbers and 
dimensions. For example the effective axial current (time component) 
is given by
\bes
  \label{e:ahqet}
 \Ahqet(x)&=& \zahqet\,[\Astat(x)+ \cahqet\delta\Astat(x)]\,, \\
 \delta\Astat(x) &=& \lightb(x){1\over2}
            (\lnab{i}\!+\!\lnabstar{i})\gamma_i\gamma_5\heavy(x)\,.
\ees
Before entering into more details on the 
renormalization, we show some examples how the $\minv$-expansion works.

\subsubsection{$\minv$-expansion of correlation functions and matrix elements \label{s:corr}}
For now we assume that the coefficients
\bes
  \rmO(1)    \,:&& \dmstat\,,\; \zahqet\,,  \nonumber \\[-1ex]
  \label{e:counting} \\[-1ex] \nonumber
  \rmO(\minv)\,:&& \omegakin\,, \; \omegaspin\,, \; \cahqet\,, \;
\ees
are known as a function of the bare coupling $g_0$ and the quark mass $\mbeauty$.
Their non-perturbative determination will be discussed later.

The rules of the $\minv$-expansion are illustrated for
of $\caa(x_0)$, \eq{e:caa}. One uses \eq{e:exp} and the HQET representation
of the composite field \eq{e:ahqet}. Then the expectation value is expanded
consistently in $\minv$, counting powers of $\minv$ as in \eq{e:counting}. 
At order $\minv$, terms proportional to $\omegakin \times \cahqet$ etc. 
are to be dropped.
As a last step, we have to take the energy shift between HQET and QCD into account.
Therefore the correlation function obtains an extra factor $\exp(-x_0\mbeauty)$,
where the scheme dependence of $\mbeauty$ is compensated by a
corresponding one in
$\dmstat$. One arrives at the expansion 
\bes
   \label{e:caahqet}
   \caa(x_0) &=& \rme^{-\mbeauty x_0} (\zahqet)^2 \,\Big[
                 \caa^\mrm{stat}(x_0) + \cahqet\,\cdaa^\mrm{stat}(x_0) \\
           && \qquad +\,\omegakin\,\caa^\mrm{kin}(x_0)+\omegaspin\,\caa^\mrm{spin}(x_0)  
                 \Big] \nonumber
\ees
with (remember the definitions in \eq{e:exp})
\bes
   \cdaa^\mrm{stat}(x_0) &=& 
        \langle \Astat(x) (\delta\Astat(0))^\dagger  \rangle_\mrm{stat}
  \,+\, \langle \delta\Astat(x) (\Astat(0))^\dagger  \rangle_\mrm{stat} \,,
        \nonumber \\
   \caa^\mrm{kin}(x_0) &=& \langle  \Astat(x) (\Astat(0))^\dagger  \rangle_\mrm{kin} 
        \,, \\
   \caa^\mrm{spin}(x_0) &=& \langle  \Astat(x) (\Astat(0))^\dagger  \rangle_\mrm{spin}
        \,. \nonumber
\ees
It is now a straight forward exercise to 
obtain the expansion of the B-meson mass\footnote{It follows  from the simple form of
the static propagator that there is no dependence on $\dmstat$ 
except for the explicitly shown energy shift $\dmstathat$.} 
\bes
    \mB &=& \mbeauty + \dmstathat + \Estat+ \omegakin \Ekin + \omegaspin \Espin \label{e:mBexp}
     \,,\\
    \Estat&=&  \left.- \lim_{x_0\to\infty} \dzero\,\ln\caa^\mrm{stat}(x_0)
               \right|_{\dmstat=0}\,, 
    \label{e:estat}\\
    \Ekin &=& - \lim_{x_0\to\infty} \dzero\, \rho_\mrm{kin}(x_0)\,,\quad
         \rho_\mrm{kin}(x_0) = {\caa^\mrm{kin}(x_0)\over\caa^\mrm{stat}(x_0)}\,,\\
    \Espin &=& - \lim_{x_0\to\infty} \dzero\, \rho_\mrm{spin}(x_0)\,,\quad
         \rho_\mrm{spin}(x_0) = {\caa^\mrm{spin}(x_0)\over\caa^\mrm{stat}(x_0)}\,,
\ees
and its decay constant
\bes 
 \fb\sqrt{\mB}&=& \lim_{x_0\to\infty} \big\{2\exp(\mB x_0)\,\caa(x_0)\big\}^{1/2} \,,
  \\
  &=& \zahqet\, \lim_{x_0\to\infty} 
      \Fhatstat \, \big\{1 + \frac12 x_0 \big[ \omegakin \Ekin + \omegaspin\Ekin\big]
  \nonumber \\
  && + \frac12 \cahqet\rho_{\delta\rm A}(x_0) 
                          + \frac12 \omegakin \rho_\mrm{kin}(x_0)
                          + \frac12 \omegaspin \rho_\mrm{spin}(x_0) 
        \big\}\,, \label{e:fBexp}
  \\
         \Fhatstat &=& \lim_{x_0\to\infty} 
	               \big\{2\exp(\Estat x_0)\,\caa^\mrm{stat}(x_0)\big\}^{1/2}
	\,,\quad 
         \rho_{\delta\rm A}(x_0) = {\cdaa^\mrm{stat}(x_0)\over\caa^\mrm{stat}(x_0)}\,.
	\nonumber
\ees
Using the transfer matrix formalism 
(with normalization $\langle B | B \rangle = 1$), 
one further observes that
\bes
  \label{e:ekin}
    \Ekin  &=& - \langle B | a^3\sum_{\vecz} \Okin(0,\vecz)| B \rangle_\mrm{stat} \,,\quad
  \label{e:espin} \nonumber \\[-1ex] \\[-1ex] \nonumber
    \Espin &=& - \langle B | a^3\sum_{\vecz} \Ospin(0,\vecz)| B \rangle_\mrm{stat} \,.
\ees
As expected, only the parameters of the action are relevant
in the expansion of hadron masses. 

A correct split of the terms in \eq{e:mBexp} and  \eq{e:fBexp} into leading 
order and next to leading order pieces which 
{\em separately have a continuum limit} requires
more thought on the renormalization of the $\minv$-expansion. We 
turn to this now. 

\subsubsection{Renormalization and continuum limit\label{s:1omren}}

For illustration we first check the self consistency of \eq{e:caahqet}. 
The relevant question concerns
renormalization, namely: are the ``free'' parameters 
$\dmstat \ldots \cahqet$ sufficient to absorb
all divergences on the r.h.s.? We consider the most difficult term, 
$\caa^\mrm{kin}(x_0)$.
According to the standard rules, 
it is renormalized as
\bes \label{e:caakinr}
  \Big(\caa^\mrm{kin}\Big)_\mrm{R}(x_0) &=&  % \rme^{-\dmstat\, x_0} 
    \big(\zastat\big)^2  \times \\\nonumber&&a^7  \sum_{\vecx,\, z}\Big\langle 
    \Astat(x)\, (\Astat(0))^\dagger  \,\big(\Okin\big)_\mrm{R}(z)\Big\rangle_\mrm{stat} + \mbox{C.T.}\,, 
\ees
where C.T. denotes contact terms to be discussed shortly. The operator 
$\big(\Okin\big)_\mrm{R}(z)$
involves a subtraction of lower dimensional ones,
\bes \label{e:okinr}
  \big(\Okin\big)_\mrm{R}(z) = Z_{\Okin} \big( \Okin(z) + {c_1\over a}\, \heavyb(z) D_0 \heavy(z) + 
  {c_2\over a^2}\, \heavyb(z)\heavy(z) \big)\,, \nonumber \\[-1ex]
\ees
written here in terms of dimensionless $c_i$. 
Since we are interested in on-shell observables
($x_0>0$ in \eq{e:caahqet}), 
we may use the equation of motion
$D_0 \heavy(z)=0$ to eliminate the second term. The third one, 
${c_2\over a^2} \heavyb(z)\heavy(z)$, is equivalent
to a mass shift and only changes  
$\dmstat$, which is hence quadratically
divergent~\footnote{Using the explicit form of the static propagator, 
\eq{e:stat_prop}, one can check that indeed
$a^3\sum_{\vecx}\,\Big\langle \Astat(x)\, 
  (\Astat(0))^\dagger a^4\sum_z \heavyb(z)\heavy(z) \Big\rangle_\mrm{stat}
  = x_0\caa^\mrm{stat}(x_0)
$, which can be absorbed by a $\minv$ correction to $\dmstat$.  
}.
 Thus all terms which are needed for the renormalization of $\Okin$ are present 
in \eq{e:caahqet}. 

It remains to consider the contact terms in \eq{e:caakinr}. 
They originate from singularities in the operator products
$\Okin(z)\Astat(x)$ as $z\to x$ (and $\Okin(z)\big(\Astat\big)^\dagger(0)$  as $z\to 0$),
in complete analogy to the discussion in \sect{s:fe}.
Using the operator product expansion 
they can be represented as linear combinations of $\Astat(x)$ and $\delta\Astat(x)$. 
Such terms are contained in
\eq{e:caahqet} in the form of $\caa^\mrm{stat}$ and $\cdaa^\mrm{stat}$. 
Indeed 
$\Astat(x)$ and $\delta\Astat(x)$ are the only operators of dimension 3 and 4 
with the correct quantum numbers.  
Higher dimensional operators contribute only terms of order $a$. 

Note that 
the coefficient of $\Astat(x)$ in the expansion of the operator product 
$a^4\sum_z \Okin(z)\Astat(x)$ is power divergent $\sim 1/a$, for simple dimensional
reasons.  This means that there is
a power divergent contribution to $\zahqet$. 
As this happens only at order $\minv$, not at the lowest order, this
contribution to $\zahqet$ behaves like $\sim 1/(a\mbeauty)$ for small 
lattice spacing.

We conclude that all terms which are needed
for the renormalization of $\caa^\mrm{kin}(x_0)$ are present in \eq{e:caahqet}; 
the parameters
may thus be adjusted to absorb all infinities and with properly chosen coefficients 
the continuum
limit of the r.h.s. is expected to exist. The basic assumption of the effective 
field theory is that once the
finite parts of the coefficients have been 
determined by matching a set of observables to QCD, these coefficients are applicable
to any other observables.

\subsubsection{The flavor currents in the effective theory\label{s:fields}}

For later use we here give the expressions for the heavy-light currents.
They are relevant in weak B-meson decays. For better readability, we include
again the time component of the axial current. 

Following our general rules 
for finding the HQET fields which represent the QCD ones we find
\bes
 \Ahqet(x)&=& \zahqet\,[\Astat(x)+ \cahqet\delta\Astat(x)]\,, \\ 
 \Vhqet(x)&=& \zvhqet\,[\Vstat(x) + \cvhqet\delta\Vstat(x)]\,,\\
  \label{e:vhqet}
 \Vkhqet(x)&=& \zvkhqet\,[\Vkstat(x) + \cvkhqet\delta\Vkstat(x)]\,,\\
 \Akhqet(x)&=& \zakhqet\,[\Akstat(x) + \cakhqet\delta\Akstat(x)]\,.
\ees
Ignoring that one can in principle simplify,
for instance $\lightb(x)\gamma_0 \gamma_5\heavy(x) = -\lightb(x)\gamma_5\heavy(x)$ 
due to $P_+\heavy=\heavy$,
the basis fields are written in full analogy to the 
ones in QCD: 
\bes
  \Vstat&=&\lightb\gamma_0\heavy\;,\quad 
  \Astat=\lightb\gamma_0 \gamma_5\heavy\,,
  \\
  \Vkstat&=&\lightb\gamma_k\heavy\;, \quad
  \Akstat=\lightb\gamma_k\gamma_5\heavy\;, 
  \\[1ex]
 \delta\Vstat &=& \lightb \frac{\lnab{i}\!+\!\lnabstar{i}}{2}
	  \gamma_i \heavy \,,
        \quad 
 \delta\Astat = \lightb \frac{\lnab{i}\!+\!\lnabstar{i}}{2}
	  \gamma_i \gamma_5 \heavy \,,
 \\
 \label{e:dvk}
 \delta\Vkstat &=& -\lightb \frac{\lnab{i}\!+\!\lnabstar{i}}{2}
	  \gamma_i \gamma_k \heavy \,,
        \quad 
 \delta\Akstat = \lightb \frac{\lnab{i}\!+\!\lnabstar{i}}{2}
	  \gamma_i \gamma_k\gamma_5 \heavy \,, 
\ees
We have chosen the bare fields such that they are exactly related 
by spin rotations,\footnote{The unnatural $-$ sign in \eq{e:dvk} 
is present because we remain with the definition of 
$\delta\Astat$ in \cite{zastat:pap1} and do not want to introduce
signs in \eq{e:ddvk}.\\
The other rotations look like 
$\delsig^j \Akstat = i\,\eps_{kjl} \Vlstat -i\,\delta_{kj}\Astat$. }
\bes
  \delsig^k \big[\Vstat\big] &=& i \Akstat\,, \quad
  \delsig^k \big[\delta\Vstat\big] = i \delta\Akstat\,, \nonumber\\[-1ex] 
	\label{e:ddvk} \\[-1ex] \nonumber
  \delsig^k \big[\Astat\big] &=& i \Vkstat\,, \quad
  \delsig^k \big[\delta\Astat\big] = i \delta\Vkstat\,.  
\ees
In general, all $Z$-factors and coefficients $c^\mrm{HQET}$ are  
functions of $g_0$ and $a\mbeauty$ which are to be determined
by matching non-perturbatively to QCD. 
\Eq{e:ddvk} will be needed in \sect{s:fb} when we
discuss the static limit of the currents.

\subsection{\SF correlation functions \label{s:sfcf}}

For an understanding of the details of the tests of HQET (\sect{s:hqett}) 
as well as the non-perturbative matching to QCD (\sect{s:hqetstrat})
we will also need some
\SF correlation functions and their HQET expansion, which have
not been defined yet.
We give these details\cite{hqet:pap4} now. The reader who is only interested 
in the general concepts may skip this section.

In \cite{zastat:pap1} static quarks in the  \SF 
were discussed including Symanzik $\Oa$-improvement. 
It turns out 
that there are no dimension four 
composite fields which involve static 
quarks fields and which are compatible with the symmetries of the 
static action and the \SF boundary conditions and which do not vanish by the
equations of motion. Thus there are no $\rmO(a)$ boundary counter terms with
static quark fields. For the same reason there are also no $\rmO(\minv)$ boundary
terms in HQET. 
This then means the HQET expansion of the boundary quark fields $\zeta,\bar\zeta$ 
is trivial up to and including $\minv$ terms. 
% There are no additional $\minv$ terms due to the \SF boundaries.

In the spatial boundary condition of the fermion fields, 
\bes
  \psi(x+\hat{k}L) = \rme^{i\theta} \psi(x) \,, \quad \psibar(x+\hat{k}L) = \rme^{-i\theta} \psibar(x)\,,
\ees
the same phase $\theta$ is taken for all quark fields, whether relativistic
or described by HQET.\footnote{In principle we can easily have different 
$\theta$ for different fields, as long as they are quenched, so in particular
for the heavy field. Since this freedom has not yet been used,
we do not discuss it here. In applications with dynamical
fermions it may well become relevant.}   
In QCD, relevant correlation functions in the pseudo-scalar and vector channel are
\bes
  \fa(x_0,\theta) &=& -{a^6 \over 2}\sum_{\vecy,\vecz}\,
  \left\langle
  (\aimpr)_0(x)\,\zetabar_{\rm b}(\vecy)\gamma_5\zeta_{\rm l}(\vecz)
  \right\rangle  \,, \label{e_fa} \\
  \kv(x_0,\theta) &=& -{a^6 \over 6}\sum_{\vecy,\vecz,k}\,
  \left\langle
  (\vimpr)_k(x)\,\zetabar_{\rm b}(\vecy)\gamma_k\zeta_{\rm l}(\vecz)
  \right\rangle  \,, \label{e_kv}
\ees
with the $\Oa$ improved currents\cite{hqet:pap1} 
\bes
   (\aimpr)_\mu &=& \lightb\gamma_\mu\gamma_5\psi_\beauty +a \ca
	\tilde{\partial}_\mu \lightb\gamma_5\psi_\beauty \\
   (\vimpr)_\mu &=& \lightb\gamma_\mu\psi_\beauty +a \cv
	\tilde{\partial}_\mu \lightb i\sigma_{\mu\nu} \psi_\beauty  \,.
\ees
Furthermore we consider boundary to boundary correlation functions
\bes
  \fone(\theta) &=&
  -{a^{12} \over 2L^6}\sum_{\vecu,\vecv,\vecy,\vecz}
  \left\langle
  \zetalbprime(\vecu)\gamma_5\zzetaprime_{\rm b}(\vecv)\,
  \zetabar_{\rm b}(\vecy)\gamma_5\zetal(\vecz)
  \right\rangle\,, \\
  \kone(\theta) &=&
  -{a^{12} \over 6L^6}\sum_{\vecu,\vecv,\vecy,\vecz,k}
  \left\langle
  \zetalbprime(\vecu)\gamma_k\zzetaprime_{\rm b}(\vecv)\,
  \zetabar_{\rm b}(\vecy)\gamma_k\zetal(\vecz)
  \right\rangle\,.
\ees
Their renormalization is standard~\cite{impr:pap5}, for example
at vanishing light quark masses
\bes
   \left[\fa\right]_\mrm{R}(x_0,\theta) &=& \za\, (1+\frac12 \ba a m_{\rm q,b}) \zzeta^2\,
      (1+\bzeta a m_{\rm q,b}) \fa(x_0,\theta)\,, \quad \\
   \left[\fone\right]_\mrm{R}(\theta) &=& \zzeta^4\,  (1+\bzeta a m_{\rm q,b})^2\fone(\theta)\,,
\ees
with $\zzeta$ a renormalization factor of the relativistic 
boundary quark fields and $\bzeta$ another improvement coefficient. 

Their expansions to first order in $\minv$ read
\bes
  \left[\fa\right]_\mrm{R} &=& \zahqet \zzetah\zzeta \rme^{-\mbeauty x_0}
        \left\{ \fastat + \cahqet \fdeltaastat + \omegakin \fakin
                + \omegaspin \faspin
        \right\}\, \nonumber\\[-0.5ex]\label{e:faexp} \\
  \left[\kv\right]_\mrm{R} &=& \zvhqet  \zzetah\zzeta \rme^{-\mbeauty x_0}
        \left\{ \kvstat + \cvhqet \kdeltavstat + \omegakin \kvkin
                + \omegaspin \kvspin
        \right\}\,  \label{e:kvexp}  \nonumber\\[-0.5ex]\\ 
      &=& -\zvhqet  \zzetah\zzeta \rme^{-\mbeauty x_0}
        \big\{ \fastat + \cvhqet \fdeltaastat + \omegakin \fakin 
               \nonumber \\&& \qquad\qquad\qquad\qquad\qquad\qquad
                -\frac13 \omegaspin \faspin
        \big\}\,, \nonumber \\
  \label{e:foneexp}
  \left[\fone\right]_\mrm{R} &=& \zzetah^2\zzeta^2 \rme^{-\mbeauty T}
        \left\{ \fonestat + \omegakin \fonekin
                + \omegaspin \fonespin
        \right\}\,, \\
  \label{e:koneexp}
  \left[\kone\right]_\mrm{R} &=& \zzetah^2\zzeta^2 \rme^{-\mbeauty T}
        \left\{ \fonestat + \omegakin \fonekin
                -\frac13 \omegaspin \fonespin
        \right\}\,.
\ees
Apart from
\bes
  \fdeltaastat(x_0,\theta) =  -{a^6 \over 2}\sum_{\vecy,\vecz}\,
  \left\langle
  \delta\Astat(x)\,\zetabar_{\rm h}(\vecy)\gamma_5\zeta_{\rm l}(\vecz)
  \right\rangle
\ees
the labeling of the different terms follows directly the one
introduced in \eq{e:exp}. We have used identities such as
$\fakin=-\kvkin\,,\; \faspin=3\kvspin$.
As a simple consequence of the spin symmetry of the static action,
these are valid at any lattice spacing.

%%% Local Variables: 
%%% mode: latex
%%% TeX-master: mb_nf0
%%% TeX-master: "sec2"
%%% End: 

%%% Local Variables: 
%%% mode: latex
%%% TeX-master: "Nara"
%%% End: 

\section{The scope of the theory \label{s:hqets}}

We are now in the position to discuss what can be 
done in the effective theory. 
This concerns also the continuum 
effective theory and in particular the question, 
where perturbation theory is sufficient to give an answer 
and where it is not. 

%%%%%%%%%%%%%%%%%%%%%%%%%%%%%%%%%%%%%%%%%%%%%%%%%%%%%%%%%%%%%%%%%%%%%%%%%%%%%%%%%%%%%%%%%
\subsection{A first example: the decay constant \label{s:fb}}

\subsubsection{Renormalization and matching in perturbation theory}

The matrix element $\Phiqcd$, \eq{e:fb}, 
is scale independent, due to the chiral symmetry of QCD
in the massless limit. But of course it depends
on the mass of the b-quark. In the effective
theory this symmetry is absent and, remaining a while at the lowest
order in $\minv$, $\zastat$ depends on the renormalization
scale, $\mu$, used in the renormalization condition which defines the
finite current. On the other hand, as long as the
effective theory is considered by itself, $\zastat$ does not depend on
the mass of the b-quark in an obvious way. The mass-dependence comes in by 
choosing an appropriate finite renormalization such that the 
matrix elements of the current in the effective theory are
equal to the ones of the QCD current (up to $\rmO(\minv)$). 
This step is called matching.
Choosing to renormalize the current in the effective theory 
in the $\msbar$ scheme (any other scheme would of course be possible) 
we therefore have
\be \label{e:match1}
   \Phiqcd =C_\mrm{match}(\mbeauty,\mu)\times\PhiMSbar(\mu)
                                + \rmO(1/\mbeauty)
\ee
with
\be
 C_\mrm{match}(\mbeauty,\mu) = 1 + c_1(\mbeauty/\mu) \gbar_{\msbar}^2(\mu) +\ldots
\ee
The finite renormalization factor $C_\mrm{match}$ 
is determined (usually in perturbation theory)
such that \eq{e:match1} holds for some particular matrix element
of the current and will then be valid {\em for all matrix elements}.
Obviously, the $\mu$-dependence in \eq{e:match1} is artificial, 
since we have a scale-independent
quantity in QCD. Only the $\mbeauty$-dependence is for real.

The $\mu$-dependence in \eq{e:match1} is removed explicitly by changing 
from $\PhiMSbar(\mu)\equiv\zastatMSbar \langle0| A_0 |B \rangle_\mrm{stat}$
to the RGI matrix element
\bes \label{e:phirgi}
  \PhiRGI
                   = \lim_{\mu\to\infty} \left[\,2b_0 \gbar^2(\mu)\,\right]^{-\gamma_0/2b_0}
                   \PhiMSbar(\mu)\,.
\ees
Here,
the lowest order coefficient of the $\beta$-function, $b_0$ enters as well
as $\gamma_0 = -1/(4\pi^2)$  defined by
\bes
   \gamma(\gbar) \equiv  {\mu \over  \zastat} {\rmd \over \rmd \mu} \zastat
   = - \gamma_0 \gbar^2 + \rmO(\gbar^4)\,.
\ees
We can now write down the HQET-expansion of the QCD matrix element
\be \label{e:matchPhi}
  \Phiqcd = \Cps(M_\beauty/\Lambda_\msbar)\times
                   \PhiRGI\,+\, \rmO(\minv) \,.
\ee
The relation between \eq{e:matchPhi} and \eq{e:match1} 
is easily seen by using
\bea
   {\PhiRGI \over \PhiMSbar(\mu)} = \left[\,2b_0 \gbar^2(\mu)\,\right]^{-\gamma_0/2b_0}
                   \exp\left\{-\int_0^{\gbar(\mu)} \rmd g
                     \left[\,{ \gamma_{\MSbar}(g) \over\beta_{\MSbar}(g)}
                           -{\gamma_0 \over b_0 g}\,\right]
                     \right\} \,
\eea
setting the arbitrary renormalization point $\mu$
to $\mbeauty$ and identifying
\bea \label{e:cps}
      \Cps\left({M_\beauty \over\Lambda_{\MSbar}}\right) &=&
                  C_\mrm{match}(1){\PhiMSbar(\mbeauty) \over \PhiRGI} \\
    &=& \left[\,2b_0 \gbar^2(\mbeauty)\,\right]^{\gamma_0/2b_0}
                   \exp\left\{\int_0^{\gbar(\mbeauty)} \rmd g
                     \left[\,{ \gamma_\mrm{match}(g) \over\beta_{\MSbar}(g)}
                           -{\gamma_0 \over b_0 g}\,\right]  \right\} \,, \nonumber
\eea
where $\gbar$ is taken in the $\msbar$ scheme.
The last equation provides a definition of the
anomalous dimension $\gamma_\mrm{match}$ in the ``matching scheme''.
Perturbatively, it has contributions from $\gamma_{\MSbar}$ as well as 
from  $C_\mrm{match}$, namely
\bes \label{e:gammamatch}
   \gamma_\mrm{match}(\gbar) = -  \gamma_0\gbar^2 - [\gamma_1^{\msbar} + 2b_0 c_1(1)]\gbar^4
   + \ldots \,.
\ees
Replacing the $\msbar$ coupling by a non-perturbative one,
$\gamma_\mrm{match}$ may also be defined beyond perturbation theory
through eqs.~(\ref{e:cps},\ref{e:matchPhi}).\footnote{Clearly the r.h.s.
of \eq{e:cps} is a function of $\gbar^2(\mbeauty)$, i.e. a function
of $\mbeauty/\Lambda_\msbar$. We prefer to write it as a function
of the ratio of renormalization group invariants, $M_\beauty /\Lambda_{\MSbar}$.}
Another advantage of \eq{e:cps} (compared to \eq{e:match1})
is that $\Cps$ is independent of the arbitrary choice of renormalization
scheme for the composite operators in the effective theory. Apart from the
choice of the QCD coupling, the ``convergence'' of the series
\eq{e:gammamatch} is dictated by the physics, nothing else.

Note further that (at leading order in $\minv$) the conversion function
$\Cps$ contains the full (logarithmic) mass-dependence.
The non-perturbative effective theory matrix elements, $\PhiRGI$, 
are mass independent numbers.
Conversion functions such as $\Cps$ are universal for all
(low energy) matrix elements of their associated operator. Thus
\bea
    \caa(x_0)
                    &\simas{x_0 \gg \minv }&
                    [\Cps(\frac{M_\beauty}{\Lambda_\msbar})\,\zastatRGI]^2
                    \langle \Astat(x)^\dagger \Astat(0)\rangle
                    +\rmO(\frac1\mbeauty)\, \nonumber\\[-1ex]
\eea
%%%%%%%%%%%%%%%%%%%%%%%%%%%%%%%%%%%%%%%%%%%%%%%%%%%%%%%%%%%%%%%%%%%%%%%%%%%%%%%%%%%%%%%%%%
\begin{figure}[tb]
\vspace{0pt}
\centerline{\epsfig{file=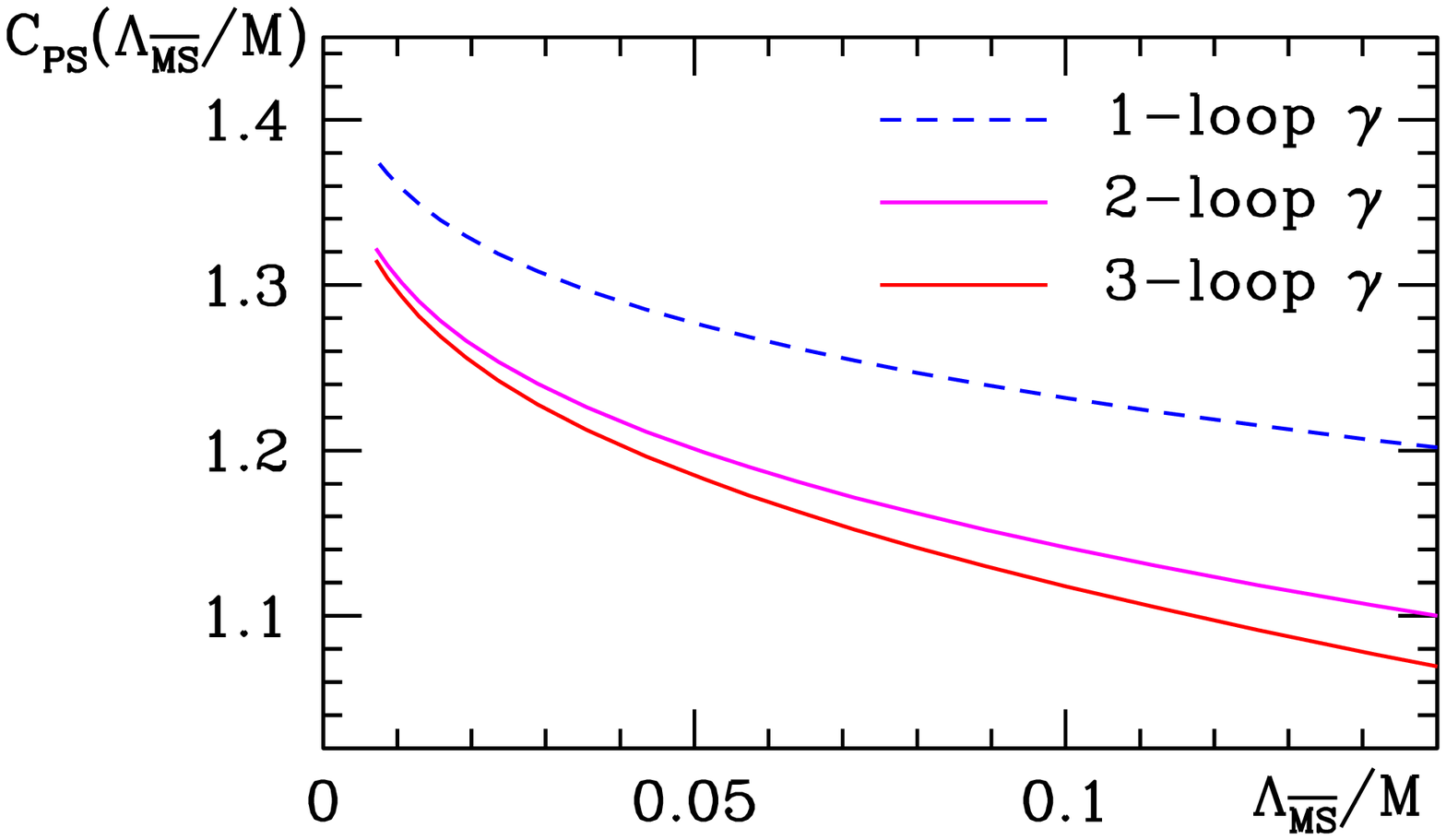,
       width=8.0cm} }
  \caption{\footnotesize $\Cps$ estimated in perturbation
        theory. For B-physics
	we have $\Lambda_\msbar/\Mbeauty\approx0.04$.
        }\label{f:cps}
\end{figure}
%%%%%%%%%%%%%%%%%%%%%%%%%%%%%%%%%%%%%%%%%%%%%%%%%%%%%%%%%%%%%%%%%%%%%%%%%%%%%%%%%%%%%%%%%%
is a straight forward generalization of \eq{e:matchPhi}.
Here, \\$\zastatRGI=\lim_{\mu\to\infty} \left[\,2b_0 \gbar^2(\mu)\,\right]^{-\gamma_0/2b_0}
                   \zastatMSbar(\mu)$.

Analogous expressions for the conversion functions
are valid for the time component of the
axial current replaced by other composite fields, for example
the space components of the vector current.
Based on the work of \cite{BroadhGrozin,Shifman:1987sm,Politzer:1988wp}
and recent efforts
their perturbative expansion is known including the 3-loop anomalous dimension
$\gamma_\mrm{match}$ obtained from the
3-loop anomalous dimension  $\gamma_{\MSbar}$ \cite{ChetGrozin}
and the 2-loop matching function $C_\mrm{match}$
\cite{Ji:1991pr,BroadhGrozin2,Gimenez:1992bf}.
\Fig{f:cps}, taken from \cite{hqet:pap3},
illustrates that the remaining $\rmO(\gbar^6(\mbeauty))$ errors in $\Cps$
seem to be relatively small.

We return to the full set of heavy-light flavor currents \sect{s:fields}.
The bare fields satisfy the symmetry relations \eq{e:ddvk}. The same
is then true for the RGI fields
in static approximation. Furthermore, the axial currents are
related to the vector ones by a chiral rotation of the light
quark fields\cite{zvstat:onogi}. It then follows that {\em in static
approximation} the effective currents are given by
\bes
  \Ahqet &=& C_{\rm PS}(\Mbeauty/\Lambda_\msbar)\, \zastatrgi(g_0)\,\Astat\,, \\
  \Vkhqet &=& C_{\rm\bf V}(\Mbeauty/\Lambda_\msbar)\, \zastatrgi(g_0)\,\,
	 \Vkstat\,,\\  
  \Vhqet &=& C_{\rm V}(\Mbeauty/\Lambda_\msbar)\, \zastatrgi(g_0)\,Z^\mrm{stat}_{\rm V/A}(g_0)\,
	 \Vstat\,,\\   
  \Akhqet &=& C_{\rm\bf A}(\Mbeauty/\Lambda_\msbar)\, \zastatrgi(g_0)Z^\mrm{stat}_{\rm V/A}(g_0)
         \,\Akstat\,.
\ees
The finite renormalization $Z^\mrm{stat}_{\rm V/A}(g_0)$ can either be
computed by a Ward identity \cite{zvstat:onogi} or, if one has a regularization
with exact chiral symmetry, it is one. The factor $\zastatrgi(g_0)$ 
is analogous to $\zM(g_0)$ in \eq{e_M_final} and has been computed in a similar
way \cite{zastat:pap3}. It can be split into
\be
   \label{e:zastatrgi}
   \zastatrgi(g_0) = \left.{\Phi_\mrm{RGI} \over \Phi(\mu)} \times \zastat(g_0,2\lmax/a)
	\right|_{\mu=1/(2\lmax)}\,,
\ee
where the second factor depends on the lattice action,
the first one does not. The first factor has been shown in \fig{f_phibar} 
for a series of $\mu$ in the quenched approximation with a smallest 
scale of $\mu=1/(2\Lmax)$.

We finally note that the bare pseudo scalar density, $P$, and the scalar density, $S$,
are  identical to $-\Astat$ and $\Vstat$, respectively. This then also holds for
the RGI ones; they are given by  $-\zastatrgi(g_0) \Astat$ and 
$\zastatrgi(g_0)\,Z^\mrm{stat}_{\rm V/A}(g_0)\Vstat$. 

For the renormalized $\Phqet$ and $\Scalarhqet$, there is in principle an arbitrariness,
since in QCD they depend on a scale and a scheme. 
This arbitrariness is fixed by considering the RGI fields
in QCD. They satisfy the PCAC and PCVC relation with the RGI quark masses, 
for example ($M_\mrm{l}$ is the mass of the light quark)
\be
  \partial_\mu (\ar)^\mrm{QCD}_\mu = (\Mbeauty+M_\mrm{l}) P_\mrm{RGI} \,. 
\ee
Taking the vacuum-to-B-meson matrix element of this relation 
(in static approximation) fixes
\bes
  \Phqet &=& -C_{\rm PS}(\Mbeauty/\Lambda_\msbar)\, {\mb \over \Mbeauty}\, \zastatrgi(g_0) \Astat\,, \\
  \Scalarhqet &=& C_{\rm V}(\Mbeauty/\Lambda_\msbar)\, {\mb \over \Mbeauty}\, Z^\mrm{stat}_{\rm V/A}(g_0)
                  \,\zastatrgi(g_0)\,\Vstat\, 
\ees
{\em in static approximation}. 
(Choosing a different matrix element would change $\mb\to\mb+\rmO(\Lambda_\mrm{QCD})$.)
We leave it as an exercise to verify these relations.

\begin{figure}[tb]
\vspace{0pt}
\centerline{\epsfig{file=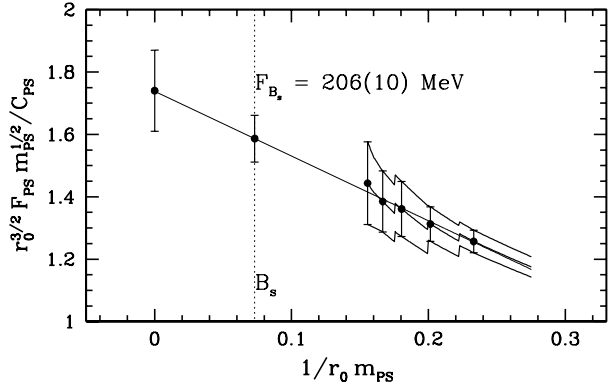,
       width=8.0cm} }
  \caption{\footnotesize Example of an interpolation between a 
        static result and results with $m_\mrm{h} < \mbeauty$. 
        Continuum extrapolations are done before the 
        interpolation \protect\cite{lat03:juri}.
	The point at $1/r_0\mp = 0$ is given by $r_0^{3/2}\Phi_\mrm{RGI}$.
        }\label{f:fbinterpol}
\end{figure}

All conversion functions $C$ are known up to
(presumably small) $\alpha(\mbeauty)^3$ errors.  At least when the conversion functions 
are known with such high 
precision, the knowledge of the leading term in expansions
such as \eq{e:matchPhi}
is very useful to constrain the large mass behavior 
of QCD observables, computed on the lattice with unphysical 
quark masses $m_\mrm{h} < \mbeauty$, typically 
$m_\mrm{h} \approx m_\mrm{charm}$. As illustrated in \fig{f:fbinterpol},
one can then, with a reasonable smoothness assumption,  
interpolate to the physical point.

%%%%%%%%%%%%%%%%%%%%%%%%%%%%%%%%%%%%%%%%%%%%%%%%%%%%%%%%%%%%%%%%%%%%%%%%%%%%%%%%%%%%%%%%%%
%%%%%%%%%%%%%%%%%%%%%%%%%%%%%%%%%%%%%%%%%%%%%%%%%%%%%%%%%%%%%%%%%%%%%%%%%%%%%%%%%%%%%%%%%%

The relation 
between the RGI fields and the bare fields has so far  been obtained
for 
\bi
\item $\zastatrgi$ with both 
      $\nf=0$  \cite{zastat:pap3} and $\nf=2$ \cite{zastat:nf2};
\item the parity violating $\Delta B=2$ four fermion operators 
        \cite{stat:zbb_pert,stat:zbb_nf0} for $\nf=0$ while 
        $\nf=2$ has been started.
\ei
In the second case, the matrix elements of the two operators, 
evaluated
in twisted mass QCD will give the standard model B-parameter
for B-$\rm\overline{B}$ mixing. 

Thus, soon one will be able to do interpolations such as
\fig{f:fbinterpol} also for semi-leptonic decays such as
$B\to\pi\;l\;\nu$ (close to zero recoil) and for the 
B-parameter.

\subsubsection{Beyond the leading order: the need for non-perturbative
        conversion functions} \label{s:need}

Still, getting the continuum extrapolations of the data
at finite heavy quark masses in \fig{f:fbinterpol}
under control may represent a challenge
when dynamical fermions are included. 
Furthermore one should not forget that the functional 
form of the interpolation in that figure does essentially
assume that the $1/m$ expansion remains reasonably accurate 
also significantly below $\mbeauty$.

It is therefore natural to try to compute the $\minv$ correction 
directly in HQET.
However, if one wants to do this consistently, the leading
order conversion functions such as  $\Cps$ have to be known
non-perturbatively. This general problem in the
determination of power corrections in QCD is seen in the following
way. Consider
the error made in \eq{e:cps} (or \eq{e:match1}) when the
anomalous dimension has been computed at $l$ loops and
$C_\mrm{match}$ at $l-1$ loop order. The conversion
function $\Cps$ is then known up to an {\em error}
\be \label{e:deltacps}
     \Delta(\Cps) \propto  [\gbar^2(\mbeauty)]^{l} \sim
        \left\{{1 \over 2b_0\ln(\mbeauty/\Lambda_\mrm{QCD})}\right\}^{l}
        \ggas{\mbeauty\to\infty}\; {\Lambda_\mrm{QCD} \over \mbeauty} \,.
\ee
As $\mbeauty$ is made large, this perturbative error becomes 
dominant over the power correction one wants to determine. 
Taking a
perturbative conversion function and adding power corrections
to the leading order effective theory is thus
a phenomenological approach, where one assumes
that the coefficient of the $[\gbar^2(\mbeauty)]^{l}$ term
(as well as higher order ones)
is small, such that the $\Lambda/\mbeauty$ corrections dominate 
over a certain mass interval. In such a phenomenological 
determination of a power correction, its size depends on the
order of perturbation theory considered.
A theoretically
consistent evaluation of power corrections requires
a fully non-perturbative formulation of the theory including
a non-perturbative matching to QCD, see \sect{s:hqetstrat}.

{\bf Note} for experts. \Eq{e:deltacps} has little to do with renormalons. 
Rather it is due to the truncation of perturbation theory at a fixed 
order. Of course a renormalon-like growth of the perturbative coefficients 
does not help.

\subsubsection{Splitting leading order (LO) and next to leading order (NLO)}

We just learned that the very definition of a NLO correction
to $\fb$ means to take \eq{e:fBexp} with all coefficients $\zahqet \ldots \cahqet$
determined non-perturbatively. We want to briefly explain that the split
between  LO and NLO is not unique. This is
fully analogous to the case of standard perturbation theory in $\alpha$, where the 
split depends on the renormalization scheme used, or better on the experimental
observable used to determine $\alpha$ in the first place.

Consider the lowest order. The only coefficient needed in \eq{e:fBexp} is
then $\Cps\zastatrgi = \lim_{\mbeauty\to\infty} \zahqet$. 
It has to be fixed by matching some matrix element of $\Astat$ to the 
matrix element of $A_0$ in QCD. For example one may choose $\langle B'|A_0^\dagger|0\rangle$,
with $|B'\rangle$ denoting the first pseudo-scalar excited state.  Or one
may take a finite volume matrix element $\langle B(L)|A_0^\dagger|\Omega(L)\rangle$, 
see \sect{s:hqett}. Since the matching involves the QCD matrix element,
there are higher order in $\minv$ ``pieces'' in these equations. There is no reason
for them to be independent of the particular matrix element. So from 
matching condition to matching condition,
$\Cps\zastatrgi$ differs by $\rmO(\Lambda_\mrm{QCD}/\mbeauty)$ terms.

The matrix element $\fb$ in static approximation inherits
this $\rmO(\Lambda_\mrm{QCD}/\mbeauty)$ ambiguity.
The $\Lambda_\mrm{QCD}/\mbeauty$ corrections are hence not unique. 
Fixing a matching condition, the leading order $\fb$ as well as the one including
the corrections can be computed and has a continuum limit.  Their difference
can be defined as the $\minv$ correction.  However, what matters is not the 
ambiguous NLO term, but the fact that the uncertainty is reduced
from $\rmO(\Lambda_\mrm{QCD}/\mbeauty)$ in the LO term
to $\rmO(\Lambda^2_\mrm{QCD}/\mbeauty^2)$ in the sum.

%%%%%%%%%%%%%%%%%%%%%%%%%%%%%%%%%%%%%%%%%%%%%%%%%%%%%%%%%%%%%%%%%%%%%%%%%%%%%%%%%%%%%%%%%%
\subsection{A second example: mass formulae}

Let us represent the formula for the meson mass,
\eq{e:mBexp}, including also the vector meson as
\be
   \label{e:mBav}
   {\mbav} \equiv {1\over 4} [\mB+3\mBstar] = \mbeauty+\dmstathat + \Estat +\omegakin \Ekin 
        +\rmO(1/\mbeauty^2)
\ee
and 
\be
   \label{e:mBsplitt}
   {\mbsplitt} \equiv \mBstar-\mB = \omegaspin \Espin +\rmO(1/\mbeauty^2)\,,
\ee
where the fact that $\Ospin$ does not contribute in
\eq{e:mBav} is a consequence of the exact spin symmetry of the 
lowest order. It is a simple exercise to verify the above equations
explicitly by considering correlation functions of $V_k=\lightb\gamma_k \psi_\beauty$
instead of $A_0$. 

Usually these equations are written as \cite{hqet:FalkNeub1}
\be \label{e:massformulae}
   \mbav \sim \mbeauty + \overline{\Lambda} 
    + {1\over 2\mbeauty} \lambda_1 %+\rmO(1/\mbeauty^2) 
    \,,\quad
   \mbsplitt \sim -{2\over \mbeauty} \lambda_2 %+\rmO(1/\mbeauty^2) 
   \,.
\ee
We arrive at a similar form by defining power divergent subtractions 
for $\Ekin$ and $\Estat$ vs.
\be
  \label{e:mbavsplit}
  \mbav = \mbeauty + \mb^{(0a)} +\mb^{(0b)} + \mb^{(1a)} +\mb^{(1b)} + 
  \rmO(\Lambda^3_\mrm{QCD}/\mbeauty^2)\,, 
\ee
where we have split up as
\bes
  \label{e:mb0ab}
    \rmO(\Lambda_\mrm{QCD}) &:&\; 
  \mb^{(0a)} =\dmstathat^\mrm{stat}+E_\mrm{stat}^\mrm{sub}\,,\quad
  \mb^{(0b)} = \Estat-E_\mrm{stat}^\mrm{sub}\,,\\
  \rmO(\Lambda^2_\mrm{QCD}/\mbeauty) &:&\;
  \mb^{(1a)} = \dmstathat^\mrm{(1)}+\omegakin\Ekin^\mrm{sub}\,,\quad
  \mb^{(1b)} = \omegakin [\Ekin-\Ekin^\mrm{sub}]\,. \nonumber \\[-0.5ex]
  \label{e:mb1ab}
\ees
The subtraction terms (the Hamiltonian should be defined as
$\ham = - \frac1{a} \ln(\trans)$ in terms of the transfer matrix $\trans$)
\bes
   \Estat^\mrm{sub} = \left.\langle \beta | \ham | \beta \rangle_\mrm{stat}
                      \right|_{\dmstat=0}\,, \quad
   \Ekin^\mrm{sub} = \langle \beta | -a^3\sum_\vecx \Okin(x) | \beta \rangle_\mrm{stat}\,.
\ees
are chosen such that $\mb^{(0a)} \ldots \mb^{(1b)}$ are finite and
have a continuum limit.
This follows from our discussion of renormalization.  By $\dmstathat^\mrm{stat}$ we
denote $\dmstathat$ in static approximation and by $\dmstathat^\mrm{(1)}$ the piece that
is to be added when $\omegakin\ne0$; it accounts for the term $c_2/a^2$ in \eq{e:okinr}.

This  rewriting exposes that the split up of
the meson formulae into various orders of $\minv$ is not unique.
In the chosen form it depends on the arbitrary state $|\beta\rangle$.
In other words: non-perturbatively, $\overline{\Lambda}$ and $\lambda_1$ can 
be defined exactly, but in many ways.\footnote{
In phenomenological applications\cite{CKM:CERN},
one works in dimensional regularization 
where the subtraction terms can be omitted.
One defines a perturbative 
scheme for $\mbeauty$. It is then clear
that (for example) $\overline{\Lambda}$ depends on the order of perturbation
theory as $ [\gbar^2(\mbeauty)]^{l+1} \mbeauty] \ggas{\mbeauty\to\infty} \Lambda_\mrm{QCD}$
as in \eq{e:deltacps}. Quantities such as
$\overline{\Lambda}$ are then {\em effective parameters}, not directly related to the
asymptotic $\minv$ expansion.}

Toward the end of this lecture we will see
how the subtraction terms and $\mbeauty+\dmstathat$ can be defined such
that they are computable in practice and the remaining 
error is reduced to  $\rmO(\Lambda^3_\mrm{QCD}/\mbeauty^2)$.

%%% Local Variables: 
%%% mode: latex
%%% TeX-master: "Nara"
%%% End: 

\section{Non-perturbative tests of HQET \label{s:hqett}}

Although it is generally accepted that HQET is an effective theory of
QCD, tests of this equivalence
are rare and mostly based on phenomenological analysis of experimental
results. A pure theory test can be performed if
QCD including a heavy enough quark can be simulated on the lattice
at lattice spacings which are small enough to be able to take the continuum limit.
This has recently been achieved~\cite{hqet:pap3}
and will be summarized below.
We start with the QCD side of such a test.
Lattice spacings such that $a \mbeauty \ll 1$ can be reached if
one puts the theory in a finite volume, $L^3 \times T$ with $L,T$ not
too large. We shall use $T=L$.
For various practical reasons,
\SF boundary conditions are chosen. Equivalent
boundary conditions are imposed in the effective theory. 
We then consider correlation functions such as $\fa$ and $\fone$. 
The first one is the correlator
of boundary quark fields $\zeta$ (located at $x_0=0$) and 
the time component of the axial current in the bulk ($0<x_0<T$).
The second one describes the propagation of
a quark-antiquark pair from the $x_0=0$ boundary to the
 $x_0=T$ boundary. See \sect{s:sfcf} for details.

We then take a ratio for which the renormalization
factors of the boundary fields cancel,
\bes
\Yr(L,\Mbeauty) &\equiv& \za \left.{\fa(L/2) \over \sqrt{f_1}}\right|_{T=L} =
            {\langle \Omega(L)| A_0 |B(L)\rangle
              \over
              || \,| \Omega(L)\rangle \,||\; ||\, | B(L)\rangle \, ||}
            ,\\[1ex] \nonumber
            && |B(L)\rangle = \rme^{ -L \ham/2 } |\varphi_{\rm B}(L)\rangle \,,\;
            |\Omega(L)\rangle =\rme^{ -L \ham/2 } |\varphi_{0}(L)\rangle \,.
\ees
As shown in the above equations, $\Yr$ can be represented as
a matrix element of the axial current between a
normalized state $ |B(L)\rangle$
with the quantum numbers of a B-meson and  $|\Omega(L)\rangle$ which
has vacuum quantum numbers. The time evolution $ \rme^{-L\ham /2 }$ ensures
that both of these states are dominated by energy eigenstates with energies
around $2/L$ and less (above $\mbeauty$). In other words, HQET is applicable 
if $1/L \ll \mbeauty$ (and of course $\Lambda_\mrm{QCD} \ll \mbeauty$).

One then expects (for fixed $L \Lambda_\mrm{QCD}$)
\bes \label{e:yrequiv}
        \Yr(L,\Mbeauty) / \Cps(\Mbeauty/\Lambda) = \XRGI + \rmO(1/z)\,, 
        \quad z=\Mbeauty L\,,
\ees
where the $1/\mbeauty$ corrections are written in the dimensionless
variable $1/z$ and $\XRGI$ is defined as $\Yr$ but
at lowest order in the effective theory and normalized as in \eq{e:phirgi}.
%%%%%%%%%%%%%%%%%%%%%%%%%%%%%%
\begin{figure}[tb]
%\centering
  \includegraphics[width=0.49\textwidth]{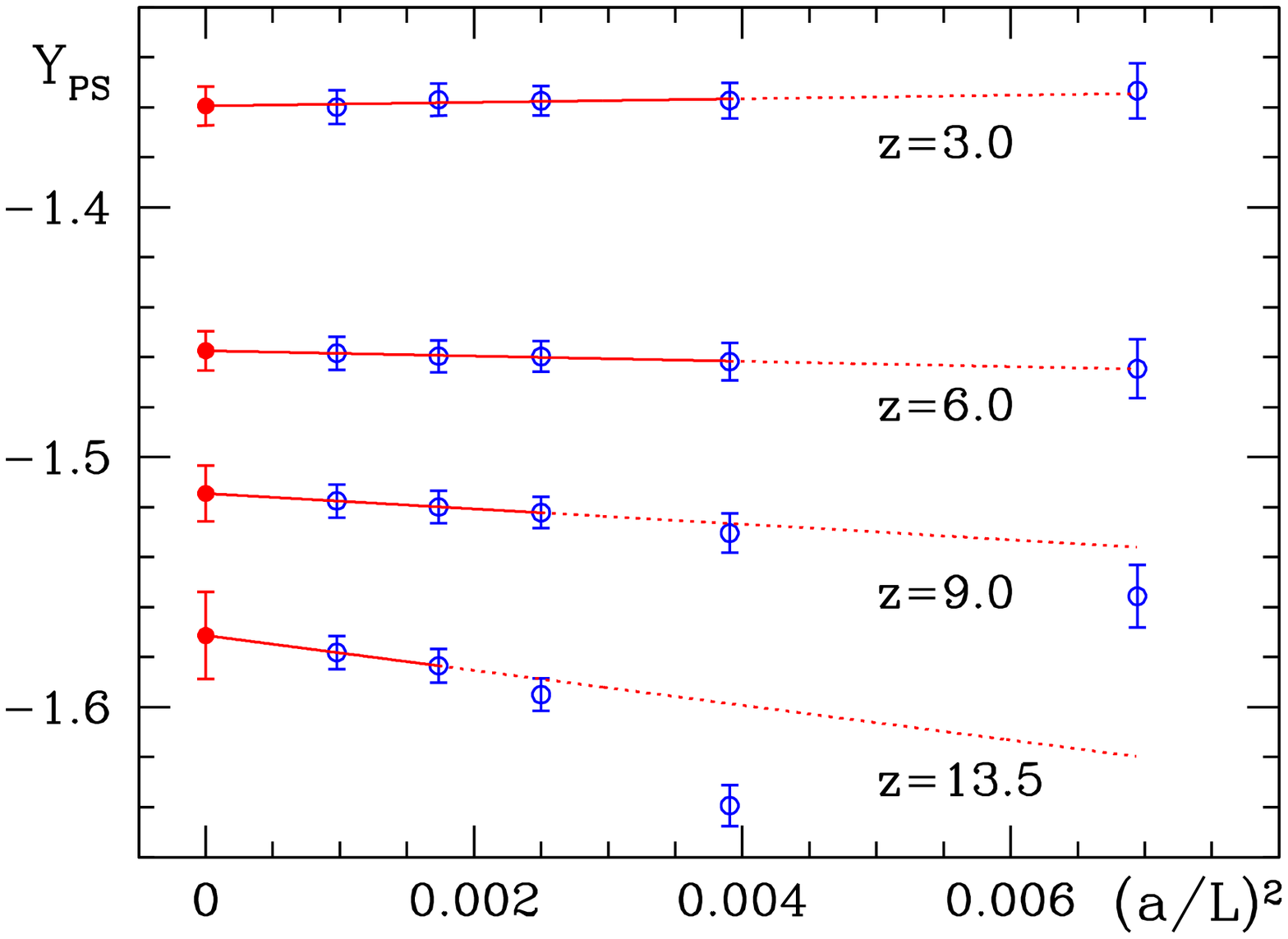}\hfill
  \includegraphics[width=0.49\textwidth]{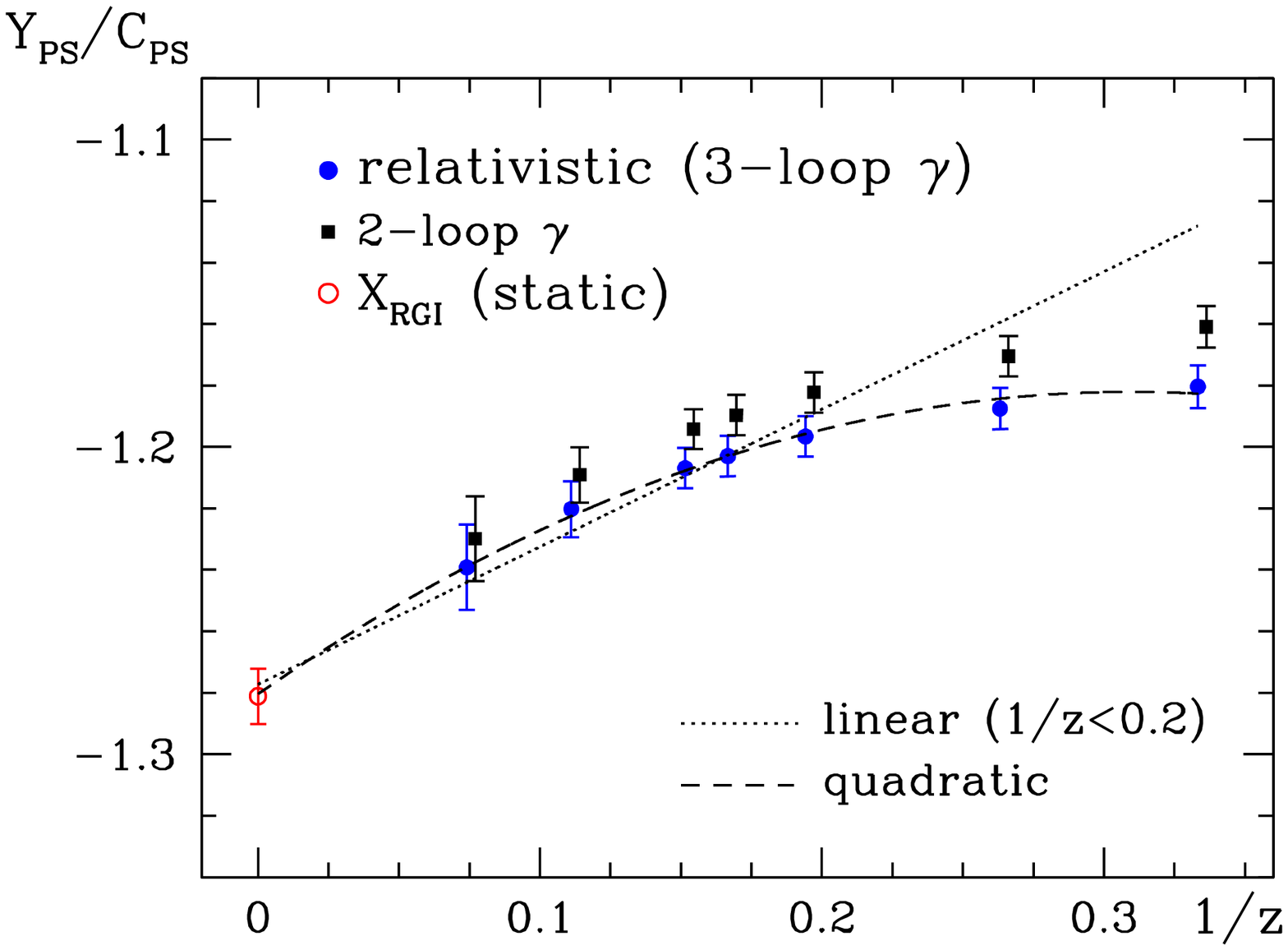}
\caption{ \footnotesize
Testing \eq{e:yrequiv} through numerical simulations in
the quenched approximation and for $L\approx0.2\,\fm$ \protect\cite{hqet:pap3}.
The physical mass of the b-quark corresponds to $z\approx5$.
}\label{f:yrmatch}
\end{figure}
%%%%%%%%%%%%%%%%%%%%%%%%%%%%%%
Of course such relations are expected after the continuum limit
of both sides has been taken separately. For the case of
$\Yr(L,\Mbeauty)$, this is done by the following steps:
\begin{itemize}
\item Fix a value $u_0$ for the renormalized coupling
        $\gbar^2(L)$ (in the \SF scheme) at
        vanishing quark mass. In \cite{hqet:pap3}
        $u_0$ is chosen such that
        $L\approx0.2\,\fm$.
\item For a given resolution $L/a$, determine the bare coupling
        from the condition $\gbar^2(L)=u_0$. This can easily be done since
        the relation between bare and renormalized coupling is known~\cite{mbar:pap1}.
\item Fix the bare quark mass $\mq$ of the heavy quark such that
        $LM=z$  using the known renormalization
        factors $\zM,Z$ in $M =\zM Z\, (1+a\bm \mq)\,\mq$,
        where $Z$ was introduced in \eq{e:mri} and 
        $\zM$ in \eq{e_M_final}.
\item Evaluate $\Yr$ and repeat for better resolution $a/L$.
\item Extrapolate to the continuum as shown in \fig{f:yrmatch}, left.

\end{itemize}
As can be seen in the figure, the continuum extrapolation becomes
more difficult as the mass of the heavy
quark is increased and $\rmO((a\,m)^2)$ discretization errors
become more and more important. In contrast the continuum
extrapolation in the static effective theory (\fig{f:XRGI}) is much easier
(once the renormalization factor relating bare current and RGI current
%%%%%% figure: XRGI %%%%%%%%%%%%%%%%%%%%%%%%%%%%%%%%%%%%%%%%%%%%%
%
\begin{figure}
  \centerline{\includegraphics[width=7.5cm]{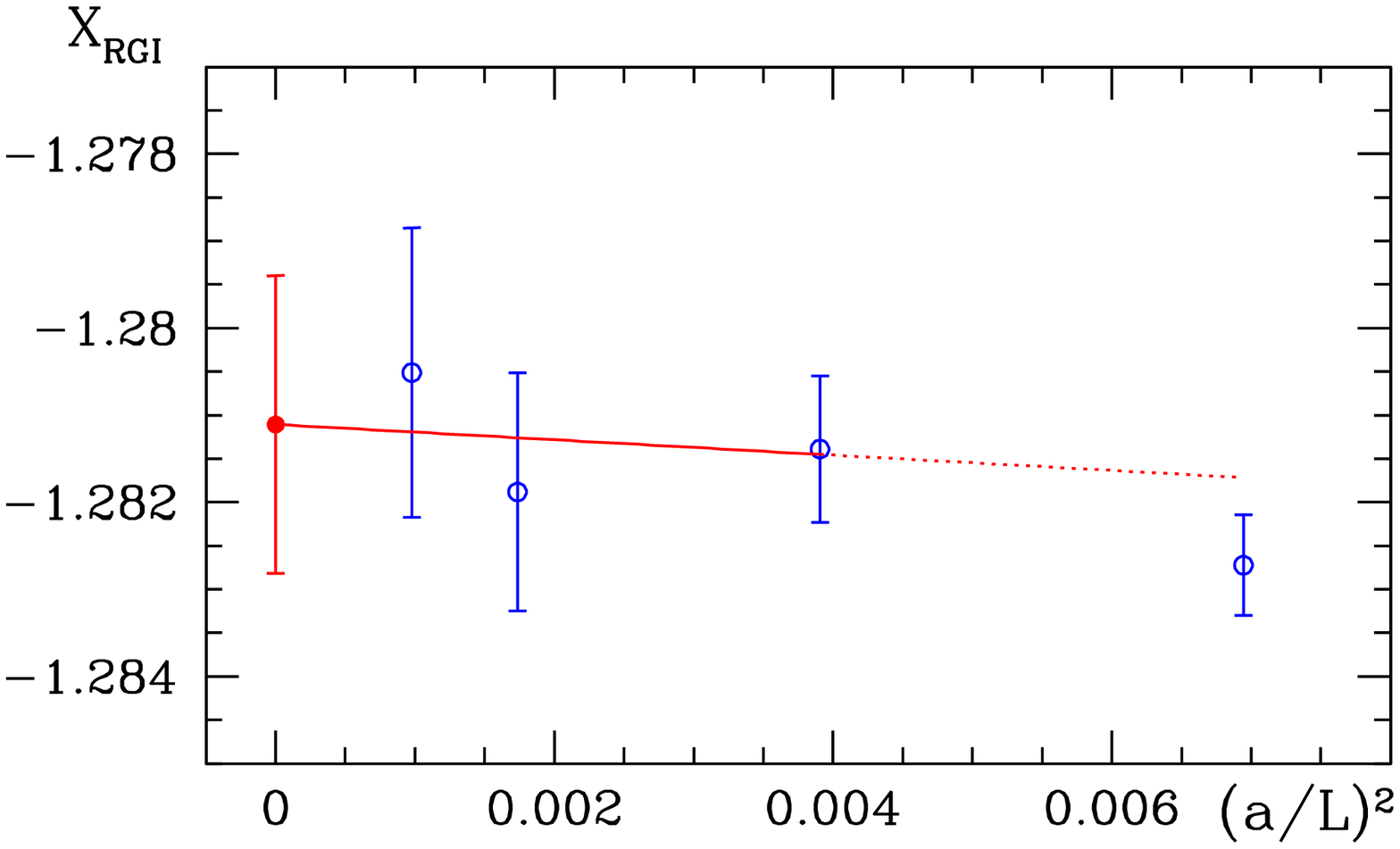}}
\caption{
\footnotesize
Continuum extrapolation of $\XRGI$ \protect\cite{hqet:pap3}.
}
\label{f:XRGI}
%\end{figure} %%%%%%%%%%%%%%%%%%%%%%%%%%%%%%%%%%%%%%%%%%%%%
\end{figure}
is known~\cite{zastat:pap3}). After the continuum limit
has been taken, the finite mass QCD observable $\Yr(L,M)$ turns
smoothly into the prediction from the effective theory
as illustrated in the r.h.s. of \fig{f:yrmatch}. Several such successful
tests were performed in \cite{hqet:pap3},
two of them with the static ($M\to\infty$) limit known from
the spin symmetry of HQET. For lack of space we do
not show more examples but only note that the coefficient
of the $1/z^n$ terms in fits to the finite mass results 
together with the static limit are roughly
of order unity.

Of course, finite mass lattice QCD results have been compared to
the static limit over the years, see for example
\cite{fb_wupp,reviews:beauty,reviews:hartmut97,El-Khadra:1998hq,Aoki:1998ji,Bernard:1998xi,heavylight:Bec98,AliKhan:2000eg,Bowler:2000xw,Lellouch:2000tw,lat01:ryan,romeII:fb,romeII:mb}
and references therein. So what is new in the tests just discussed?
The composite
fields were renormalized non-perturbatively throughout
and, by considering a small volume, the continuum limit could be taken
at large quark masses.

%%% Local Variables: 
%%% mode: latex
%%% TeX-master: "Nara"
%%% End: 

\section{Strategy for non-perturbative matching \label{s:hqetstrat}}

Following \sect{s:need}, the missing piece for a general 
computation including $\minv$ corrections is a practical
strategy for determining the parameters in the Lagrangian and in the
effective fields beyond perturbation theory.
Let us denote the number of parameters which have to be 
determined, not counting the parameters in the light 
sector of QCD, by  $\Neff$. For instance, including $\minv$ terms
but not considering matrix elements of any composite fields,
we have $\Neff=3$, namely $\omegakin,\omegaspin$ and $\mbeauty+\dmstathat$.
Given these three parameters, all masses can be computed.
If in addition
we want to compute matrix elements of $A_0$, such as $\fb$,
we have $\Neff=5$, since also $\zahqet,\cahqet$ are parameters
of the theory.

\subsection{Matching in small volume}

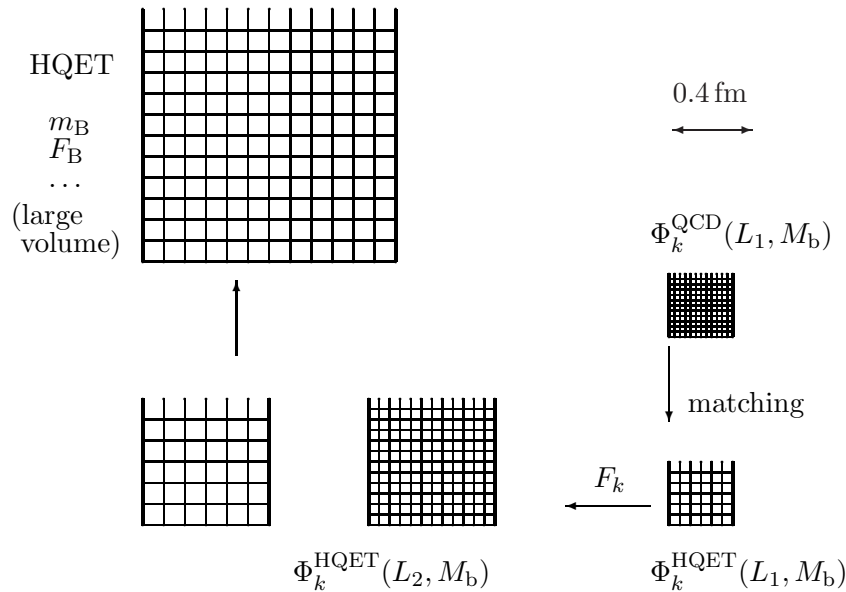
\begin{figure}[tb]
\vspace{0pt}
\centering
 \definecolor{Red}{gray}{0}
 \definecolor{PineGreen}{gray}{0}
 \definecolor{Blue}{gray}{0}
 \definecolor{Magenta}{gray}{0}
 \definecolor{Black}{gray}{0}
 \definecolor{Gray}{gray}{0}
 \definecolor{Lavender}{gray}{0}
\newcommand{\mgt}{\cmag}

\newcommand{\cred}{\color{Red}}
\newcommand{\cblu}{\color{Blue}}
\newcommand{\cmag}{\color{Magenta}}
\newcommand{\cgre}{\color{PineGreen}}
\newcommand{\cbla}{\color{Black}}

\newcommand{\bla}{\cbla}
\newcommand{\red}{\cred}
\newcommand{\gre}{\cgre}
\newcommand{\blu}{\cblu}
\newcommand{\asylat}[1]  % \unitlength 1mm for 8 cm lattice!
{\begin{picture}(160,160)
\unitlength #1
\linethickness{0.1mm}
\blu\multiput(0,0)(0,10.0){8}{\line( 1, 0){80.0}}
\blu\multiput(0,0)(2.5,0){32}{\line( 0, 1){80.0}}
\end{picture}
}
\newcommand{\vbiglat}[1]  % \unitlength 1mm for 16 cm lattice!
{\begin{picture}(160,160)
\unitlength #1
\linethickness{0.1mm}
\blu\multiput(0,0)(0,10.0){16}{\line( 1, 0){160.0}}
\blu\multiput(0,0)(10.0,0){16}{\line( 0, 1){160.0}}
\end{picture}
}
\newcommand{\vvbiglat}[1]  % \unitlength 1mm for 32 cm lattice!
{\begin{picture}(320,280)
\unitlength #1
\linethickness{0.1mm}
\blu\multiput(0,0)(0,10.0){28}{\line( 1, 0){320.0}}
\blu\multiput(0,0)(10.0,0){32}{\line( 0, 1){280.0}}
\end{picture} 
}

\newcommand{\bwave}[1]  % \unitlength 1mm for a=1 cm lattice!
{\unitlength #1
\begin{picture}(50,20)
\linethickness{0.1 mm}
\mgt \qbezier(0,0)(1.2,20)(1.9,20)     % 0,pi/2 (pi=120)
     \qbezier(1.9,20)(2.5,20)(3.75,0)     % pi/2,pi
      \qbezier(3.75,0)(4.95,-20)(5.6,-20)
      \qbezier(5.6,-20)(6.3,-20)(7.5,0)
\end{picture}
}
\newcommand{\bwavelnod}[1]  % \unitlength 1mm for a=1 cm lattice!
{\unitlength #1
\begin{picture}(50,20)
\linethickness{0.7 mm}
\mgt \qbezier(0,0)(24,20)(37.5,20)     % 0,pi/2 (pi=120)
     \qbezier(37.5,20)(50,20)(75,0)     % pi/2,pi
      \qbezier(75,0)(99,-20)(112.5,-20)
      \qbezier(112.5,-20)(126,-20)(150,0)
\end{picture}
}
\newcommand{\bwaveasy}[1]  % \unitlength 1mm for a=1 cm lattice!
{\unitlength #1
\begin{picture}(50,20)
\linethickness{0.7 mm}
\mgt \qbezier(0,0)(12,20)(19,20)     % 0,pi/2 (pi=120)
     \qbezier(19,20)(25,20)(37.5,0)     % pi/2,pi
      \qbezier(37.5,0)(50,-20)(56,-20)
      \qbezier(56,-20)(69,-20)(75,0)
\end{picture}
}

\newcommand{\piwave}[1]  % \unitlength 1mm for a=1 cm lattice!
{\unitlength #1
\begin{picture}(50,20)
\linethickness{0.2 mm}
\gre \qbezier(0,0)(40,20)(70,20)      % 0,pi/2 (pi=140)
     \qbezier(70,20)(100,20)(140,0)   % pi/2,pi
      \qbezier(140,0)(170,-20)(210,-20)
      \qbezier(210,-20)(240,-20)(280,0)
\end{picture}
}
\newcommand{\pihwave}[1]  % \unitlength 1mm for a=1 cm lattice!
{\unitlength #1
\begin{picture}(50,20)
\linethickness{0.7 mm}
\gre \qbezier(0,0)(40,20)(70,20)      % 0,pi/2 (pi=140)
     \qbezier(70,20)(100,20)(140,0)   % pi/2,pi
\end{picture}
}

%%% Local Variables: 
%%% mode: latex
%%% TeX-master: t
%%% End: 
\input plots/hqetstrat.tex
  \caption{\footnotesize The strategy for a non-perturbative determination of 
        the HQET-parameters from QCD simulations in a small volume. 
        Steps indicated by arrows are to be repeated at smaller
        lattice spacings to reach a continuum limit. 
        }\label{f:strat}
\end{figure}

Observables,
e.g. dimensionless renormalized correlation functions or energies 
are denoted by $\Phihqet$  in the 
effective theory and by $\Phiqcd$ in QCD. The $\Neff$ unknown
parameters can be determined from $\Phihqet_k=\Phiqcd_k\,,k=1\ldots\Neff$,
provided the sensitivity of these conditions to the desired parameters
is sufficient.  In general the determination of
$\Phiqcd_k$ will be very difficult because 
a b-quark has to be simulated and $\rmO((a\mbeauty)^2)$ cutoff effects 
will be large. The way around is once again to 
consider a finite volume, where small lattice spacings are accessible.
In practice using furthermore \SF boundary conditions is a good idea, since
then the simulations can easily be done also with dynamical fermions.
Given the experience of the tests of HQET (\sect{s:hqett}),
a good choice is $L=L_1\approx 0.4\,\fm$. Then 
$1/z = 1/(L\Mbeauty)\approx 1/10$ and the HQET expansion is very
accurate. So for the matching step we impose
\bes
  \label{e:match}
   \Phihqet_k(L_1,\Mbeauty) = \Phiqcd_k(L_1,\Mbeauty)\,, \quad k=1,\ldots,\Neff\,.
\ees
to determine the $\Neff$ parameters in the effective theory 
(right hand side of \fig{f:strat}). 
We assume that the observables $\Phi_k(L,\Mbeauty)$ have been made
dimensionless by multiplication with appropriate
powers of $L$. They should be chosen with care (e.g. 
no large momenta should appear) but the effect of variations in 
the matching conditions on the final results is in any case
of a higher order in the $\minv$ expansion.

%%%%%%%%%%%%%%%%%%%%%%%%%%%%%%%%%%%%%%%%%%%%%%%%%%%%%%%%%%%%%%%%%%%%%%%%%%%%%%%%%%%%%%%%%%
%%%%%%%%%%%%%%%%%%%%%%%%%%%%%%%%%%%%%%%%%%%%%%%%%%%%%%%%%%%%%%%%%%%%%%%%%%%%%%%%%%%%%%%%%%

\subsection{Step scaling functions}

The matching conditions, \eq{e:match}, define the HQET parameters
for any value of the lattice spacing (or equivalently
bare coupling). 
In practice, for $L_1\approx0.4\,\fm$,  
the parameters of the effective theory are then determined
at rather small lattice spacings in a range of $a\approx0.02\,\fm$
to $a\approx0.05\,\fm$.
Large volumes as they are needed to compute
the physical mass spectrum or matrix elements then require
very large lattices ($L/a > 50$). A further step is needed to bridge 
the gap to practicable lattice spacings. A well-defined procedure is 
as follows (bottom part of \fig{f:strat}).
We define step scaling
functions~\cite{alpha:sigma}, $F_k$, by
\bes
\label{e:ssfcont}
   \Phihqet_k(sL,M) = F_k(\{\Phihqet_j(L,M)\,,\,j=1\ldots\Neff\})\,,\;\; 
  %\nonumber \\\ && 
  k=1\ldots\Neff \,,
\ees
where usually one uses scale changes of $s=2$.
These dimensionless functions describe the change of
the complete set of observables $\{\Phihqet_k\}$ under a scaling of
$L\to sL$.
In order to compute them one 
\bi
\item[i] selects a lattice with a certain resolution $a/L$. 
\item[ii] The specification of $\Phihqet_j(L,M)$, $j=1,\ldots,\Nn$, then
          fixes all (bare) parameters of the theory. 
\item[iii] The l.h.s.~of \eq{e:ssfcont} is now computed,  
           keeping the bare parameters fixed
           while changing $L/a \to L'/a = sL/a$.
\item[iv]   The values for the continuum $F_k$ are reached by 
            extrapolating the resulting lattice numbers to $a/L\rightarrow 0$.
\ei
Starting from $L=L_1$ it turns out that a single step going 
to $L_2=2L_1$ is sufficient~\cite{hqet:pap4}. Then one
can switch at fixed bare parameters to a large volume
where finite size effects are negligible (left part of \fig{f:strat}). 
This is done 
in full analogy to the steps i\ldots iv above. One only computes
the large volume quantities for the bare parameters fixed in step ii
(with $L=L_2$). Again a continuum extrapolation can be carried out.

\subsection{Example: The mass of the b quark} \label{s:mb}

For illustration purposes we consider  a simple 
example, the computation of the b-quark mass, starting
from the observed (spin averaged) B-meson mass. 
For this calculation one obviously has to consider 
a range of masses $\Mbeauty$ in \eq{e:match} and determine the physical 
point from the requirement $\mbav=\mbav|_\mrm{experiment}$.
It is thus the first computation to be carried out. Subsequently,
one may directly choose the physical point in \eq{e:match}.

\subsubsection{Static approximation}

Remembering \eq{e:mb0ab}, we are after a precise
definition and calculation of 
$E_\mrm{stat}^\mrm{sub}$ and $\mb^{(0a)} = \dmstathat^\mrm{stat}+E_\mrm{stat}^\mrm{sub}$.
Already in the static approximation, a non-perturbative 
matching is required, if one wants to 
take the continuum limit. (Perturbatively one  
determines $\dmstathat^\mrm{stat}$ 
with an in the continuum limit divergent error term 
of order $\Delta(\dmstathat^\mrm{stat}) = c_{l+1} g_0^{2l+2}/a$.) 

In \eq{e:match}
we have the simple case $\Neff=1$.  We omit 
the discussion of fixing 
the bare light quark masses and coupling. 
Obviously any finite volume energy in the b-sector,
denoted by $\Gamma$, will do to fix  $\dmstathat$. Two precise
definitions are given in the following four equations.
 The reader who is only interested in the
general concept may skip this detail. 

The first choice which comes to 
mind is\cite{hqet:pap1,hqet:pap2,hqet:pap4}
\newcommand{\atxhalf}{\;\mbox{at}\;x_0=L/2\,,\;T=L}
\newcommand{\athalf}{\;\mbox{at}\;T=L/2}
\be \label{e:defgam}
  \meffav(L,\theta_0) = -{\partial_0 +  \partial_0^* \over 2} F_\mrm{av}(x_0,\theta_0) 
          \atxhalf\,,
\ee
with (see \sect{s:sfcf})
\be
 F_\mrm{av}(x_0,\theta) = {1\over4}\ln \big[-\fa(x_0,\theta)\, (\kv(x_0,\theta))^3\big]\,.
\label{e:fav}
\ee
However, at order $\minv$ the energy $\meffav$ depends on $\cahqet,\cvhqet$. This is clearly 
inconvenient and  can be avoided by choosing instead
\be\label{e:meff1}
        \meffone(L,\theta_0) = -{\partial_T +  \partial_T^* \over 2}  F_1(L,\theta_0) 
        \athalf\,,
\ee
with
\be \label{e:f1av}
  F_1(L,\theta) = {1\over 4} \ln\big[ \fone(\theta)\, (\kone(\theta))^3 \big]\,.
\ee
A spin average is taken which will be relevant when we include
the first order in $\minv$. Both $\meffav$ and $\meffone$ turn into $\mbav$ when 
$L$ and $T$ are large.

%%%%%%%%%%%%%%%%%%%%%%%%%%%%%%%%%%%%%%%%%%%%%%%%%%%%%%%%%%%%%%%%%%%%%%%%%%%%%%%%%%%%%%%%%%
\begin{figure}[tb]
\vspace{0pt}
\centerline{\includegraphics*[width=0.42\textwidth]{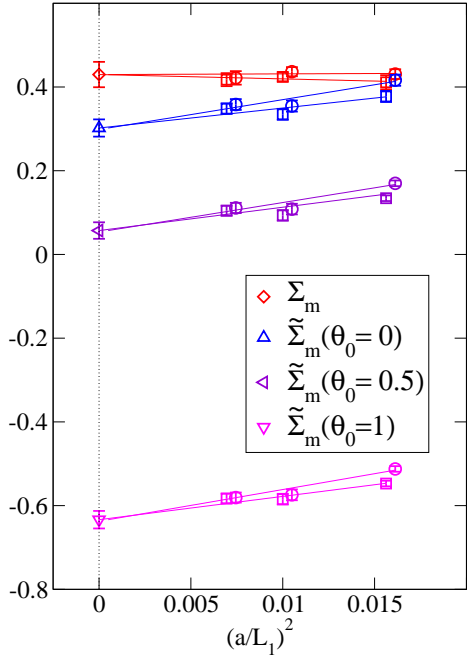}}
\caption{\footnotesize
Continuum extrapolation of $\sigmam = \lim_{a/L\to0}\Sigmam$
and $\sigmamtilde$. Two different discretizations for the 
static quark action are used and extrapolated to a common
continuum limit\protect\cite{hqet:pap4}. Dividing by $L_2$, the $y$-axes 
covers about an energy range of $300\,\MeV$.
}\label{f:sigmam}
\end{figure}
%%%%%%%%%%%%%%%%%%%%%%%%%%%%%%%%%%%%%%%%%%%%%%%%%%%%%%%%%%%%%%%%%%%%%%%%%%%%%%%%%%%%%%%%%%

In the numerical evaluation we chose $\theta_0=0$. 
We require matching, $ \Phiqcd_{1}(L_1,\Mbeauty)=\Phihqet_{1}(L_1,\Mbeauty)$, with 
\bes
  \Phiqcd_{1}(L_1,\Mbeauty)&\equiv& L_1\,\meffone(L_1,\theta_0)\,,\quad   \\
 \Phihqet_{1}(L_1,\Mbeauty)&\equiv& L_1\,(\meffonestat(L_1,\theta_0) + \mbeauty) \,.
\ees
Here  $\meffonestat$ refers to \eq{e:meff1} at the lowest order 
in $\minv$. \Eq{e:ssfcont} can then be written in the simple form,
\bes
    \Phihqet_{1}(2L,\Mbeauty) &=&   2 \Phihqet_{1}(L,\Mbeauty) + \sigmam\left(\gbar^2(L)\right)\,,
    \\ \label{e:ssfmass}
    \sigmam\left(\gbar^2(L)\right) &\equiv& 
	2L\, [\,\meffonestat(2L,\theta_0) - \meffonestat(L,\theta_0)\,]\,.
\ees
In $\sigmam$ the divergent $\dmstat$ as well as the mass shift $\mbeauty$ cancel. 
Its continuum extrapolation is illustrated in \fig{f:sigmam}.

%%%%%%%%%%%%%%%%%%%%%%%%%%%%%%%%%%%%%%%%%%%%%%%%%%%%%%%%%%%%%%%%%%%%%%%%%%%%%%%%%%%%%%%%%%
\begin{figure}[tb]
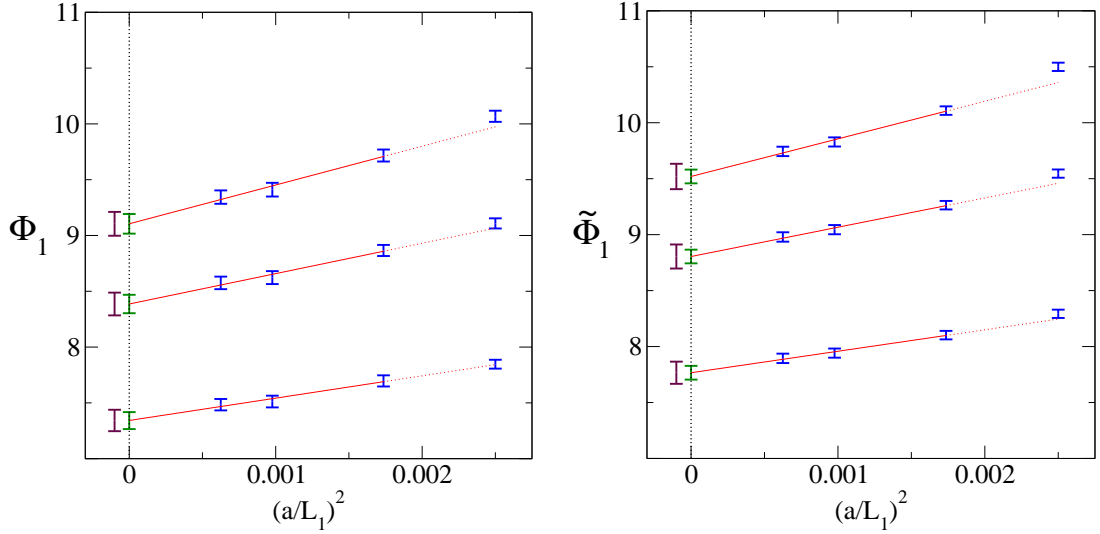

\includegraphics*[width=0.48\textwidth]{plots/Phi1}
\hspace{0.015\textwidth}
\includegraphics*[width=0.48\textwidth]{plots/Phi1tilde}
\caption{\footnotesize
Continuum extrapolation of $\Phi_1(L_1,\Mbeauty)$,
for $z=10.4\,,\; 12.1\,,\; 13.3$ from bottom to top\protect\cite{hqet:pap4}. 
On the right, the equivalent in the alternative strategy
is shown with $\theta_0=1/2$. Dividing by $L_1$, the $y$-axes 
covers about a range of $2\,\GeV$.
}\label{f:contphi1}
\end{figure}
%%%%%%%%%%%%%%%%%%%%%%%%%%%%%%%%%%%%%%%%%%%%%%%%%%%%%%%%%%%%%%%%%%%%%%%%%%%%%%%%%%%%%%%%%%

We now see immediately that 
\def\text#1{\mbox{#1}}
\bes \label{e:master}
  \mbav &=& \mb^{(0a)} +  \mb^{(0b)} + \rmO(\Lambda_\mrm{QCD}^2/\mbeauty)\\
   \mb^{(0b)}    &=& \underbrace{\Estat - \meffonestat(L_2,\theta_0)}_{a\to0 \text{ in HQET}} \\
   \mb^{(0a)}    &=& 
        \underbrace{\meffonestat(L_2,\theta_0)-\meffonestat(L_1,\theta_0)}_{a\to0 \text{ in HQET}}
       +\frac1{L_1}\underbrace{\Phiqcd_1(L_1,{ \Mb})}_
                {a\to0 \text{ in QCD} }  \,, \label{e:mb0aexp}
\ees 
where the first term in \eq{e:mb0aexp} (in dimensionless form) is given
by \eq{e:ssfmass} and
$\Estat$ is the infinite volume
energy of a B-meson in static approximation introduced earlier. 
It is often called the static binding energy. 
As indicated, the continuum
limit can be taken in each individual step; a numerical example
for the last term is shown in \fig{f:contphi1}.

\begin{figure}[tb]
\centering
\includegraphics*[width=0.5\textwidth]{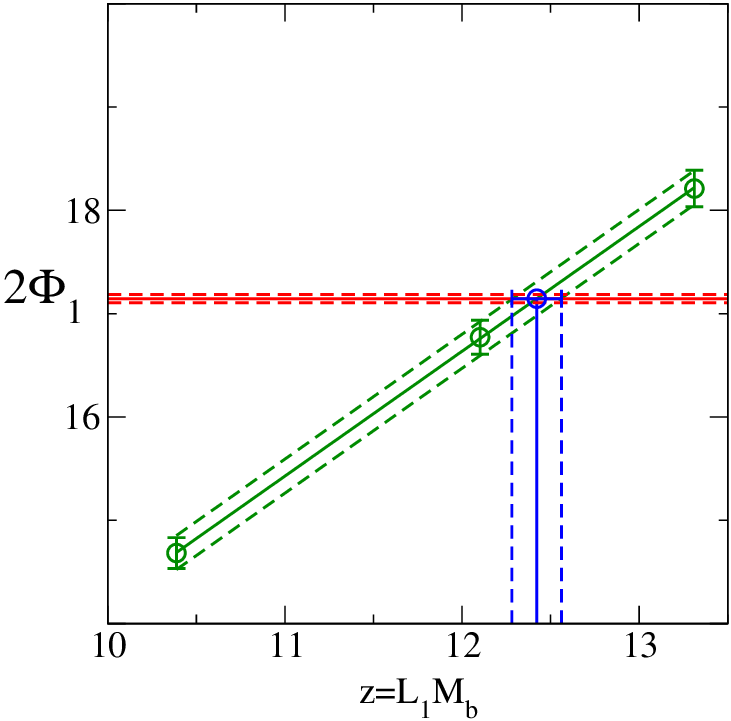}
\caption{\footnotesize
Graphical solution of \eq{e:master}. Data points are $2\Phi_1$ and
the horizontal error band is $ L_2 {\mbav} - \sigmam -  L_2 [E^\mrm{stat} - \meffstat_1(L_2,\theta_0)]$.
% On the right hand side, the analogous terms are shown for the alternative strategy.
}\label{f:staticmb}
\end{figure}

After obtaining all pieces in \eq{e:master},
the equation is numerically solved for $z_\beauty=\Mbeauty L_1$,
see \fig{f:staticmb}.
Since also the size of  $L_1$ in units of $r_0$ \cite{pot:r0} is known, 
one can quote 
\bes \label{e:Mb0}
 r_0 \Mbeauty^{(0)}= 17.25(20)\; \to\;\Mbeauty^{(0)}=  6.806(79)\, \GeV\,,
\ees
where $r_0 = 0.5\,\fm$ is used.
This result is in the quenched approximation but includes
the lowest non-trivial order in $1/\mbeauty$ and a continuum limit. 
In the form of \eq{e:mbavsplit}, we have defined the subtraction
$E_\mrm{stat}^\mrm{sub}=\meffonestat(L_2)$. The term $\mb^{(0a)}$ is then given 
entirely by the finite volume computations while $\mb^{(0b)}$ results from
the large volume $\Estat$ and $\meffonestat(L_2)$.
We proceed to discuss the $\minv$ corrections.

%%%%%%%%%%%%%%%%%%%%%%%%%%%%%%%%%%%%%%%%%%%%%%%%%%%%%%%%%%%%%%%%%%%%%%%%%%%%%%%%%%%%%%%%%%
%%%%%%%%%%%%%%%%%%%%%%%%%%%%%%%%%%%%%%%%%%%%%%%%%%%%%%%%%%%%%%%%%%%%%%%%%%%%%%%%%%%%%%%%%%

\subsubsection{Including $\minv$ corrections.}

Both the spin average $\mbav$ and the finite volume
energy $\meffone$ are constructed such that 
$\omegaspin$ drops out of their $\minv$ expansion. 
Furthermore,  $\meffone$ does not obtain any $\minv$-terms
due to the expansion of composite fields, a convenient
property of the \SF boundary fields. Thus, for
the particular problem of relating $\mbav$ to the mass
of the quark, only one more matching observable has to be defined,
to determine $\omegakin$.

The choice in \cite{hqet:pap4} was
\bes
       \Phi_2(L,\Mbeauty) &=& 
       \ratone(L,\theta_1,\theta_2) - \ratonestat(L,\theta_1,\theta_2)\,, \\
        R_1(L,\theta_1,\theta_2) &=&  F_1(L,\theta_1) - F_1(L,\theta_2) \, \athalf \,, \\
       \ratonestat(L,\theta_1,\theta_2) &=& 
       \ln \big[\fonestat(\theta_1)/\fonestat(\theta_2)\big]\,,
        \athalf
\ees
where the static piece $\ratonestat$ is subtracted such that $\Phi_2$ is proportional
to $\omegakin$. It is now a matter of simple algebra to compute the step 
scaling functions and finally the $\minv$-correction to \eq{e:Mb0}.
In the numerical evaluation three combinations 
$(\theta_1,\theta_2) \,=\, (0,1/2)\,,\,(1/2,1)\,,\,(0,1) $ were chosen.
The $\minv$-correction, split as in \eq{e:mbavsplit}, but for the quark mass, 
turned out to be about independent of $\theta_1,\theta_2$ with 
(after continuum extrapolation)
\be
   r_0\,\Mb^{(1a)}= -0.06(3) \,,\quad  r_0\,\Mb^{(1b)} = -0.06(8)\;\to \;r_0 \Mbeauty= 17.12(22).
\ee
Indeed since for our choice of matching condition the static approximation
is independent of $\theta_1,\theta_2$, the values for $r_0\,\Mb^{(1)}$
have to be independent up to small $r_0\,\Lambda_\mrm{QCD}^3/\mbeauty^2$ terms.
This is so because the result including all terms has this precision. 

With $\Lambda_\msbar r_0=0.602(48)$~\cite{alpha:SU3,pot:intermed},
the 4-loop $\beta$ function and the mass anomalous
dimension~\cite{MS:4loop1,MS:4loop2,MS:4loop3,MS:4loop4},
we translate
$\Mbeauty= \Mbeauty^{(0)}+\Mbeauty^{(1)}$
to the mass in the $\msbar$ scheme,
\bes
  \mbar_\beauty(\mbar_\beauty)= 4.347(48) \GeV\,;
  \label{e:msbarmass}
\ees
the associated perturbative uncertainty can safely be neglected.
In the $\msbar$ scheme the $\minv$ term amounts to $ -27(22)\MeV$.
\subsubsection{An alternative strategy.}

Also an alternative strategy has been tested. It is based on $\meffav$, \eq{e:defgam}.
It hence also involves the $\minv$-corrections to the currents, but only in
the combination $\cavhqet=(\cahqet+3\cvhqet)/4$. Three matching observables are 
necessary. They can be found in \cite{hqet:pap4}. Here we only mention that 
again three angles $\theta_0,\theta_1,\theta_2$ appear, but now in nine different combinations.
Together with the above discussed strategy, there are 12 different matching conditions.

\TABLE{\small
{\small \begin{center}\begin{tabular}{cccccc}
\hline\hline \\[-1.75ex]
$\theta_0$ &  $r_0\,\Mb^{(0)}$ && \multicolumn{3}{c}{$r_0\,\Mb=r_0\,(\Mb^{(0)} + \Mb^{(1a)} + \Mb^{(1b)})$} \\[1ex] 
\hline\hline
& &&
$\theta_1=0$   &  $\theta_1=1/2$ & $\theta_1=1$ \\
& &&
$\theta_2=1/2$ &  $\theta_2=1$   & $\theta_2=0$ \\
\hline
 && \multicolumn{4}{c}{Main strategy} \\
0   & 17.25(20) && 17.12(22)  & 17.12(22)  & 17.12(22) \\
\hline
 && \multicolumn{4}{c}{Alternative strategy} \\
0   & 17.05(25) && 17.25(28)  & 17.23(27)  & 17.24(27) \\
1/2 & 17.01(22) && 17.23(28)  & 17.21(27)  & 17.22(28) \\
1   & 16.78(28) && 17.17(32)  & 17.14(30)  & 17.15(30) \\
\hline\hline
\end{tabular}\end{center}}
\caption{
RGI results of $\Mb$ in the static approximation and of the $\minv$ correction
for the alternative strategy.%, in units of $r_0$.
}\label{t:Mb}

%%% Local Variables: 
%%% mode: latex
%%% TeX-master: "Nara"
%%% End: 
 
}

At lowest order in $\minv$, the equations relating quark mass and 
the spin-averaged B-meson mass remain unchanged.
Only the step scaling function $\sigmam$ is replaced by
$\sigmamtilde$, and $\Phi_1$ is replaced by $\tilde\Phi_1$.
Both depend on the angle $\theta_0$, but not
on $\theta_1,\theta_2$. 
A comparison is shown in \fig{f:sigmam}, \fig{f:contphi1}.
Dividing by $L_1$ and $L_2$ respectively, 
one observes differences of about $250\,\MeV$ in these quantities. 
According to the HQET counting they are expected to be of order
$\Lambda_\mrm{QCD}$. 
In the combination, which leads to $\Mbeauty^{(0)}$, these differences
are reduced to $\rmO(\Lambda_\mrm{QCD}^2/\mbeauty)$.

In % \tab{t:Mb}
the table shown, such differences are barely visible in the lowest order
results, where one would estimate $\rmO(\Lambda_\mrm{QCD}^2/\mbeauty^2)\sim1\%$ effects. 
After including the $\minv$ corrections no  
signs of differences remain. Indeed one would estimate them to be
$\rmO((\Lambda_\mrm{QCD}/\mbeauty)^3) \sim 0.1\%$, quite below our statistical 
precision. We conclude that the expansion works very well, as expected.

We finally note that the overall uncertainty of the b-quark mass is 
dominated by the renormalization
factors in relativistic QCD, i.e. by the relation between the bare mass and 
the RGI mass
in QCD. The second most important contribution to the errors is the
large volume $\Ekin$. This quantity contains a quadratic divergence. 
Once $\Ekin^\mrm{sub}$ is subtracted, the remaining finite piece
has errors which grow fast as one approaches the continuum limit. 
However, the methods of \cite{alltoall:dublin} have not yet been applied 
and we expect to reduce this source of error.

%%% Local Variables: 
%%% mode: latex
%%% TeX-master: "Nara"
%%% End: 

\section{More literature  \label{s:hqetlit}}

Before summarizing and discussing the perspectives for
HQET on the lattice, let us attempt to give a brief guide to literature
on subjects which we did not cover.

Early numerical computations have been summarized in 
reviews~\cite{reviews:beauty,reviews:wittig97,reviews:FS}.
The power divergent mixing of operators of different
dimensions was seen in an explicit perturbative
computation~\cite{stat:MMS}. The $\rm B^*-B$ mass splitting,
which is a $\minv$ effect given at lowest order
by $\Espin$ has been investigated with perturbative
$\omegaspin$ by~\cite{hqet:spinsplitt_boch,stat:ospin_flynn,stat:ukqcd,hqet:spinsplitt_gms}.

In static approximation, the four fermion operators responsible
for $\rm B-\overline{B}$ mixing with Wilson fermions 
require the subtraction of 
operators with different chiralities, which was attempted with
perturbatively estimated mixing coefficients~\cite{bbar:interpol_roma1,stat:ukqcd}.
Later it was realized that the mixing is completely avoided with an action with exact chiral
symmetry (and spin symmetry)~\cite{stat:chirsymm} and a calculation of
$\rm B-\overline{B}$ mixing was carried out with perturbative 
renormalization~\cite{stat:bbar_orsay}.

The dependence of heavy-light meson properties on the mass of the
light quark can be described by a suitable chiral effective theory \cite{chpt:b1,chpt:b2}.
Its Lagrangian involves the $\rm B^*B\pi$ coupling as a low energy constant.
This was estimated in a static computation~\cite{stat:ghat}. Also an exploratory
computation of the 
Isgur-Wise functions $\tau_{1/2},\tau_{3/2}$ at zero recoil
has been carried out~\cite{stat:taus,stat:derop}. These functions
give the form factors for transitions between heavy-light mesons
of different orbital angular momentum.

Finally, the formulation of HQET at a finite velocity has
been investigated by Aglietti et al. We refer to \cite{hqet:agliettif,hqet:agliettiff} 
and references therein. Both perturbative investigations of the 
renormalization~\cite{hqet:CDM,stat:derop} and numerical 
simulations~\cite{hqet:MO,stat:taus} have been carried out.

%%% Local Variables: 
%%% mode: latex
%%% TeX-master: "Nara"
%%% End: 

\section{Summary and perspectives \label{s:hqetp}}

Non-perturbative HQET at the leading order in $\minv$ has reached a 
satisfactory status. To underline this statement, we comment briefly 
on the progress made in recent years. 
The ground breaking work of 
\cite{stat:eichten,fbstat:old1,fbstat:old2} was followed by intense activity 
leading to a computation where (within the quenched approximation) 
all error sources were investigated and controlled as well as possible 
in the middle of the nineties~\cite{stat:fnal2}. 
For the $B$ decay constant, the authors estimated 
three sources of errors between 7\% and 12\% each, one of them due to 
an estimated precision of the perturbative renormalization. 

It took almost 
a decade until a non-perturbative method for the renormalization in the 
effective theory was fully developed~\cite{zastat:pap1,zastat:pap3}, but meanwhile
a total error of 4\% in the B$_\strange$-meson decay constant has been 
reached~\cite{lat06:damiano}, including
the continuum extrapolation. In reaching this accuracy 
also the reduction of  
statistical errors~\cite{stat:letter,stat:actpaper} was relevant. 

Also the standard model B-parameter for B-$\overline{\mrm{B}}$ mixing is well
on its way~\cite{stat:zbb_pert,stat:zbb_nf0,lat06:carlos}.
In these cases, bare perturbation theory is now avoided 
by non-perturbative renormalization of the lattice operators.
As explained in \sect{s:fb}, all static-light bilinears require only
one common renormalization factor $\zastatrgi$. As this is
known~\cite{zastat:pap3}, also semi-leptonic decay form factors
can be computed with non-perturbative renormalization. 
For completeness we note that a source of a perturbative error remains in
$\Cps$, \fig{f:cps}, and its relatives for other bilinears,
but here high order continuum perturbation
theory is available~\cite{Ji:1991pr,BroadhGrozin2,Gimenez:1992bf,ChetGrozin} 
and is applied at the b-scale. This error 
is under reasonable control. 
Applying these methods 
to the theory with dynamical fermions is straight forward; ''only''
the usual problems of simulations with light quarks have to be solved.

By themselves such lowest order (in $\minv$) results are not expected to
have an interesting precision for phenomenological applications,
but certainly they can constrain the large mass behavior computed
with other methods \cite{fb_wupp,reviews:beauty,reviews:hartmut97,El-Khadra:1998hq,Aoki:1998ji,Bernard:1998xi,heavylight:Bec98,AliKhan:2000eg,Bowler:2000xw,Lellouch:2000tw,lat01:ryan,romeII:fb,romeII:mb}.
An interesting application has been the combination of 
the approach of \cite{romeII:fb,romeII:mb} with the large mass behavior
computed via HQET \cite{lat06:damiano}. 

%% A relevant technical advance has been the realization
%% that a change of the regularization details allows to
%% achieve much better statistical errors in HQET, while 
%% keeping the discretization errors small \cite{stat:letter,stat:actpaper}.

The $1/\mbeauty$ corrections can also be
computed directly in the effective theory. Here, the 
necessary steps have  been carried out in detail
for the mass of the b-quark in the quenched approximation.
The resulting precision is rather satisfactory and
higher order $\minv$-corrections can be neglected.
The latter has also been verified explicitly 
by comparing the results following from a number
of different matching conditions. No significant differences
were found.

An extension of this computation to full QCD has been
started by the ALPHA collaboration. 
This is of particular interest because for full QCD
it is of course more difficult to reach the small lattice
spacings needed for computations with relativistic  
quarks of masses around $m_\mrm{charm}$ -- the basis 
of the alternatives mentioned above. 
It is time to address other
observables such as the B-meson decay constant in this 
direct approach!

%%% Local Variables: 
%%% mode: latex
%%% TeX-master: "Nara"
%%% End: 

\vspace{10mm}

\noindent
{\bf Acknowledgment.} 
I am grateful to the organizers of this school for composing
a very interesting programme and a pleasant atmosphere
In particular I thank Y. Kuramashi for his efforts and his patience. 
I would like to thank my friends in the 
ALPHA-collaboration for the enjoyable
and fruitful collaboration and all I learned from them. 
In particular it is worth emphasizing that the lecture
on HQET is based on joint work with 
M.~Della~Morte, N.~Garron, J.~Heitger and M.~Papinutto. 
I am grateful to N.~Garron, J.~Heitger and S.~Takeda
for sending figures or data to be used in this
writeup and to M.~Della~Morte, D.~Guazzini and
H.~Simma and U.~Wolff for comments on the manuscript.
I finally thank NIC/DESY for allocating computer 
time for the ALPHA-projects, which was essential for most of 
the numerical investigations that were discussed in these lectures.

\bibliographystyle{nara}
\bibliography{nara}

\end{document}